\pdfoutput=1
\documentclass[acmsmall,screen]{acmart}

\newif\iftechreport
\techreporttrue

\usepackage{bm}

\setcitestyle{number}

\usepackage{xargs}
\usepackage{tabls}
\usepackage{varwidth}
\usepackage{textcomp}
\usepackage[T1]{fontenc}
\usepackage{enumitem}
\setlist[itemize]{leftmargin=2em}
\usepackage{chngcntr}
\usepackage{wasysym}
\usepackage{wrapfig}
\usepackage{graphicx,amssymb,calc}
\usepackage{pifont}
\usepackage{etoolbox}
\usepackage{thmtools}
\usepackage{thm-restate}
\usepackage{accents}
\usepackage[utf8]{inputenc}
\usepackage{multirow}
\usepackage{hhline}
\usepackage{minibox}
\usepackage{xparse}
\usepackage{xstring}
\usepackage{xspace}
\usepackage[f]{esvect}  % vector, option a
\usepackage{parskip}
\usepackage{mathtools}
\usepackage{changepage}
\usepackage{relsize}

\usepackage{listings}% http://ctan.org/pkg/listings
\usepackage{algorithm}
\usepackage[noend]{algpseudocode}
\usepackage{ebproof}
\usepackage{setspace,array}
\usepackage{lscape}
\usepackage{caption}
\usepackage{wasysym}
\usepackage{color}   %May be necessary if you want to color links
\usepackage{xcolor,colortbl}
\usepackage{hyperref}
\usepackage{cleveref}
\usepackage{nameref}
\usepackage{tikz}
\usetikzlibrary{calc,shapes.multipart,chains,arrows,positioning}
\usetikzlibrary{decorations.pathreplacing}

\usepackage[colorinlistoftodos,prependcaption,textsize=tiny]{todonotes}

\definecolor{timberwolf}{rgb}{0.86, 0.84, 0.82}
\definecolor{shadow}{rgb}{0.54, 0.47, 0.36}
\definecolor{sepia}{rgb}{0.44, 0.26, 0.08}
\definecolor{sanddune}{rgb}{0.59, 0.44, 0.09}
\definecolor{pastelbrown}{rgb}{0.51, 0.41, 0.33}
\definecolor{goldenbrown}{rgb}{0.6, 0.4, 0.08}
\definecolor{palebrown}{rgb}{0.6, 0.46, 0.33}

\definecolor{Gray}{gray}{0.85}
\definecolor{LightCyan}{rgb}{0.88,1,1}
\newcolumntype{g}{>{\columncolor{Gray}}c}
\newcolumntype{w}{>{\columncolor{white}}c}
\newcolumntype{L}{>{\columncolor{Gray}}l}
\newcolumntype{C}{>{\columncolor{Gray}}c}
\newcolumntype{M}{>{\columncolor{Gray}}m}

\newcommand{\code}[1]{\ensuremath{\mathtt{#1}}}

\newcommand{\textcomment}[1]{\ensuremath{\color{palebrown}{\text{#1}}}}
\newcommand{\textspecs}[1]{\ensuremath{\color{OliveGreen}{\text{#1}}}}
\newcommand{\codeLineNum}[1]{\scriptsize{\ensuremath{\mathsf{{\color{pastelbrown} #1}}}}}

\newcommand{\overbar}[1]{
  \mkern 1.5mu\overline{\mkern-1.5mu#1\mkern-1.5mu}\mkern 1.5mu}

\newcommand{\forms}[2]{\ensuremath{#1 {\mapsto} #2}}
\newcommand{\formss}[2]{\ensuremath{#1 {\scalebox{0.8}[1]{$\mapsto$}} #2}}
\newcommand{\formst}[3]{\ensuremath{{#2}
    {\overset{\mkern-2mu\hspace{-1pt}#1}\mapsto}{#3}}}

\newcommand{\formr}[2]{\ensuremath{{#1}{(#2)}}} % predicate
\newcommand{\formp}[2]{\ensuremath{#1(#2)}} % predicate
\newcommand{\rel}[1]{\ensuremath{\mathit{#1}}}
\newcommand{\proc}[1]{\ensuremath{\mathsf{#1}}}
\newcommand{\validlem}[1]{\ensuremath{{valid}(#1)}}
\newcommand{\validindt}[1]{\ensuremath{{valid}_{\mathsf{ID}}(#1)}}
\newcommand{\hasUnk}[1]{\ensuremath{{hasUnk}(#1)}}

\newcommand{\evalVar}[2]{\ensuremath{{#2}({#1})}}
\newcommand{\evalForm}[2]{\ensuremath{\llbracket {#1} \rrbracket_{#2}}}
\newcommand{\semPred}[1]{\ensuremath{\llbracket {#1} \rrbracket}}
\newcommand{\modelExtOne}[3]{\ensuremath{[{#1}|#2{:}#3]}}

\def\setempty{\ensuremath{\mathord{\varnothing}}}
\newcommand{\setenum}[1]{\{ #1 \}}

\newcommand{\hboxauto}[1]{         % hbox with autowidth
  \hbox{\begin{varwidth}{\textwidth} #1 \end{varwidth}}}

\def\entails{\mathrel{
    \raisebox{0.1em}{\scalebox{1}[0.85]{\ensuremath{|}}}
    \mkern-3mu\mkern-1mu
    \raisebox{0.068em}{\scalebox{0.95}[1]{\ensuremath{-}}}}}
\def\satisfies{\mathrel{
    \raisebox{0.1em}{\scalebox{1}[0.85]{\ensuremath{|}}}
    \mkern-2.7mu\mkern-1mu
    \raisebox{0.06em}{\scalebox{0.95}[1]{\ensuremath{=}}}}}

\def\synequiv{\mathbin{\cong}}

\def\ltpred{\mathbin{\prec}}

\def\lttau{\mathbin{\prec_{\tau}}}

\def\sminus{\scalebox{0.75}[1.0]{$-$}}
\def\sequal{\scalebox{0.9}[1.0]{$=$}}

\def\defrightarrow{\mathbin{\stackrel{
      \mathclap{\mkern-0mu\mbox{$\scriptscriptstyle \mathrm{def}$}}}{
      \Rightarrow}}}
\def\defeq{\mathbin{\stackrel{\mathclap{\mbox{\normalfont\tiny def}}}{=}}}
\def\definedas{\mathrel{\stackrel{\makebox[0pt]{\mbox{\normalfont\tiny def}}}{=}}}
\def\denotedas{\mathrel{~\triangleq~}}

\newcommand{\val}[1]{\ensuremath{\mathsf{#1}}}                % constant, value

\def\midwedge{\mathrel{\ensuremath{\scalebox{1}{$\bigwedge$}}}}  % medium size wedge
\def\midsum{\mathrel{\ensuremath{\scalebox{1.06}{$\sum$}}}}  % medium size
\def\hunions{\circ}  % union of two heap functions
\def\disjoins{\mathrel{\#}}   % two heap functions are disjoint

\def\mult/{\ensuremath{\hspace{0.1em}{\cdot}\hspace{0.1em}}}

\makeatletter
\newcommand{\algComment}[1]{{\small \textrm{/\!/}\textit{#1}}}
\newcommand{\algCommentColor}[2]{{\color{#1}{//\,\textit{#2}}}}
\algnewcommand{\LineComment}[1]{\Statex \hskip\ALG@thistlm \algComment{#1}}
\algnewcommand{\LineCommentIndent}[1]{\Statex \hskip\ALG@thistlm \hskip\algorithmicindent \algComment{#1}}
\algnewcommand{\CommentColor}[2]{\hfill \algCommentColor{#1}{#2}}
\algnewcommand{\CommentNofillColor}[2]{\quad\algCommentColor{#1}{#2}}
\algnewcommand{\CommentLn}[1]{\Statex \hskip\ALG@thistlm \algComment{#1}}
\algnewcommand{\CommentLnColor}[2]{\Statex \hskip\ALG@thistlm \algCommentColor{#1}{#2}}
\algnewcommand{\CommentLnIndent}[1]{\Statex \hskip\ALG@thistlm \hskip\algorithmicindent \algComment{#1}}
\algnewcommand{\CommentLnIndentColor}[2]{\Statex \hskip\ALG@thistlm \hskip\algorithmicindent \algCommentColor{#1}{#2}}

\algrenewcommand{\algorithmiccomment}[1]{\hfill \algComment{#1}}
\algrenewcommand\textproc{}

\algrenewcommand\alglinenumber[1]{\hspace{-2em}\scriptsize #1:}
\makeatother

\algrenewcommand\algorithmicindent{1em}  %% redefine indentation size

\newcommand{\Keyword}[1]{\ensuremath{\mathrel{\mathbf{#1}}}}

\newcommand{\Summary}[1]{\textsl{#1}}

\newcommand{\Assign}[2]{\ensuremath{#1}~\ensuremath{\leftarrow}~\ensuremath{#2}}
\algdef{Se}[IF]{If}{EndIf}[1]{\algorithmicif\ \ensuremath{#1}\ \algorithmicthen}{}%
\algdef{Se}[FOR]{For}{EndFor}[1]{\algorithmicfor\ \ensuremath{#1}\ \algorithmicdo}{}%
\renewcommand\Return[1]{\Keyword{return} \ensuremath{#1}}
\renewcommand{\Call}[2]{\proc{#1}\ensuremath{(#2)}}             % function call
\newcommand{\ReturnTuple}[1]{\Keyword{return} \ensuremath{(#1)}}

\newcommand{\rulename}[1]{$#1$}

\newcommand{\rulesidecond}[1]{    % side condition
  \!\!\footnotesize$#1$}

\def\mtimply{\mathrel{\rightarrow}}

\def\pureImply{\mathrel{\rightarrow}}

\newcounter{myequationNo}

\newcommand{\PrintMathFonts}{%
  \typeout{*** Math fonts list ***}
  \count255=0
  \loop\ifnum\count255<16
  \typeout{(\the\count255: \the\textfont\count255 =
  \fontname\textfont\count255)}
  \advance\count255 by 1
  \repeat
  \typeout{***}}

\newcommand{\assumptions}[1]{\ensuremath{\code{Asms}(#1)}}
\newcommand{\premises}[1]{\ensuremath{\code{Prems}(#1)}}
\newcommand{\conclusion}[1]{\ensuremath{\code{Concl}(#1)}}
\newcommand{\sound}[1]{\ensuremath{sound(#1)}}

\newenvironment{verify}
 {\par\addvspace{\topsep}
  \centering
  \begin{minipage}{\linewidth}
  \hrule\kern2pt}
 {\par\kern2pt\hrule
  \end{minipage}
  \par\addvspace{\topsep}}

\def\nil{\mathord{\val{nil}}}

\newcommand{\ValidWitness}[1]{\ensuremath{\textsc{Valid}\langle #1 \rangle}}
\def\Unknown{\ensuremath{\textsc{Unkn}}}

\def\true{\mathord{\mathit{true}}}
\def\false{\mathord{\mathit{false}}}

\def\emp{\mathord{\mathsf{emp}}}
\def\POne{\mathord{\pred{P_1}}}
\def\PTwo{\mathord{\pred{P_2}}}
\def\P{\mathord{\mathsf{P}}}
\def\Q{\mathord{\mathsf{Q}}}
\def\rUnk{\mathord{\rel{U}}}
\def\rUnkSol{\mathord{\overbar{U}}}
\def\formF{\mathord{F}}
\def\POne{\mathord{\P_1}}
\def\PTwo{\mathord{\P_2}}

\def\Ls{\mathord{\mathsf{ls}}}
\def\LsRev{\mathord{\mathsf{lsrev}}}

\def\Dll{\mathord{\mathsf{dll}}}
\def\DllRev{\mathord{\mathsf{dllrev}}}
\def\ListO{\mathord{\mathsf{ListO}}}
\def\ListE{\mathord{\mathsf{ListE}}}

\def\sepAnte{,}

\def\ruleHypo{\mathord{\mathsf{IH}}}
\def\ruleLemmaLeft{\mathord{\mathsf{LM}_\mathrm{L}}}
\def\ruleLemmaRight{\mathord{\mathsf{LM}_\mathrm{R}}}
\def\ruleInduction{\mathord{\mathsf{ID}}}
\def\rulePureEntail{\mathord{\entails_{\mathrm{\Pi}}}}
\def\ruleFalseLeftOne{\mathord{\bot^1_{\text{L}}}}
\def\ruleFalseLeftTwo{\mathord{\bot^2_{\text{L}}}}
\def\ruleStarData{\mathord{*{\mapsto}}}
\def\ruleStarPred{\mathord{*{\P}}}
\def\ruleEmpLeft{\mathord{\mathsf{E}_{\text{L}}}}
\def\ruleEmpRight{\mathord{\mathsf{E}_{\text{R}}}}

\def\ruleExistsLeft{\mathord{\exists_{\text{L}}}}
\def\ruleExistsRight{\mathord{\exists_{\text{R}}}}

\def\rulePredLeft{\mathord{\P_{\text{L}}}}
\def\rulePredRight{\mathord{\P_{\text{R}}}}
\def\ruleEqualLeft{\mathord{=_{\text{L}}}}

\def\ruleCaseAnalysis{\mathord{\mathsf{CA}}}

\def\ruleSynPiOne{\mathord{\mathsf{U}^1_{\mathrm{\Pi}}}}
\def\ruleSynPiTwo{\mathord{\mathsf{U}^2_{\mathrm{\Pi}}}}
\def\ruleSynSigmaOne{\mathord{\mathsf{U}^1_{\mathrm{\Sigma}}}}
\def\ruleSynSigmaTwo{\mathord{\mathsf{U}^2_{\mathrm{\Sigma}}}}
\def\ruleSynHypo{\mathord{\mathsf{U}_{\text{IH}}}}

\def\sInt{\mathord{\code{Int}}}

\def\sSort{\mathord{\code{Sort}}}
\def\sVar{\mathord{\code{Var}}}
\def\sVal{\mathord{\code{Val}}}
\def\Loc{\mathord{\code{Loc}}}

\def\SLUID{\ensuremath{\mathord{\mathrm{SL}^{\mathsf{U}}_{\mathsf{ID}}}}}

\newcommand{\freevars}[1]{\ensuremath{\mathtt{FV}(#1)}}
\newcommand{\funcDom}[1]{{\ensuremath{\mathrm{dom}(#1)}}}

\def\vL{\ensuremath{\mathord{L}}}
\def\vR{\ensuremath{\mathord{r}}}

\def\vRule{\mathord{\mathsf{R}}}
\def\vRuleInst{\mathord{\mathsf{\overbar{R}}}}
\def\sRule{\mathord{\mathcal{R}}}
\def\sRuleSelected{\mathord{\overbar{\mathcal{R}}}}
\def\Assumpts{\ensuremath{\mathord{\mathcal{A}}}}

\def\varSort{\iota}
\def\A{\mathord{\mathcal{A}}}

\def\Lemmas{\mathord{\mathcal{L}}}
\def\LemmasSyn{\mathord{\mathcal{L}_\mathrm{syn}}}
\def\Templates{\mathord{\mathcal{T}}}

\def\Hypos{\mathord{\mathcal{H}}}
\def\HyposPrim{\mathord{\mathcal{H}'}}
\def\Lemmas{\mathord{\mathcal{L}}}

\def\Unify{\proc{Unify}}
\def\CreateUnknownRelation{\proc{CreateUnknownRelation}}
\def\CreateWitnessProofTree{\proc{CreateWitnessProofTree}}
\def\Prove{\proc{Prove}}

\def\ProofMode{mode}
\def\res{res}
\def\SynLemma{\textsc{SynLM}}
\def\NoSynLemma{\textsc{NoSyn}}
\def\Assumptions{\proc{Assumptions}}

\def\Premises{\proc{Premises}}
\def\Prove{\proc{Prove}}
\def\SynthesizeLemma{\proc{SynthesizeLemma}}
\def\CreateLemmaTemplate{\proc{CreateTemplate}}
\def\FreeVars{\proc{FreeVars}}
\def\NeedLemmas{\proc{NeedLemmas}}
\def\IsSynthesisRule{\proc{IsSynthesisRule}}
\def\IsAxiomRule{\proc{IsAxiomRule}}

\def\Preprocess{\proc{Preprocess}}
\def\RefineAnte{\proc{RefineAnte}}
\def\RefineConseq{\proc{RefineConseq}}
\def\FineTuneConseq{\proc{FineTuneConseq}}

\def\Solve{\proc{Solve}}
\def\Invariant{\proc{Invariant}}

\def\slide{\textbf{Slide}\xspace}
\def\spen{\textbf{Spen}\xspace}
\def\sleek{\textbf{Sleek}\xspace}
\def\cyclist{\textbf{Cyclist}\xspace}
\def\songbird{\textbf{Songbird}\xspace}

\def\songbirdLS{\textbf{\textsc{SLS}}\xspace}
\def\songbirdMI{\textbf{Songbird}\xspace}

\def\slideAbr{\textsc{Sld}}
\def\spenAbr{\textsc{Spn}}
\def\sleekAbr{\textsc{Slk}}
\def\cyclistAbr{\textsc{Ccl}}
\def\songbirdMiAbr{\textsc{Sbd}}
\def\songbirdLsAbr{\textsc{Sls}}

\def\songbirdLSURL{\url{https://songbird-prover.github.io/lemma-synthesis}}

\usepackage{booktabs}
\usepackage{subcaption}

\setcopyright{rightsretained}
\acmPrice{}
\acmDOI{10.1145/3158097}
\acmYear{2018}
\copyrightyear{2018}
\acmJournal{PACMPL}
\acmVolume{2}
\acmNumber{POPL}
\acmArticle{9}
\acmMonth{1}

\begin{document}

\title[Automated Lemma Synthesis in Symbolic-Heap Separation Logic]
{Automated Lemma Synthesis in Symbolic-Heap\\Separation Logic}

%%%%%%%%%%%%%%%%%%%%%%%%%%%%%%%%%%%%%%%%%%%%%%%%
\author{Quang-Trung Ta}
\authornote{\,The first and the second author contributed equally.}
\affiliation{
  \department{Department of Computer Science}
  \institution{School of Computing, National University of Singapore}
  \country{Singapore}
}
\email{taqt@comp.nus.edu.sg}

\author{Ton Chanh Le}
\affiliation{
  \department{Department of Computer Science}
  \institution{School of Computing, National University of Singapore}
  \country{Singapore}
}
\email{chanhle@comp.nus.edu.sg}

\author{Siau-Cheng Khoo}
\affiliation{
  \department{Department of Computer Science}
  \institution{School of Computing, National University of Singapore}
  \country{Singapore}
}
\email{khoosc@comp.nus.edu.sg}

\author{Wei-Ngan Chin}
\affiliation{
  \department{Department of Computer Science}
  \institution{School of Computing, National University of Singapore}
  \country{Singapore}
}
\email{chinwn@comp.nus.edu.sg}

\begin{abstract}
  The symbolic-heap fragment of separation logic has been actively
  developed and advocated for verifying the memory-safety property of
  computer programs. At present, one of its biggest challenges is to
  effectively prove entailments containing inductive heap predicates. These
  entailments are usually proof obligations generated when verifying
  programs that manipulate complex data structures like linked lists,
  trees, or graphs.

  To assist in proving such entailments, this paper introduces a
  \emph{lemma synthesis framework}, which automatically discovers lemmas to
  serve as eureka steps in the proofs. Mathematical induction and
  template-based constraint solving are two pillars of our framework. To
  derive the supporting lemmas for a given entailment, the framework
  firstly identifies possible lemma templates from the entailment's heap
  structure. It then sets up unknown relations among each template's
  variables and \emph{conducts structural induction proof} to generate
  constraints about these relations. Finally, it \emph{solves the
    constraints} to find out actual definitions of the unknown relations,
  thus discovers the lemmas. We have integrated this framework into a
  prototype prover and have experimented it on various entailment
  benchmarks. The experimental results show that our
  lemma-synthesis-assisted prover can prove many entailments that could not
  be handled by existing techniques. This new proposal opens up more
  opportunities to automatically reason with complex inductive heap
  predicates.
\end{abstract}

%%% Local Variables:
%%% mode: latex
%%% TeX-master: "main"
%%% End:




\maketitle

%%==============

%%%%%%%%%%%%%%%%%%%%%%%%%%%%%%%%%%%%%%%%%%%%%%%%%%%%%%%%%%%%%%
%%%%%%%%%%%%%%%%%%%%%%%%%%%%%%%%%%%%%%%%%%%%%%%%%%%%%%%%%%%%%%
%%%%% RULES

\newcommand\minipagePureEntail{\hbox{
    \begin{prooftree}[label separation = 0.3em]
      \hypo{}
      \infer
      [left label = \rulename{\rulePureEntail}]
      1
      [\rulesidecond{\Pi_1 {\pureImply} \Pi_2}]
      {\Hypos \sepAnte\, \Lemmas \sepAnte\, \Pi_1 \entails \Pi_2}
    \end{prooftree}}}

%% ~~~~~~~~~~~~~~~~~~~~~~~~~~~~~~~~~~
\newcommand\minipageFalseLeftOne{\hbox{
    \begin{prooftree}[label separation = 0.3em]
      \hypo{}
      \infer
      [left label = \rulename{\ruleFalseLeftOne}]
      1
      {\Hypos \sepAnte\, \Lemmas \sepAnte\,
        F_1 * \formst{\varSort_1}{u}{\vec{v}} *
        \formst{\varSort_2}{u}{\vec{t}} \entails F_2 }
    \end{prooftree}}}

%% ~~~~~~~~~~~~~~~~~~~~~~~~~~~~~~~~~~
\newcommand\minipageFalseLeftTwo{\hbox{
    \begin{prooftree}[label separation = 0.3em]
      \hypo{}
      \infer
      [left label = \rulename{\ruleFalseLeftTwo}]
      {1}
      [\rulesidecond{\Pi_1 {\pureImply} \false}]
      {\Hypos \sepAnte\, \Lemmas \sepAnte\, F_1 \wedge \Pi_1 \entails F_2}
    \end{prooftree}}}

%% ~~~~~~~~~~~~~~~~~~~~~~~~~~~~~~~~~~
\newcommand\minipageEqualLeft{\hbox{
    \begin{prooftree}[label separation = 0.3em]
      \def\ScoreOverhang{0em}
      \hypo{\Hypos \sepAnte\, \Lemmas \sepAnte\, F_1[u/v] \entails F_2 [u/v]}
      \infer
      [left label = \rulename{\ruleEqualLeft}]
      {1}
      {\Hypos \sepAnte\, \Lemmas \sepAnte\, F_1 \wedge u{=}v \entails F_2}
    \end{prooftree}}}

%% ~~~~~~~~~~~~~~~~~~~~~~~~~~~~~~~~~~
\newcommand\minipageExistsLeft{\hbox{
    \begin{prooftree}[label separation = 0.3em]
      \hypo{\Hypos \sepAnte\, \Lemmas \sepAnte\,
        F_1 [u/x] \entails F_2}
      \infer
      [left label = \rulename{\ruleExistsLeft}]
      {1}
      [\rulesidecond{u \,{\not\in}\, \freevars{F_2}}]
      {\Hypos \sepAnte\, \Lemmas \sepAnte\, \exists x. F_1 \entails F_2}
    \end{prooftree}}}

%% ~~~~~~~~~~~~~~~~~~~~~~~~~~~~~~~~~~
\newcommand\minipageExistsRight{\hbox{
    \begin{prooftree}[label separation = 0.3em]
      \hypo{\Hypos \sepAnte\, \Lemmas \sepAnte
        F_1 \entails \exists \vec{x}.F_2[u/v]}
      \infer
      [left label = \rulename{\ruleExistsRight}]
      {1}
      {\Hypos \sepAnte\, \Lemmas \sepAnte\,
        F_1 \entails \exists \vec{x},v.(F_2 \wedge u{=}v)}
    \end{prooftree}}}

%% ~~~~~~~~~~~~~~~~~~~~~~~~~~~~~~~~~~
\newcommand\minipageEmpLeft{\hbox{
    \begin{prooftree}[label separation = 0.3em]
      \hypo{\Hypos \sepAnte\, \Lemmas \sepAnte\,
        F_1 \entails F_2}
      \infer
      [left label = \rulename{\ruleEmpLeft}]
      {1}
      {\Hypos \sepAnte\, \Lemmas \sepAnte\,
        F_1 * \emp \entails F_2}
    \end{prooftree}}}

%% ~~~~~~~~~~~~~~~~~~~~~~~~~~~~~~~~~~
\newcommand\minipageEmpRight{\hbox{
    \begin{prooftree}[label separation = 0.3em]
      \hypo{\Hypos \sepAnte\, \Lemmas \sepAnte\,
        F_1 \entails \exists \vec{x}. F_2}
      \infer
      [left label = \rulename{\ruleEmpRight}]
      {1}
      {\Hypos \sepAnte\, \Lemmas \sepAnte\,
        F_1 \entails \exists \vec{x}. (F_2 * \emp)}
    \end{prooftree}}}

%% ~~~~~~~~~~~~~~~~~~~~~~~~~~~~~~~~~~
\newcommand\minipageCaseAnalysis{\hbox{
    \begin{prooftree}[label separation = 0.3em]
      \hypo{\Hypos \sepAnte\, \Lemmas \sepAnte\,
        F_1 \wedge \Pi \entails F_2}
      \hypo{\Hypos \sepAnte\, \Lemmas \sepAnte\,
        F_1 \wedge \neg\Pi \entails F_2}
      \infer
      [left label = \rulename{\ruleCaseAnalysis}]
      {2}
      {\Hypos \sepAnte\, \Lemmas \sepAnte\,
        F_1 \entails F_2}
    \end{prooftree}}}

%% ~~~~~~~~~~~~~~~~~~~~~~~~~~~~~~~~~~
\newcommand\minipageStarData{\hbox{
    \begin{prooftree}[label separation = 0.3em]
      \def\ScoreOverhang{0em}
      \hypo{\Hypos \sepAnte\, \Lemmas \sepAnte\,
        F_1 \entails \exists \vec{x}.
        (F_2 \wedge u{=}t \wedge \vec{v}{=}\vec{w})}
      \infer
      [left label = \rulename{\ruleStarData}]
      1
      [\rulesidecond{u \,{\not\in}\, \vec{x},
          \vec{v} \,{\disjoins}\, \vec{x}} ]
      {\Hypos \sepAnte\, \Lemmas \sepAnte\,
        F_1 * \formst{\varSort}{u}{\vec{v}}
        \entails
        \exists \vec{x}.(F_2 * \formst{\varSort}{t}{\vec{w}})}
    \end{prooftree}}}

%% ~~~~~~~~~~~~~~~~~~~~~~~~~~~~~~~~~~
\newcommand\minipageStarPred{\hbox{
    \begin{prooftree}[label separation = 0.3em]
      \hypo{\Hypos \sepAnte\, \Lemmas \sepAnte\,
        F_1 \entails
        \exists \vec{x}. (F_2 \wedge \vec{u}{=}\vec{v})}
      \infer
      [left label = \rulename{\ruleStarPred}]
      {1}
      [\rulesidecond{\vec{u} \disjoins \vec{x}}]
      {\Hypos \sepAnte\, \Lemmas \sepAnte\,
        F_1 * \formp{\P}{\vec{u}}
        \entails \exists \vec{x}. (F_2 * \formp{\P}{\vec{v}})}
    \end{prooftree}}}

%% ~~~~~~~~~~~~~~~~~~~~~~~~~~~~~~~~~~
\newcommand\minipagePredIntroLeft{\hbox{
    \begin{prooftree}[label separation = 0.3em]
      \hypo{\Hypos \sepAnte\, \Lemmas \sepAnte\,
        F_1 * \formp{\formF^{\P}_1}{\vec{u}} \entails F_2}
      \hypo{\dots}
      \hypo{\Hypos \sepAnte\,
        F_1 * \formp{\formF^{\P}_m}{\vec{u}} \entails F_2}
      \infer
      [left label = \rulename{\rulePredLeft}]
      {3}
      [\rulesidecond{
        \formp{\P}{\vec{u}} \denotedas
        \formp{\formF^{\P}_1}{\vec{u}} \vee \ldots
        \vee \formp{\formF^{\P}_m}{\vec{u}}}]
      {\Hypos \sepAnte\, \Lemmas \sepAnte\,
        F_1 * \formp{\P}{\vec{u}} \entails F_2}
    \end{prooftree}}}

%% ~~~~~~~~~~~~~~~~~~~~~~~~~~~~~~~~~~
\newcommand\minipagePredIntroRight{\hbox{
    \begin{prooftree}[label separation = 0.3em]
      \hypo{\Hypos \sepAnte\, \Lemmas \sepAnte\,
        F_1 \entails \exists \vec{x}.
        (F_2 * \formp{\formF^{\P}_i}{\vec{u}})}
      \infer
      [left label = \rulename{\rulePredRight}]
      {1}
      [\rulesidecond{
        \formp{\formF^{\P}_i}{\vec{u}} \,{\defrightarrow}\,
        \formp{\P}{\vec{u}}}]
      {\Hypos \sepAnte\, \Lemmas \sepAnte\,
        F_1 \entails
        \exists \vec{x}.(F_2 * \formp{\P}{\vec{u}})}
    \end{prooftree}}}

%% ~~~~~~~~~~~~~~~~~~~~~~~~~~~~~~~~~~
\newcommand\minipageInduction{
  \begin{tabular}{l}
    \hbox{
      \begin{prooftree}[label separation = 0.3em][separation=0.8em]
        \hypo{\HyposPrim \sepAnte\, \Lemmas \sepAnte\,
          \Sigma_1 * \formp{\formF^{\P}_1}{\vec{u}} \wedge \Pi_1 \entails F_2}
        \hypo{\dots}
        \hypo{\HyposPrim \sepAnte\, \Lemmas \sepAnte\,
          \Sigma_1 * \formp{\formF^{\P}_m}{\vec{u}} \wedge \Pi_1 \entails F_2}
        \infer
        [left label = \rulename{\ruleInduction}]
        {3}
        [\rulesidecond{\formp{\P}{\vec{u}} ~\definedas~
          \formp{\formF^{\P}_1}{\vec{u}}
          \vee ... \vee
          \formp{\formF^{\P}_m}{\vec{u}}; \dagger}]
        {\Hypos \sepAnte\, \Lemmas \sepAnte\, \,
          \Sigma_1 * \formp{\P}{\vec{u}} \wedge \Pi_1 \entails F_2}
      \end{prooftree}} \\[1em]

    \hspace*{1.5em} {\small $\dagger: \HyposPrim \triangleq %{\denotedas}
      \Hypos \cup \setenum{H},$
      where $H$ is obtained by freshly renaming all variables of
      $\Sigma_1 * \formp{\P}{\vec{u}} \wedge \Pi_1
      \entails F_2$}

  \end{tabular}}

%% ~~~~~~~~~~~~~~~~~~~~~~~~~~~~~~~~~~
\newcommand\minipageHypothesis{
  \begin{tabular}{l}
    \hbox{
      \begin{prooftree}[label separation = 0.3em]
        \hypo{\Hypos \cup
          \{\Sigma_3 * \formp{\P}{\vec{v}} \wedge \Pi_3
          \entails F_4\}
          \sepAnte\, \Lemmas
          \sepAnte\,
          F_4\theta * \Sigma \wedge \Pi_1 \entails F_2}
        \infer
        [left label = \rulename{\ruleHypo}]
        {1}
        [\hboxauto{
          \rulesidecond{\formp{\P}{\vec{u}} \ltpred \formp{\P}{\vec{v}}; \\
            \exists \theta, \Sigma. (\Sigma_1 \,{*}\,
            \formp{\P}{\vec{u}}
            ~\synequiv~
            \Sigma_3\theta \,{*}\, \formp{\P}{\vec{v}}\theta
            \,{*}\, \Sigma)
            \wedge (\Pi_1 {\rightarrow} \Pi_3\theta)}}]
        {\Hypos \cup
          \{\Sigma_3 * \formp{\P}{\vec{v}} \wedge
          \Pi_3 \entails F_4\}
          \sepAnte\, \Lemmas
          \sepAnte\,
          \Sigma_1 * \formp{\P}{\vec{u}} \wedge
          \Pi_1 \entails F_2}
      \end{prooftree}}\\[1em]
  \end{tabular}}

%% ~~~~~~~~~~~~~~~~~~~~~~~~~~~~~~~~~~
\newcommand\minipageNotLeft{\hbox{
    \begin{prooftree}[label separation = 0.3em]
      \hypo{\Hypos \sepAnte\,
        F_1 \wedge \Pi \entails F_2}
      \hypo{\Hypos \sepAnte\,
        F_1 \wedge \neg \Pi \entails F_2}
      \infer
      [left label = \rulename{\ruleNotLeft}]
      {2}
      {\Hypos \sepAnte\, F_1 \entails F_2}
    \end{prooftree}}}

%% ~~~~~~~~~~~~~~~~~~~~~~~~~~~~~~~~~~
\newcommand\minipageWeakenLHS{\hbox{
    \begin{prooftree}[label separation = 0.3em]
      \hypo{\Hypos \sepAnte\, F_1 \entails F_2}
      \infer
      [left label = \rulename{\ruleAndLeft}]
      {1}
      {\Hypos \sepAnte\,
        F_1 \wedge \Pi \entails F_2}
    \end{prooftree}}}

%% ~~~~~~~~~~~~~~~~~~~~~~~~~~~~~~~~~~
\newcommand\minipageUninstVar{\hbox{
    \begin{prooftree}[label separation = 0.3em]
      \hypo{\Hypos \sepAnte\,
        F_1 [x/e] \entails F_2}
      \infer
      [left label = \rulename{\ruleNotLeft}]
      {1}
      {\Hypos \sepAnte\, F_1 \entails F_2}
    \end{prooftree}}}

%% ~~~~~~~~~~~~~~~~~~~~~~~~~~~~~~~~~~
\newcommand\minipageLemmaLeft{
  \hbox{
    \begin{prooftree}[label separation = 0.3em]
      \Hypo{
        \Hypos \sepAnte\,
        \Lemmas \cup \{\Sigma_3 \wedge \Pi_3 \entails F_4\}
        \sepAnte\,
        F_4\theta * \Sigma \wedge \Pi_1 \entails F_2
      }
      \Infer [left label = \rulename{\ruleLemmaLeft}] 1
      [$\exists{} \theta, \Sigma.~
      (\Sigma_1 \synequiv \Sigma_3\theta * \Sigma)
      \wedge (\Pi_1 {\pureImply} \Pi_3\theta)$]{
        \Hypos \sepAnte\,
        \Lemmas \cup \{\Sigma_3 \wedge \Pi_3 \entails F_4\}
        \sepAnte\,
        \Sigma_1 \wedge \Pi_1 \entails F_2
      }
    \end{prooftree}}}

%% ~~~~~~~~~~~~~~~~~~~~~~~~~~~~~~~~~~
\newcommand\minipageLemmaRight{
  \hbox{
    \begin{prooftree}[label separation = 0.3em]
      \Hypo{
        \Hypos \sepAnte\,
        \Lemmas \cup \{F_3 \entails
        \exists \vec{w}.(\Sigma_4 \wedge \Pi_4)\} \sepAnte\,
        F_1 \entails \exists \vec{x}.(F_3\theta * \Sigma \wedge \Pi_2)}
      \Infer [left label = \rulename{\ruleLemmaRight}] 1
      [$\exists{} \theta, \Sigma.~
      (\Sigma_4\theta * \Sigma \synequiv \Sigma_2)
      \wedge (\vec{w}\theta \subseteq \vec{x})$] {
        \Hypos \sepAnte\,
        \Lemmas \cup
        \{F_3 \entails
        \exists \vec{w}.(\Sigma_4 \wedge \Pi_4)\} \sepAnte\,
        F_1 \entails \exists \vec{x}.(\Sigma_2 \wedge \Pi_2)
      }
    \end{prooftree}}
}

%% ~~~~~~~~~~~~~~~~~~~~~~~~~~~~~~~~~~
\newcommand\minipagePiOne{\hbox{
    \begin{prooftree}[label separation = 0.3em]
      \Hypo{
        \Assumpts \triangleq
        \{\Pi_1 \wedge \formr{\rUnk}{\vec{x}} \pureImply \Pi_2\}
        }
      \Infer [left label = \rulename{\ruleSynPiOne}] 1 [] {
        \Hypos \sepAnte\, \Lemmas \sepAnte\,
        \Pi_1 \wedge \formr{\rUnk}{\vec{x}} \entails \Pi_2
        }
    \end{prooftree}}}

%% ~~~~~~~~~~~~~~~~~~~~~~~~~~~~~~~~~~
\newcommand\minipagePiTwo{\hbox{
  \begin{prooftree}[label separation = 0.3em]
    \Hypo{\Assumpts \triangleq \{
      \Pi_1 \pureImply
      \exists \vec{w}. (\Pi_2 \wedge \formr{\rUnk}{\vec{x}})
      \}}
    \Infer[left label = \rulename{\ruleSynPiTwo}]1{
      \Hypos \sepAnte\, \Lemmas \sepAnte\, \Pi_1 \entails
      \exists \vec{w}. (\Pi_2 \wedge \formr{\rUnk}{\vec{x}})}
  \end{prooftree}}}

%% ~~~~~~~~~~~~~~~~~~~~~~~~~~~~~~~~~~
\newcommand\minipageSigmaOne{\hbox{
    \begin{prooftree}[label separation = 0.3em]
      \Hypo{
        \Assumpts \triangleq
        \{\Pi_1 \wedge \formr{\rUnk} {\vec{x}} \pureImply \false\}
      }
      \Infer [left label = \rulename{\ruleSynSigmaOne}] 1
      [\rulesidecond{\Sigma_1 {\neq} \emp}] {
        \Hypos \sepAnte\, \Lemmas \sepAnte\,
        \Sigma_1 \wedge \Pi_1 \wedge \formr{\rUnk}{\vec{x}} \entails \Pi_2
      }
  \end{prooftree}}}

%% ~~~~~~~~~~~~~~~~~~~~~~~~~~~~~~~~~~
\newcommand\minipageSigmaTwo{\hbox{
    \begin{prooftree}[label separation = 0.3em]
      \Hypo{
        \Assumpts \triangleq
        \{\Pi_1 \wedge \formr{\rUnk} {\vec{x}} \pureImply \false\}
      }
      \Infer [left label = \rulename{\ruleSynSigmaTwo}] 1
      [\rulesidecond{\Sigma_2 {\neq} \emp}] {
        \Hypos \sepAnte\, \Lemmas \sepAnte\,
        \Pi_1 \wedge \formr{\rUnk}{\vec{x}} \entails
        \exists \vec{w}. (\Sigma_2 \wedge \Pi_2)
      }
  \end{prooftree}}}

%% ~~~~~~~~~~~~~~~~~~~~~~~~~~~~~~~~~~
\newcommand\minipageSynHypo{
  \begin{tabular}{l}
    \hbox{
      \begin{prooftree}[label separation = 0.3em]
        \Hypo{
          \Hypos \cup\setenum{
            \Sigma_3 * \formp{\P}{\vec{v}} \wedge \Pi_3
            \wedge \formr{\rUnk}{\vec{y}} \entails F_4}
          \sepAnte\, \Lemmas \sepAnte\,
          F_4\theta * \Sigma \wedge \Pi_1 \entails F_2}
        \Hypo{
          \Assumpts \triangleq
          \{\Pi_1 \wedge \formr{\rUnk} {\vec{x}}
          \pureImply (\Pi_3 \wedge \formr{\rUnk}{\vec{y}})\theta\}}
        \Infer [left label = \rulename{\ruleSynHypo}] 2
        [$\dagger$] {
          \Hypos \cup
          \{\Sigma_3 * \formp{\P}{\vec{v}} \wedge \Pi_3
          \wedge \formr{\rUnk}{\vec{y}} \entails F_4\}
          \sepAnte\, \Lemmas \sepAnte\,
          \Sigma_1 * \formp{\P}{\vec{u}} \wedge \Pi_1 \wedge
          \formr{\rUnk}{\vec{x}}
          \entails F_2
        }
      \end{prooftree}}
    \\[1.1em]
    \hspace*{1.5em}{\small
      with $\dagger:~
      \formp{\P}{\vec{u}} \ltpred \formp{\P}{\vec{v}};~
      \exists \theta, \Sigma.\,
      (\Sigma_1 * \formp{\P}{\vec{u}} \synequiv
      \Sigma_3\theta * \formp{\P}{\vec{v}}\theta
      * \Sigma) $}
  \end{tabular}}

%% ~~~~~~~~~~~~~~~~~~~~~~~~~~~~~~~~~~
\newcommand\minipageGeneralRule{
  \begin{tabular}{l}
    \hbox{
      \begin{prooftree}[label separation = 0.3em][separation=0.8em]
        \hypo{\Hypos_1 \sepAnte\, \Lemmas_1 \sepAnte\,
          F_{11} \entails F_{21}}
        \hypo{\dots}
        \hypo{\Hypos_n \sepAnte\, \Lemmas_n \sepAnte\,
          F_{1n} \entails F_{2n}}
        \hypo{\Assumpts \triangleq \{\Pi_1 \pureImply \Pi_2\}}
        \infer
        [left label = \rulename{\vRule}]
        {4}
        [$\dagger$]
        {\Hypos \sepAnte\, \Lemmas \sepAnte\, \,
          F_1 \entails F_2}
      \end{prooftree}} \\[1em]
  \end{tabular}}
%%% Local Variables:
%%% mode: latex
%%% TeX-master: "main"
%%% End:

%%=================

%%%%% pre-loaded figure
%%%%%%%%%%%%%%%%%%%%%%%%%%%%%%%%%%%%%%%%%%%%%%%%%%%%%%%%
%%%%% Some definitions

%%%%%%%%%%%%%%%%%%%%%%%%%%%%%%%
%% data structure list_head
\newcommand\minipageListHead[2]{
\begin{minipage}[t]{#1}
\begin{tabular}{m{#2}l}
  \codeLineNum{1} & \textcomment{/* data structure modeling an entry of a list */}\\
  \codeLineNum{2} & \code{struct~list\_head~\{}\\
  \codeLineNum{3} &
  \code{\quad struct~list\_head~\pointerVar{next},~\pointerVar{prev};}\\
  \codeLineNum{4} & \code{\}}\\
\end{tabular}
\end{minipage}
}

%%%%%%%%%%%%%%%%%%%%%%%%%%%%%%%
%% procedure __list_del
\newcommand\minipageListDel[2]{
\begin{minipage}[t]{#1}
\begin{tabular}{m{#2}l}
  \codeLineNum{5} &
  \textcomment{/* procedure that deletes an entry by making the prev}\\
  \codeLineNum{6} &
  \textcomment{ ~~ and the next entries (p, n) point to each other */}\\
  \codeLineNum{7} &
  \code{void~\ListDel(\structListHead/~\pointerVar{p},~}\\
  \codeLineNum{8} &
  \hspace{7.3em} \code{\structListHead/~\pointerVar{n})}\\
  \codeLineNum{9} &
  \textcomment{/* a specification for the scenario that p{$\neq$}n */} \\
  \codeLineNum{10} &
  \textspecs{\text{//@}\code{requires}~$
     \forms{p}{e{,}u} * \forms{e}{n{,}p}
     * \forms{n}{v{,}e} * \formp{\pDls}{v{,}p{,}u{,}n} $}\\
  \codeLineNum{11} &
  \textspecs{\text{//@}\code{ensures}~$
     \forms{p}{{\color{red}n}{,}u} * \forms{n}{v{,}{\color{red}p}}
     * \formp{\pDls}{v{,}p{,}u{,}n} * \forms{e}{n{,}p}$}\\
  \codeLineNum{12} & \code{\{}\\
  \codeLineNum{13} & \code{\quad n{\rightarrow}prev = p;}\\
  \codeLineNum{14} & \code{\quad p{\rightarrow}next = n;}\\
  \codeLineNum{15} & \code{\}}\\
\end{tabular}
\end{minipage}
}

%%%%%%%%%%%%%%%%%%%%%%%%%%%%%%%
%% procedure __list_del_entry
\newcommand\minipageListDelEntry[2]{
\begin{minipage}[t]{#1}
\begin{tabular}{m{#2}l}
  \codeLineNum{16} &
  \textcomment{/* procedure that deletes an entry (e) from a list */}\\
  \codeLineNum{17} &
  \code{void~\ListDelEntry(\structListHead/~\pointerVar{e})}\\
  \codeLineNum{18} &
  \textcomment{/* a specs for the scenario that
    e{$\rightarrow$}prev{$\neq$}e{$\rightarrow$}next */} \\
  \codeLineNum{19} &
  \textspecs{\text{//@}\code{requires}\,
    $\forms{e}{n{,}p} * \formp{\pDls}{n{,}e{,}p{,}e} \wedge n{\neq}p$}\\
  \codeLineNum{20} &
  \textspecs{\text{//@}\code{ensures}\,$
     {\color{red} \exists x.} (\forms{e}{n{,}p}\,{*}\,
     {\color{red} \forms{n}{x{,}p}\,{*}\,\formp{\pDls}{x{,}n{,}p{,}n}}
     \,{\wedge}\,n{\neq}p)
    $}\\
  \codeLineNum{21} & \code{\{}\\
  \codeLineNum{22} & \textspecs{\text{\quad //@}
    $\forms{e}{n{,}p} * \formp{\pDls}{n{,}e{,}p{,}e} \wedge n{\neq}p$}\\
  \codeLineNum{23} & \code{\quad {\_\_}list\_del(e{\rightarrow}prev,~e{\rightarrow}next);}\\
  \codeLineNum{24} & \code{\}}\\
\end{tabular}
\end{minipage}
}

%%%%%%%%%%%%%%%%%%%%%%%%%%%%%%%
%% procedure list_move_tail
\newcommand\minipageListMoveTail[1]{
\begin{minipage}[t]{#1}
  \[
  \begin{array}{l}
  \textcomment{/* procedure that deletes an entry from one list (\code{l}),}\\
  \textcomment{ * and add it as tail of another list (\code{h}) ({\it include/linux/list.h})}\\
  \code{void~\ListMoveTail/(\structListHead/~\pointerVar{l}},\\
  \hspace{9.75em}         \code{\structListHead/~\pointerVar{h})}\\
  \textcomment{/* a specification for the scenario that \code{l{\neq}h} and both of}\\
  \textcomment{ * them reside in the same circular doubly linked-list */}\\
%   \text{//@~} \code{requires}~~
%      \forms{l}{u{,}v} * \formp{\pDls}{u{,}h{,}w{,}l}
%      * \formp{\pDls}{h{,}l{,}v{,}w} \\
%   \text{//@~} \code{ensures}~~~~
%      \formp{\pDls}{u{,}l{,}w{,}v} * \forms{l}{h{,}w}
%      * \formp{\pDls}{h{,}u{,}v{,}l} \wedge l{\neq}h \\
%   \text{//}\\
  \text{//@~} \code{requires}~~
     \formp{\pDls}{l{,}h{,}u{,}v} * \formp{\pDls}{h{,}l{,}v{,}u}
     \wedge h{\neq}l\\
  \text{//@~} \code{ensures}~~~~
     \formp{\pDls}{h{,}l{,}u{,}l} * \forms{l}{h{,}u} \\
  \code{\{}\\
  \text{\quad //@~} \formp{\pDls}{l{,}h{,}u{,}v}
     * \formp{\pDls}{h{,}l{,}v{,}u} \wedge h{\neq}l \\
  \text{\quad //@pre:~} \forms{l}{x{,}y}
     * \formp{\pDls}{x{,}y{,}z{,}l}
     * \forms{y}{l{,}z} \wedge h{\neq}l \\
  \code{\quad {\_\_}list\_del\_entry(l);}\\
  \code{\quad list\_add\_tail(l,~h);}\\
  \code{\}}\\
  \end{array}
  \]
\end{minipage}
}

%%%%%%%%%%%%%%%%%%%%%%%%%%%%%%%
%% procedure list_rotate_left
\newcommand\minipageListRotateLeft[1]{
\begin{minipage}[t]{#1}
  \[
  \begin{array}{l}
  \textcomment{/* procedure that rotates a non-empty circular doubly}\\
  \textcomment{ * linked-list to the left (from: \textit{include/linux/list.h})}\\
  \textcomment{ * \code{h}: the head of the list */}\\
  \code{void~ \ListRotateLeft/(\structListHead/~\pointerVar{h})}\\
%   \textcomment{// a complete specification}\\
  \text{//@~} \code{requires}~
    \forms{h}{x{,}z} * \forms{x}{y{,}h} * \formp{\pDls}{y{,}h{,}z{,}x}\\
  \text{//@~} \code{ensures} \hspace{0.8em}
    \forms{h}{y{,}x} * \formp{\pDls}{y{,}x{,}z{,}h} * \forms{x}{h{,}z}\\
  \code{\{}\\
  \code{\quad \structListHead/~\pointerVar{first};}\\
  \code{\quad if~(~!list\_empty(h)~)~\{}\\
  \code{\quad\quad first~=~h{\rightarrow}next;}\\
%   \text{\quad\quad //@~} \forms{h}{x{,}z} * \forms{x}{y{,}h}
%     * \formp{\pDls}{y{,}h{,}z{,}x} \wedge first{=}x\\
%   \text{\quad\quad //}\equiv~ \forms{x}{y{,}h} * \formp{\pDls}{y{,}h{,}z{,}x}
%     * \forms{h}{x{,}z} \wedge first{=}x\\
%   \text{\quad\quad //}\equiv~ \forms{x}{y{,}h} * \formp{\pDls}{y{,}h{,}z{,}x}
%     * \formp{\pDls}{h{,}x{,}h{,}z} \wedge first{=}x\\
%   \text{\quad\quad//}\\
  \text{\quad\quad //@~} \forms{h}{x{,}z} * \forms{x}{y{,}h}
    * \formp{\pDls}{y{,}h{,}z{,}x} \wedge first{=}x\\
  \text{\quad\quad //pre:}~ \formp{\pDls}{h{,}x{,}h{,}z}
    * \formp{\pDls}{x{,}h{,}z{,}h} \wedge x{\neq}h \wedge first{=}x\\
  \code{\quad\quad list\_move\_tail(first,~h);}\\
%   \text{\quad\quad //@~} \formp{\pDls}{y{,}x{,}z{,}h} * \forms{x}{h{,}z}
%     * \formp{\pDls}{h{,}y{,}h{,}x} \wedge first{=}x\\
%   \text{\quad\quad//}\\
  \text{\quad\quad //post:~} \formp{\pDls}{h{,}x{,}z{,}x}
    * \forms{x}{h{,}z} \wedge x{\neq}h \wedge first{=}x\\
  \code{\quad \}}\\
  \code{\}}
  \end{array}
  \]
\end{minipage}
}

\newcommand\minipageMotivatingExample[2]{
\begin{minipage}[t]{#1}
  \vspace{-2em}
  \minipageListHead{#1}{#2}
  \\[1em]
  \minipageListDel{#1}{#2}
  \\[1em]
  \minipageListDelEntry{#1}{#2}
\end{minipage}
}

\newcommand\minipageMotivatingExampleFigure[2]{
\begin{figure}[#2]
\begin{tabular}{l}
  \\[0.5em]
  \hspace{-1.1em}\minipageMotivatingExample{#1}{0.3em}
  \\
\end{tabular}
\caption{Circular doubly linked-list used in Linux}
\label{figureMotivatingExample}
\end{figure}
}

%%%%%%%%%%%%%%%%%%%%%%%%%%%%%%%%%%%%%%%%%%%%%%%%%%%%%%%%%%
%%%%%%%%%%%%%%%%%%%%%%%%%%%%%%%%%%%%%%%%%%%%%%%%%%%%%%%%%%

\newcommand\minipageDls[1]{
\begin{minipage}[t]{#1}
  \vspace{-2em}
  \begin{minipage}[t]{#1}
  \begin{adjustwidth}{1.5em}{}
  \[
  \begin{array}{lcl}
    \formp{\pDls}{u{,}v{,}p{,}q}
    & \triangleq
    & (\pred{emp} {\wedge} u{=}v {\wedge} p{=}q)\\
    & & \, \vee \,
      \exists w.(\forms{u}{w{,}q} \, {*}  \,
      \formp{\pDls}{w{,}v{,}p{,}u})
    \\
  \end{array}
  \]
  \end{adjustwidth}
  \end{minipage}

  \begin{minipage}[t]{#1}
  \begin{center}
  \begin{tikzpicture}[
    rectnode/.style={rectangle split, rectangle split parts=2, inner sep=5pt,
      draw, fill=timberwolf, rectangle split horizontal},
    hiddennode/.style={rectangle, fill=white, minimum size = 3mm,
      text width = 1.8mm},
    label/.style={rectangle,minimum size = 8mm},
  ]
    \node[hiddennode]  (nodeA)                                          {};
    \node[label]                  at ($(nodeA.center) + (-0.22,-0.15)$) {$q$};
    \node[rectnode]    (nodeB)    [right=5pt of nodeA]                  {};
    \node[label]                  at ($(nodeB.center) + (-0.38,0.38)$)  {$u$};
    \node[rectnode]    (nodeC)    [right=14pt of nodeB]                 {};
    \node[label]                  at ($(nodeC.center) + (-0.38,0.38)$)  {$w$};
    \node[rectnode]    (nodeD)    [right=14pt of nodeC]                 {};
    \node[hiddennode]  (nodeE)    [right=14pt of nodeD]                 {};
    \node[label]                  at ($(nodeE.center) + (0,0)$)         {$...$};
    \node[rectnode]    (nodeF)    [right=14pt of nodeE]                 {};
    \node[label]                  at ($(nodeF.center) + (-0.38,0.38)$)  {$p$};
    \node[hiddennode]  (nodeG)    [right=14pt of nodeF]                 {};
    \node[label]                  at ($(nodeG.center) + (-0.06,0.12)$)  {$v$};

    \draw[->] ($(nodeB.east) + (0,0.12)$) -- ($(nodeC.west) + (0,0.12)$);
    \draw[->] ($(nodeC.east) + (0,0.12)$) -- ($(nodeD.west) + (0,0.12)$);
    \draw[->] ($(nodeD.east) + (0,0.12)$) -- ($(nodeE.west) + (0,0.12)$);
    \draw[->] ($(nodeE.east) + (0,0.12)$) -- ($(nodeF.west) + (0,0.12)$);
    \draw[->] ($(nodeF.east) + (0,0.12)$) -- ($(nodeG.west) + (0,0.12)$);

    \draw[->] ($(nodeB.west) + (0,-0.12)$) -- ($(nodeA.east) + (-0.3,-0.12)$);
    \draw[->] ($(nodeC.west) + (0,-0.12)$) -- ($(nodeB.east) + (0,-0.12)$);
    \draw[->] ($(nodeD.west) + (0,-0.12)$) -- ($(nodeC.east) + (0,-0.12)$);
    \draw[->] ($(nodeE.west) + (0,-0.12)$) -- ($(nodeD.east) + (0,-0.12)$);
    \draw[->] ($(nodeF.west) + (0,-0.12)$) -- ($(nodeE.east) + (0,-0.12)$);
  \end{tikzpicture}
  \end{center}
  \end{minipage}
\end{minipage}
}

\newcommand\minipageDlsFigure[2]{
\begin{figure}[#2]
\begin{tabular}{|l|}
 \hline
 \\[1.2em]
 \hspace{-1.3em}\minipageDls{#1}
 \\
 \hline
\end{tabular}
\captionof{figure}{{Doubly linked list segment $\pDls$}}
\label{figurePredDls}
\end{figure}
}

\newcommand\minipageLinkedList[1]{
\begin{minipage}[t]{#1}
\begin{tikzpicture}[
  listhead/.style={rectangle split, rectangle split parts=2,
    text height=1.5ex, text depth=.25ex, text width=11em, text centered,
    draw, fill=timberwolf, rectangle split horizontal, text width = 6.5mm, inner sep=4.5pt},
  listentry/.style={rectangle split, rectangle split parts=2,
    text height=1.5ex, text depth=.25ex, text width=11em, text centered,
    draw, rectangle split horizontal, text width = 8mm, inner sep=4.5pt},
  label/.style={rectangle,minimum width = 20mm},
]
  \node[listhead] (nodeA)  {\code{prev} \nodepart{second} \code{next}};
  \node[label] at ($(nodeA.center) + (0,-0.6)$)  {\code{struct~list\_head}};

  \node[listhead] (nodeB)  [below right=14pt and 70pt of nodeA] {\code{prev} \nodepart{second} \code{next}};
  \draw[dashed] ($(nodeB.west) + (0.008,0.29)$) -- ($(nodeB.west) + (0.008,1.4)$);
  \draw[dashed] ($(nodeB.west) + (0,1.4)$) -- ($(nodeB.east) + (0,1.4)$);
  \draw[dashed] ($(nodeB.east) + (-0.008,1.4)$) -- ($(nodeB.east) + (-0.008,0.29)$);
  \node[label] at ($(nodeB.center) + (0,-0.6)$)  {\it a custom structure};
  \node[label] at ($(nodeB.center) + (0,-1)$)  {\it including \code{list\_head}};

  \node[listhead] (nodeC)  [below=33pt of nodeA]   { \nodepart{second} };
  \draw [<->] ($(nodeC.west)$) to [out=160,in=20,looseness=3.6] ($(nodeC.east)$);
  \node[label] at ($(nodeC.center) + (0.15,-0.6)$)  {{\it a circular list of 1 element}};

  \node[listhead] (nodeD)  [below left = 25pt and -43pt of nodeC]   { \nodepart{second} };
%   \node[label] at ($(nodeD.west) + (5pt,13pt)$)  {\it x};
  \node[label] at ($(nodeD.center) + (0.55,-2.1)$)  {{\it a circular list of 3 elements}};

  \node[listhead] (nodeE)  [below right = 14pt and 20pt of nodeD]   { \nodepart{second} };
%   \node[label] at ($(nodeE.west) + (5pt,13pt)$)  {\it y};
  \draw[dashed] ($(nodeE.west) + (0.007,0.29)$) -- ($(nodeE.west) + (0.007,1.4)$);
  \draw[dashed] ($(nodeE.west) + (0,1.4)$) -- ($(nodeE.east) + (0,1.4)$);
  \draw[dashed] ($(nodeE.east) + (-0.007,1.4)$) -- ($(nodeE.east) + (-0.007,0.29)$);

  \node[listhead] (nodeF)  [right = 20pt of nodeE]   { \nodepart{second} };
%   \node[label] at ($(nodeF.west) + (5pt,13pt)$)  {\it z};
  \draw[dashed] ($(nodeF.west) + (0.007,0.29)$) -- ($(nodeF.west) + (0.007,1.4)$);
  \draw[dashed] ($(nodeF.west) + (0,1.4)$) -- ($(nodeF.east) + (0,1.4)$);
  \draw[dashed] ($(nodeF.east) + (-0.007,1.4)$) -- ($(nodeF.east) + (-0.007,0.29)$);

  \draw [<->] ($(nodeD.east)$) to [out=0,in=-180] ($(nodeE.west)$);
  \draw [<->] ($(nodeE.east)$) to ($(nodeF.west)$);
  \draw [<->] ($(nodeF.east)$) to [out=-30,in=-150,looseness=1.15] ($(nodeD.west)$);
  \vspace{-3em}
\end{tikzpicture}
\end{minipage}
}

\newcommand\minipageListHeadFigure[2]{
\begin{figure}[#2]
\framebox{
\parbox[c][6.3cm]{0.465\textwidth}{
\begin{tabular}{l}
 \\[-1.8em]
 \hspace{-7.9em}\minipageLinkedList{#1}
 \\[-3em]
\end{tabular}
}}
\caption{Circular doubly linked list data structure in Linux}
\label{figureCircularDoublyLinkedList}
\end{figure}
}

%%% Local Variables:
%%% mode: latex
%%% TeX-master: "main"
%%% End:

\section{Introduction}
\label{sec:Intro}

Having been actively developed in the recent two decades, separation logic
appears as one of the most popular formalisms in verifying the
memory-safety property of computer programs \cite{Reynolds02,OHearnRY01}.
It combines {\em spatial operations}, which describe the separation of the
memory, and {\em inductive heap predicates} to expressively model the shape
of complex data structures, such as variants of linked lists, trees, or
graphs. This modeling has been successfully realized in both academic
research and industrial use. For example, it is implemented by the static
analysis tools \textsf{SLAYer} \cite{BerdineCI11} and \textsf{Infer}
\cite{CalcagnoDDGHLOP15} to find memory bugs in system code or mobile
applications in large scale.

Both \textsf{SLAYer} and \textsf{Infer} employ a logic fragment called
\emph{symbolic-heap separation logic}, which differentiates \emph{spatial
  formulas} describing the memory shape of a program's states from
\emph{pure formulas} representing Boolean constraints of the program's
variables. This fragment is also utilized by other academic verification
systems such as HIP \cite{ChinDNQ12}, jStar \cite{DistefanoP08}, and
Smallfoot \cite{Smallfoot05}. Primarily, the verification of a program in
such systems involves three phases: (i) specifying desired properties of
the program using separation logic, (ii) analyzing the program's behavior
against its specification to obtain a set of verification conditions,
mostly in the form of \emph{separation logic entailments}, and (iii)
proving the collected entailments to determine whether the program behaves
accordingly to its specification.

There have been several efforts to prove separation logic entailments, and
one of the biggest challenges is to \emph{effectively} handle their
inductive heap predicates. A commonly practised approach is to restrict the
predicates to only certain classes, such as: predicates whose syntax and
semantics are defined beforehand \cite{BerdineCO04, BerdineCO05,
  PiskacWZ13, PiskacWZ14, BozgaIP10, PerezR11, PerezR13}, predicates
describing only variants of linked lists \cite{EneaLSV14}, or predicates
satisfying the particular {\em bounded tree width property}
\cite{IosifRS13, IosifRV14}. On the one hand, such predicate restrictions
lead to the invention of effective entailment proving techniques; on the
other hand, they \emph{prevent} the predicates from modeling complex
constraints, which involve not only the shape, but also the size or the
content of data structures.

A more general approach to handle inductive heap predicates is to use proof
techniques which are \emph{similar to or based on mathematical induction}.
These techniques include cyclic proof \cite{BrotherstonDP11}, structural
induction proof \cite{ChuJT15}, and mutual induction proof \cite{TaLKC16}.
In general, induction-based techniques are capable of reasoning about
structures which are recursively defined \cite{Bundy01}. Therefore, they
can be used to deal with inductive heap predicates in separation logic
entailments. However, it is well-known that the induction-based proof
techniques are {\em incomplete}, due to the \emph{failure of
  cut-elimination} in inductive theories. Consequently, their successes
often depend on an eureka step: \emph{discovering supporting lemmas}
\cite{Bundy01}.

Not only are the \emph{supporting lemmas} directly important to
\emph{induction-based} proof systems, but they are also highly needed by
verification systems that use \emph{non-induction-based} back-end provers,
as in the works of \citet{NguyenC08} and \citet{QiuGSM13}. These systems
cannot automatically discover the desired lemmas but \emph{require the
  users to manually provide} them. Consequently, the aforementioned
verification systems are not fully automated.

Although any \emph{valid entailment} can become a lemma, the lemmas that
are really important are the ones that can convert between inductive heap
predicates, or can combine many inductive heap predicates into another one,
or can split an inductive heap predicate into a combination of many others.
These lemmas are helpful for \emph{rearranging inductive heap predicates}
in a goal entailment so that the entailment proof can be quickly
discovered. They are also called \emph{inductive lemmas}, as induction
proofs are often needed to prove them, i.e., there are derivation steps
that record and apply induction hypotheses in the lemmas' proofs.

In this work, we propose a novel framework to \emph{automatically
  synthesize inductive lemmas} to assist in proving entailments in the
fragment of symbolic-heap separation logic. Our framework is developed
based on structural induction proof and template-based constraint solving.
Given a goal entailment, the framework first analyzes the entailment's heap
structure to identify essential lemma templates. It then sets up unknown
relations among each template's variables and conducts structural induction
proof to generate constraints about the relations. Finally, it solves the
constraints to find out actual definitions of the unknown relations, thus
discovers the desired inductive lemmas.

In summary, our work makes the following contributions:
\vspace{-0.2em}

\begin{itemize}[label=--,noitemsep,leftmargin=1.2em]
\setlength\itemsep{0.1em}
\item We propose a novel framework to automatically synthesize lemmas to
  assist in proving entailments in the fragment of symbolic-heap separation
  logic with inductive heap predicates.

\item We integrate the lemma synthesis framework into an entailment proof
  system, which allows the lemmas to be flexibly discovered {\em
    on-the-fly} to support the entailment proofs.

\item We implement a prototype prover and experiment it with numerous
  entailments from various benchmarks. The experimental result is promising
  since our tool can prove most of the valid entailments in these
  benchmarks and outperforms all existing separation logic provers.

\end{itemize}

%%% Local Variables:
%%% mode: latex
%%% TeX-master: "main"
%%% End:

%%%%%%%%%%%%%%%%%%%%%%%%%%%%%%%%%%%%%%%%%%%
%%%%%%%%%%%%%%%%%%%%%%%%%%%%%%%%%%%%%%%%%%%
\vspace{-0.6em}
\section{Motivating example}
\label{sec:MotivatingExample}
\vspace{-0.5em}

In this section, we present a running example to illustrate our work and
result. Here, we consider an inductive heap predicate $\Dll$ and its
variant $\DllRev$, both modeling a doubly linked list data structure with a
length property.

\vspace{-0.2em}
\begin{center}
\begin{small}
\begin{tabular}{rll}

  $\formp{\Dll}{hd,pr,tl,nt,len}$
  & $~\defeq~$
  & $\forms{hd}{pr,nt} \wedge hd{=}tl \wedge len{=}1$
  \\
  {} & {}
  & $\vee~ \exists u. (\forms{hd}{pr,u} * \formp{\Dll}{u,hd,tl,nt,len{-}1})$
  \\

  $\formp{\DllRev}{hd,pr,tl,nt,len}$
  & $~\defeq~$
  & $\forms{hd}{pr,nt} \wedge hd{=}tl \wedge len{=}1$
  \\
  {} & {}
  & $\vee~ \exists u. (\formp{\DllRev}{hd,pr,u,tl,len{-}1} * \forms{tl}{u,nt})$
  \\

\end{tabular}
\end{small}
\end{center}
\vspace{-0.1em}

Each element in the linked list is modeled by a singleton heap predicate
$\forms{x}{pr,nt}$, where $x$ is its memory address, $pr$ and $nt$
respectively point to the \emph{previous} and the \emph{next} element in
the list. Moreover, both $\formp{\Dll}{hd,pr,tl,nt,len}$ and
$\formp{\DllRev}{hd,pr,tl,nt,len}$ denote a non-empty doubly linked list
from the first element pointed-to by $hd$ (head) to the last element pointed-to
by $tl$ (tail). The \emph{previous} and the \emph{next} element of the
entire list are respectively pointed-to by $pr$ and $nt$, and $len$ denotes
the linked list's length. The only difference of the two predicates is that
$\Dll$ is recursively defined from the linked list's head to the tail, whereas
$\DllRev$ is defined in the reversed direction.

Suppose that when verifying a program manipulating the doubly linked list
data structure, a verifier needs to prove an entailment relating to an
extraction of the last element from a concatenation of two linked lists
with certain constraints on their lengths, such as the following entailment
$E_1$.

\vspace{-0.2em}
\begin{center}
\begin{small}
  $E_1 \triangleq \formp{\DllRev}{x,y,u,v,n} * \formp{\Dll}{v,u,z,t,200}
  \wedge n{\ge}100 \entails
  \exists r. (\formp{\Dll}{x,y,r,z,n{+}199} * \forms{z}{r,t})$
\end{small}
\end{center}
\vspace{-0.2em}

Unfortunately, the existing entailment proving techniques could not prove
$E_1$. Specifically, the predicate-restriction approaches either consider
only singly linked list \cite{BerdineCO05, BozgaIP10, CookHOPW11, PerezR11,
  PerezR13}, or do not handle linear arithmetic constraints
\cite{IosifRS13, IosifRV14, EneaLSV14}, or require a pre-defined
translation of heap predicates into other theories \cite{PiskacWZ13,
  PiskacWZ14}. Moreover, the non-induction-based approaches require users
to provide supporting lemmas \cite{ChinDNQ12}. Finally, the induction-based
techniques \cite{BrotherstonDP11, ChuJT15} fail to prove $E_1$ because this
entailment is {\em not general enough} to be an effective induction
hypothesis, due to its \emph{specific constant values}.

We briefly explain the \emph{failure of induction proof} on $E_1$ as
follows. Typically, induction is performed on an inductive heap predicate
in $E_1$'s antecedent. Suppose the chosen predicate is
$\formp{\DllRev}{x,y,u,v,n}$; $E_1$ is then recorded as an induction
hypothesis $H$, whose variables are renamed to avoid confusion:

\begin{center}
\begin{small}
  $H \triangleq \formp{\DllRev}{a,b,p,q,m} *
  \formp{\Dll}{q,p,c,d,200} \wedge m{\geq}100 \entails \exists k.
  (\formp{\Dll}{a,b,k,c,m{+}199} * \formss{c}{k,d})$
\end{small}
\end{center}

Subsequently, the predicate $\formp{\DllRev}{x,y,u,v,n}$ of $E_1$ is
unfolded by its recursive definition to derive new sub-goal entailments. We
consider an interesting case when the below entailment $E'_1$ is obtained
from an inductive case unfolding of $\formp{\DllRev}{x,y,u,v,n}$:

\begin{center}
\begin{small}
  $E'_1 \triangleq \formp{\DllRev}{x,y,w,u,n{-}1} * \forms{u}{w,v} *
  \formp{\Dll}{v,u,z,t,200} \wedge n{\geq}100
  \entails
  \exists r. (\formp{\Dll}{x,y,r,z,n{+}199} * \forms{z}{r,t})$
\end{small}
\end{center}

When applying the induction hypothesis $H$ to prove $E'_1$, the predicates
of the same symbols ($\DllRev$ or $\Dll$) in the antecedents of $H$ and
$E'_1$ need to be unified by a substitution. However, such substitution
does not exist since $q$ is mapped to $u$ when unifying
$\formp{\DllRev}{a,b,p,q,m}$ vs. $\formp{\DllRev}{x,y,w,u,n{-}1}$, whereas
$q$ is mapped to the \emph{different} variable $v$ when unifying
$\formp{\Dll}{q,p,c,d,200}$ vs. $\formp{\Dll}{v,u,z,t,200}$.

Alternatively, we can weaken the spatial formula $\forms{u}{w,v} *
\formp{\Dll}{v,u,z,t,200}$ of $E'_1$'s antecedent into
$\formp{\Dll}{u,w,z,t,201}$, w.r.t.\ the definition of $\Dll$, to derive
a new entailment $E''_1$:
\begin{center}
\begin{small}
  $E''_1 \triangleq \formp{\DllRev}{x,y,w,u,n{-}1} * \formp{\Dll}{u,w,z,t,201}
  \wedge n{\geq}100
  \entails
  \exists r. (\formp{\Dll}{x,y,r,z,n{+}199} * \forms{z}{r,t})$
\end{small}
\end{center}

Again, no substitution can unify the antecedents of $H$ and $E''_1$. For
example, the substitution $\theta_1 \triangleq [x/a,y/b,w/p,u/q,n{-}1/m]$
might be able to unify $\formp{\DllRev}{a,b,p,q,m}$ with
$\formp{\DllRev}{x,y,w,u,n{-}1}$, that is,
$\formp{\DllRev}{a,b,p,q,m}\theta_1 \equiv \formp{\DllRev}{x,y,w,u,n{-}1}$.
However, the constraint $n{\geq}100$ in the antecedent of $E''_1$
\emph{cannot prove} that $n{-}1{\geq}100$, which is obtained from applying
$\theta_1$ on the constraint $m{\geq}100$ of $H$. In addition, the
substitution $\theta_2 \triangleq [u/q,w/p,z/c,t/d]$ could not unify
$\formp{\Dll}{q,p,c,d,200}$ with $\formp{\Dll}{u,w,z,t,201}$ since the
two constants $201$ and $200$ are \emph{non-unifiable}. In short, the above
inability in unifying heap predicates makes induction proof fail on $E_1$.

Nevertheless, the entailment $E_1$ is provable if necessary lemmas can be
discovered. For instance, the following lemmas $L_1$, $L_2$, and $L_3$ can
be introduced to assist in proving $E_1$. More specifically, $L_1$
\emph{converts} a linked list modeled by the predicate symbol $\DllRev$
into a linked list modeled by the variant predicate $\Dll$; $L_2$
\emph{combines} two linked lists modeled by $\Dll$ into a new one (similar
in spirit to the ``composition lemma'' introduced by \citet{EneaSW15});
lastly, $L_3$ \emph{splits} a linked list modeled by $\Dll$ into two parts
including a new linked list and a singleton heap:

\begin{center}
\begin{small}
\begin{tabular}{lll}
  $L_1$ & $\triangleq$
  & $\formp{\DllRev}{a,b,c,d,m} \entails \formp{\Dll}{a,b,c,d,m}$\\

  $L_2$ & $\triangleq$
  & $\formp{\Dll}{a,b,p,q,m} * \formp{\Dll}{q,p,c,d,l} \entails
  \formp{\Dll}{a,b,c,d,m+l}$\\

  $L_3$ & $\triangleq$
  & $\formp{\Dll}{a,b,c,d,m} \wedge m {\ge} 2
  \entails \exists w. (\formp{\Dll}{a,b,w,c,m {-} 1} * \forms{c}{w,d})$
\end{tabular}
\end{small}
\end{center}

The three lemmas $L_1$, $L_2$, and $L_3$ can be used to prove $E_1$ as
shown in Figure~\ref{fig:ProofTreeMotiv}. They are successively utilized by
the lemma application rules $\ruleLemmaLeft, \ruleLemmaRight$ to finally
derive a new entailment $E_4$, which can be easily proved by standard
inference rules in separation logic. We will explain the details of these
lemma application rules $\ruleLemmaLeft, \ruleLemmaRight$ and other
inference rules in Section~\ref{sec:InferenceRules}.

\vspace{-0.3em}
\begin{figure}[H]
  \captionsetup{justification=centering}
  \begin {small}
    \begin{prooftree}[rule margin = 0.1em]
      \Hypo{}
      \Infer1[$\rulePureEntail$]{
        n{\geq}100
        \entails
        (x{=}x \wedge y{=}y \wedge z{=}z \wedge t{=}t
        \wedge n{+}200{=}n{+}200)}
      \Infer1[$\ruleStarPred$]{
        E_4 \triangleq
        \formp{\Dll}{x,y,z,t,n{+}200} \wedge n{\geq}100
        \entails
        \formp{\Dll}{x,y,z,t,n{+}200}}
      \Infer1[$\ruleLemmaRight$ with $L_3$ and $\theta_3$]{
        E_3 \triangleq
        \formp{\Dll}{x,y,z,t,n{+}200} \wedge n{\geq}100
        \entails
        \exists r. (\formp{\Dll}{x,y,r,z,n{+}199} * \forms{z}{r,t})}
      \Infer1[$\ruleLemmaLeft$ with  $L_2$ and $\theta_2$] {
        E_2 \triangleq\,
        \formp{\Dll}{x,y,u,v,n} * \formp{\Dll}{v,u,z,t,200}
        \wedge n{\geq}100
        \entails
        \exists r. (\formp{\Dll}{x,y,r,z,n{+}199} * \formss{z}{r,t})}
      \Infer1[$\ruleLemmaLeft$ with $L_1$ and $\theta_1$]{
        E_1 \triangleq
        \formp{\DllRev}{x,y,u,v,n} * \formp{\Dll}{v,u,z,t,200}
        \wedge n{\geq}100
        \entails
        \exists r. (\formp{\Dll}{x,y,r,z,n{+}199} * \formss{z}{r,t})}

    \end{prooftree}
  \end{small}

    \vspace{-0.6em}
    \caption{Proof tree of $E_1$ using the lemmas $L_1$, $L_2$, and $L_3$,
      with the substitutions
      $\theta_1 \equiv [x/a, y/b, u/c, v/d, n/m]$,
      $\theta_2 \equiv [x/a, y/b, u/p, v/q, n/m, z/c, t/d, 200/l]$, and
      $\theta_3 \equiv [x/a, y/b, z/c, t/d, r/w, n{+}200/m]$}
    \label{fig:ProofTreeMotiv}
\end{figure}
\vspace{-1.2em}

In the proof tree in Figure~\ref{fig:ProofTreeMotiv}, the entailment $E_1$
can only be concluded valid if all the lemmas $L_1$, $L_2$, and $L_3$ are
also valid. Furthermore, this proof tree is not constructed by any
induction proofs. Instead, induction is performed in the proof of each
lemma. We illustrate a partial induction proof tree of $L_1$ in
Figure~\ref{fig:ProofTreeLemmaL1}, where an induction hypothesis is
recorded by the induction rule $\ruleInduction$ and is later utilized by
the induction hypothesis application rule $\ruleHypo$. As discussed earlier
in Section~\ref{sec:Intro}, these three lemmas are called \emph{inductive
  lemmas}. In general, the inductive lemmas help to modularize the proof of
a complex entailment like $E_1$: induction is not directly performed to
prove $E_1$; it is alternatively utilized to prove simpler supporting
lemmas such as $L_1$, $L_2$, and $L_3$. This modularity, therefore,
increases the success chance of proving complex entailments.

\vspace{-0.5em}
\begin{figure}[H]
  \begin{small}
    \begin{prooftree}[separation = 1em, label separation = 0.3em,
        rule margin = 0.2em]
      \Hypo{\ldots}
      \Infer[rule margin = 0.35em]1[$\rulePredRight$]{\hboxauto{
          $\forms{a}{b,d} \wedge a{=}c \wedge m{=}1$\\
          $\hspace*{1.5em} \entails \formp{\Dll}{a,b,c,d,m}$}}
      \Infer[rule margin=0em, rule style = no rule]1{\hspace*{9.5em}}
      \Hypo{\ldots}
      \Hypo{}
      \Infer1[$\rulePureEntail$]{
        true \entails
        \exists t. (v{=}t \wedge a{=}a \wedge c{=}c \wedge d{=}d
        \wedge m{-}1{=}m{-}1)}
      \Infer1[$\ruleStarPred$]{
        \formp{\Dll}{v,a,c,d,m{-}1}
        \entails \exists t. (\formp{\Dll}{t,a,c,d,m{-}1} \wedge v{=}t)}
      \Infer1[$\ruleHypo (2)$]{
        \formp{\Dll}{v,a,u,c,m{-}2} * \forms{c}{u,d}
        \entails \exists t. (\formp{\Dll}{t,a,c,d,m{-}1} \wedge v{=}t)}
      \Infer1[$\ruleStarData$]{
        \forms{a}{b,v} * \formp{\Dll}{v,a,u,c,m{-}2} * \forms{c}{u,d}
        \entails
        \exists t. (\forms{a}{b,t} * \formp{\Dll}{t,a,c,d,m{-}1})}
      \Infer1[$\rulePredRight$]{
        \forms{a}{b,v} * \formp{\Dll}{v,a,u,c,m{-}2} *
        \forms{c}{u,d} \entails \formp{\Dll}{a,b,c,d,m}}
      \Infer2[$\ruleInduction(2)$]{
        \formp{\Dll}{a,b,u,c,m{-}1} * \forms{c}{u,d}
        \entails \formp{\Dll}{a,b,c,d,m}}
      \Infer1[$\ruleHypo (1)$]{
        \formp{\DllRev}{a,b,u,c,m{-}1} * \forms{c}{u,d} \entails
        \formp{\Dll}{a,b,c,d,m}}
      \Infer[separation = 1em,
        rule margin=0em, rule style = no rule]1{\hspace*{20em}}
      \Infer2[$\ruleInduction(1)$]{
        L_1 \triangleq \formp{\DllRev}{a,b,c,d,m}
        \entails \formp{\Dll}{a,b,c,d,m}}
    \end{prooftree}
  \end{small}
    \vspace{-0.5em}
    \caption{Partial induction proof tree of the inductive lemma $L_1$}
    \vspace{0.3em}

  \begin{small}
    \noindent where $\ruleInduction(1), \ruleInduction(2)$
    are performed on
    $\formp{\DllRev}{a,b,c,d,m}$,
    $\formp{\Dll}{a,b,u,c,m{-}1}$
    to obtain IHs $H_1, H_2$,\\
    $\ruleHypo (1)$ with
    $H_1 \triangleq \formp{\DllRev}{a',b',c',d',m'}
    \entails \formp{\Dll}{a',b',c',d',m'}$ and
    $\theta_1 \equiv [a/a',b/b',u/c',c/d',m{-}1/m']$,\\
    $\ruleHypo (2)$ with
    $H_2 \,{\triangleq}\, \formp{\Dll}{a'\!,b'\!,u'\!,c'\!,m'{-}1} {*}
    \forms{c'}{u'\!,d'}
    \,{\entails}\, \formp{\Dll}{a'\!,b'\!,c'\!,d'\!,m'}$,
    $\theta_2 \equiv [v/a'\!,a/b'\!,u/u'\!,c/c'\!,d/d'\!,m{-}1/m']$
    \label{fig:ProofTreeLemmaL1}
  \end{small}
\end{figure}
\vspace{-0.8em}

On the other hand, we are \emph{not} interested in
\emph{trivial\,\,lemmas}, which can be proved \emph{directly without using}
any induction hypotheses: the rule $\ruleHypo$ is not used in their proofs.
For example, $L'_3 \triangleq \formp{\Dll}{a,b,c,d,m} \wedge m {\ge} 2
\entails \exists w. (\forms{a}{b,w} * \formp{\Dll}{w,a,c,d,m {-} 1})$ is a
trivial lemma, whose direct proof tree is presented in Figure
\ref{fig:ProofTree:TrivialLemma}. Obviously, if a trivial lemma can be
applied to prove an entailment, then the same sequence of inference rules
utilized in the lemma's proof can also be directly applied to the goal
entailment. For this reason, the \emph{trivial~lemmas} do not help to
modularize the induction proof.

\vspace{-0.5em}
\begin{figure}[H]
\begin{small}
  \centering
  \begin{prooftree}[rule margin = 0.2em, separation = 0.2em,
    rule separation = 0.1em, label separation = 0.1em]
    \Hypo{}
    \Infer1[$\ruleFalseLeftTwo$]{\parbox{13.2em}{
      $\forms{a}{b,d} \wedge a{=}c \wedge m{=}1 \wedge m{\ge}2 \entails$\\
      $\exists w. (\forms{a}{b,w} *
      \formp{\Dll}{w,a,c,d,m {\sminus} 1})$}}
    \Infer[rule margin=0em, rule style = no rule]1{\hspace*{13em}}
    \Hypo{ }
    \Infer1[$\rulePureEntail$]{
      m{\ge}2 \entails \exists w. (u{=}w \wedge a{=}a
      \wedge c{=}c \wedge d{=}d \wedge m{\sminus}1{=}m{\sminus}1)}
    \Infer1[$\ruleStarPred$]{
      \formp{\Dll}{u,a,c,d,m {\sminus} 1} \wedge m{\ge}2 \entails
      \exists w. (\formp{\Dll}{w,a,c,d,m {\sminus} 1} \wedge u{=}w)}
    \Infer1[$\ruleStarData$]{
      \forms{a}{b,u} \,{*}\, \formp{\Dll}{u,a,c,d,m {\sminus} 1}
      {\wedge}\, m{\ge}2 \,{\entails}
      \exists w. (\forms{a}{b,w} \,{*}\,
      \formp{\Dll}{w,a,c,d,m {\sminus} 1})}
    \Infer[rule margin=0em, rule style = no rule]1{\hspace*{28em}}
    \Infer2[$\ruleInduction$]{
      L'_3 \triangleq \formp{\Dll}{a,b,c,d,m} \wedge m{\ge}2
      \entails \exists w. (\forms{a}{b,w} * \formp{\Dll}{w,a,c,d,m{\sminus}1})}
  \end{prooftree}
\end{small}
  \vspace{-0.6em}
  \caption{Direct proof tree of the trivial lemma $L'_3$}
  \label{fig:ProofTree:TrivialLemma}
\end{figure}
\vspace{-0.8em}
\vspace{-0.8em}

In this work, we propose a novel framework to synthesize \emph{inductive
  lemmas}, such as $L_1$, $L_2$, and $L_3$, to assist in proving separation
logic entailments. For a given goal entailment $E$, we first identify all
possible lemma templates, which essentially are entailments constructed
from heap predicate symbols appearing in $E$. The templates will be refined
with more Boolean constraints on their variables until \emph{valid}
inductive lemmas are discovered. We will explain the details in Section
\ref{sec:LemmaSynthesis}.

%%% Local Variables:
%%% mode: latex
%%% TeX-master: "main"
%%% End:

\vspace{-0.5em}
\section{Theoretical background}
\label{sec:Background}
\vspace{-0.3em}

Our lemma synthesis framework is developed to assist in proving entailments
in the fragment \emph{of symbolic-heap separation logic} with inductive
heap predicates and linear arithmetic. This fragment is similar to those
introduced in \cite{BrotherstonDP11, BrotherstonGKR16, AlbarghouthiBCK15,
  TaLKC16}. It is also extended with \emph{unknown relations} to represent
Boolean constraints of the desired lemmas. We will present related
background of the entailment proof in the following subsections.

\vspace{-0.8em}
\subsection{Symbolic-heap separation logic with unknown relations}
\vspace{-0.5em}

{\bf Syntax.} We denote our \emph{symbolic-heap separation logic} fragment
with \emph{inductive heap predicates} and \emph{unknown relations} as
$\SLUID$ and present its syntax in Figure \ref{fig:SyntaxMSSL}. In
particular, $x$ is a variable; $c$ and $e$ are respectively an integer
constant and an integer expression\footnote{We write $c\,e$ to denote the
  multiplication by constant in linear arithmetic.}; $\nil$ is a constant
denoting a {\em dangling memory address (null)} and $a$ is a spatial
expression modeling a memory address. Moreover, $\sigma$ denotes a {\em
  spatial atom}, which is either (i) a predicate $\emp$ modeling an {\em
  empty} memory, (ii) a {\em singleton} heap predicate
$\formst{\varSort}{x}{x_1,\ldots,x_n}$ describing an $n$-field data
structure of sort $\iota$ in the memory, pointed-to by $x$, and having
$x_1,\ldots,x_n$ as values of its fields\footnote{Each {\em sort}
  $\varSort$ represents a unique data type. For brevity, we omit using it
  when presenting the motivating example.}, or (iii) an {\em inductive}
heap predicate $\formp{\P}{x_1,\ldots,x_n}$ modeling a recursively defined
data structure (Definition \ref{def:InductivePredicate}). These spatial
atoms constitute a {\em spatial formula} $\Sigma$ via the separating
conjunction operator $*$. On the other hand, $\pi$ denotes a {\em pure
  atom} comprising equality constraints among spatial expressions and
linear arithmetic constraints among integer expressions. These pure atoms
compose a {\em pure formula} $\Pi$ using standard first-order logic
connectives and quantifiers. Moreover, $\Pi$ may contain an {\em unknown
  relation} $\rUnk$ (Definition \ref{def:UnknownRelation}). We will utilize
the unknown relation in various phases of the lemma synthesis.

\vspace{-0.6em}
\vspace{-0.2em}
\begin{figure}[H]
  \begin{tabular}{c}
    \small
    \begin{tabular}{m{1em} m{1.5em} l}

      $e$
      & $\Coloneqq$
      & $c ~|~ x ~|~ -e ~|~ e_1 + e_2 ~|~ e_1 - e_2 ~|~ c\,e$
      \\[1pt]

      $a$
      & $\Coloneqq$
      & $\nil ~|~ x$
      \\[1pt]

      $\pi$
      & $\Coloneqq$
      & $\true ~|~ \false
      ~|~ a_1 \,{=}\, a_2 ~|~ a_1 \,{\neq}\, a_2
      ~|~ e_1 \,{=}\, e_2 ~|~ e_1 \,{\neq}\, e_2
      ~|~ e_1 \,{>}\, e_2 ~|~ e_1 \,{\geq}\, e_2
      ~|~ e_1 \,{<}\, e_2 ~|~ e_1 \,{\leq}\, e_2$
      \\[1pt]

      $\sigma$
      & $\Coloneqq$
      & $\emp ~|~ \formst{\varSort}{x}{x_1,\ldots,x_n} ~|~
         \formp{\P}{x_1,\ldots,x_n}$
      \\[1pt]

      $\Pi$
      & $\Coloneqq$
      & $\pi ~|~ \formr{\rUnk}{x_1,\ldots,x_n} ~|~
        \neg\Pi ~|~ \Pi_1 \wedge \Pi_2 ~|~ \Pi_1 \vee \Pi_2 ~|~
        \Pi_1 \pureImply \Pi_2 ~|~ \forall x .\Pi ~|~ \exists x .\Pi$
      \\[1pt]

      $\Sigma$
      & $\Coloneqq$
      & $\sigma ~|~ \Sigma_1 *  \Sigma_2$
      \\[1pt]

      $F$
      & $\Coloneqq$
      & $\Sigma ~|~ \Pi ~|~ \Sigma \wedge \Pi ~|~ \exists x.\,F$
    \end{tabular}
  \end{tabular}
  \vspace*{-0.8em}
  \caption{Syntax of formulas in \SLUID}
  \label{fig:SyntaxMSSL}
\end{figure}
\vspace{-1.2em}

%%%%%%%%%%%%%%%%%%%%%%%%%%%%%%%%
\vspace{-0.3em}
\begin{definition}[Inductive heap predicate]\cite{TaLKC16}
  \label{def:InductivePredicate}
  A system of $k$ inductive heap predicates $\P_i$ of arity $n_i$, with $i
  = 1, \ldots, k$, is defined as follows:

  \vspace{-0.3em}
  \begin{center}
  \begin{small}
    $\big\{\, \formp{\P_i}{x^i_1, \ldots, x^i_{n_i}}$
    \hspace{2em}$\defeq$\hspace{2em}
    $\formF^i_1(x^i_1, \ldots, x^i_{n_i})
    ~\vee\, \dots \,\vee~
    \formF^i_{m_i}(x^i_1, \ldots, x^i_{n_i}) \,\big\}^k_{i\,=\,1}$
  \end{small}
  \end{center}

  \vspace{-0.3em} where $\formF^i_j$ is a {\em definition case} of $\P_i$.
  This fact is also denoted by $\formF^i_j(x^i_1, \ldots, x^i_{n_i})
  \defrightarrow \formp{\P_i}{x^i_1, \ldots, x^i_{n_i}}$, with $1\leq j
  \leq m_i$. Moreover, $\formF^i_j$ is a {\em base case} if it does not
  contain any predicates recursively defined with $\P_i$; otherwise, it is
  an {\em inductive case}.
\end{definition}

\vspace{-0.5em}
\begin{example}
  The two predicates $\formp{\Dll}{hd,pr,tl,nt,len}$ and
  $\formp{\DllRev}{hd,pr,tl,nt,len}$ in Section \ref{sec:MotivatingExample}
  are two examples of inductive heap predicates. Moreover, their
  definitions are \emph{self-recursively} defined.
\end{example}

\vspace{-0.5em}
\begin{example}
  \label{example:ListEListO}
  The two predicates $\ListE$ and $\ListO$ \cite{BrotherstonDP11} are
  \emph{mutually recursively defined} to model linked list segments which
  respectively contain \emph{even} and \emph{odd} number of elements.

  \vspace{-0.3em}
  \begin{center}
  \begin{small}
    $(1)~\formp{\ListO}{x,y}~\defeq~
    \forms{x}{y} ~\vee~ \exists u. (\forms{x}{u} * \formp{\ListE}{u,y})$
    \hspace{3em}
    $(2)~\formp{\ListE}{x,y} ~\defeq~
    \exists u. (\forms{x}{u} * \formp{\ListO}{u,y})$
  \end{small}
  \end{center}

\end{example}

%%%%%%%%%%%%%%%%%%%%%%%%%%%%%%%%
\vspace{-0.5em}
\begin{definition}[Unknown relation]
  \label{def:UnknownRelation}
  An unknown relation $\formp{\rUnk}{u_1,\ldots.,u_n}$ is an $n$-ary
  pure predicate in first-order logic, whose definition is undefined.

\end{definition}

\vspace{-0.5em}

\textbf{Semantics.} Figure \ref{fig:SemanticsMSSL} exhibits the
semantics of formulas in \SLUID. Given a set $\sVar$ of variables, $\sSort$
of sorts, $\sVal$ of values, $\Loc$ of memory addresses ($\Loc \subset
\sVal$), a model of a formula consists of: a {\em stack} model $s$, which
is a function $s{:} ~ \sVar \rightarrow \sVal$, and a {\em heap} model $h$,
which is a partial function $h{:} ~ (\Loc \times \sSort) \rightharpoonup
\sVal^+$. We write \evalForm{\Pi}{s} to denote the valuation of a pure
formula $\Pi$ under the stack model $s$. Moreover, $\funcDom{h}$ denotes
the domain of $h$; $h \disjoins h'$ indicates that $h$ and $h'$ have
disjoint domains, i.e., $\funcDom{h} \,{\cap}\, \funcDom{h'} \,{=}\,
\setempty$; and $h \hunions h'$ is the union of two disjoint heap models
$h$ and $h'$. Besides, $[f\,|\,x{:}y]$ is a function like $f$ except that
it returns $y$ for the input $x$. To define the semantics of the inductive
heap predicates, we follow the standard \emph{least fixed point semantics}
by interpreting an inductive predicate symbol $\P$ as the least prefixed
point $\semPred{\P}$ of a monotone operator constructed from its inductive
definition. The construction is standard and can be found in many places,
such as \cite{BrotherstonS11}.

%%%%%%%%%%%%%%%%%%%%%%%%%%%%%%%%%%%%
\vspace{-0.5em}
%% Semantics of SepLogic

\begin{figure}[H]
\begin{tabular}{c}
\small
\begin{minipage}{\textwidth}
  \begin{tabular}{lm{8.3em}m{1.5em}l}

  {\hspace*{3em}}
  & $s,h \satisfies \Pi$
  & iff
  & $\evalForm{\Pi}{s}\,{=}\,\true$ and $\funcDom{h}\,{=}\,\setempty$
  \\ [0.2em]

  {} & $s,h \satisfies \emp$
  & iff
  & $\funcDom{h}\,{=}\,\setempty$
  \\ [0.2em]

  {} & $s,h \satisfies \formst{\varSort}{x}{x_1,...,x_n}$
  & iff
  & $\funcDom{h}\,{=}\,\{ \evalVar{x}{s} \}$
  and $h(\evalVar{x}{s}, \varSort) = (\evalVar{x_1}{s}, ...,\evalVar{x_n}{s})$
  \\[0.2em]

  {} & $s,h \satisfies \formp{\P}{x_1,...,x_n}$
  & iff
  & $(h, \evalForm{x_1}{s},...,\evalForm{x_n}{s}) \in \semPred{\P}$
  \\[0.2em]

  {} & $s,h \satisfies \Sigma_1 * \Sigma_2$
  & iff
  & $\exists h_1, h_2:$  $h_1 \,{\disjoins}\, h_2$ and
  $h_1 \,{\hunions}\, h_2\,{=}\,h$ and
  $s,h_1 \,{\satisfies}\, \Sigma_1$ and
  $s,h_2 \,{\satisfies}\, \Sigma_2$
  \\[0.2em]

  {} & $s,h \satisfies \Sigma \wedge \Pi$
  & iff
  & $\evalForm{\Pi}{s}\,{=}\,\true$ and $s,h \satisfies \Sigma$
  \\[0.2em]

  {} & $s,h \satisfies \exists x.\,F$
  & iff
  & $\exists v \,{\in}\, \sVal : [s\,|\,x{:}v],h \satisfies F$
  \\[0.2em]

\end{tabular}
\end{minipage}
\end{tabular}
\vspace*{-1em}
\caption{Semantics of formulas in $\SLUID$}
\vspace*{-0.5em}
\label{fig:SemanticsMSSL}
\end{figure}

\vspace{-0.8em}

\textbf{Entailments.} Given the syntax and semantics of formulas, we can
define entailments as follows. This definition is similar to those in
separation logic's literature, such as \cite{BerdineCO04}.

\begin{definition}[Entailment]
  \label{def:Entailment}
  An entailment between two formulas $F_1$ and $F_2$,
  denoted as $F_1 \entails F_2$,
  is said to be {\em valid}, iff
  $s,h \satisfies F_1$ implies that $s,h \satisfies F_2$,
  for all models $s,h$.
  Formally,\\[0.3em]
  \centerline{$F_1 \entails F_2$ is valid, \quad iff\quad
    $\forall s,h.\,(s,h \satisfies F_1 \mtimply s,h \satisfies F_2)$}
\end{definition}
\vspace{-0.5em}

Here, $F_1$ and $F_2$ are respectively called the antecedent and the
consequent of the entailment. In general, separation logic entailments satisfy a
\emph{transitivity property} as stated in Theorems
\ref{thm:EntailmentTransitivity}.

\begin{theorem}[Entailment transitivity]
  \label{thm:EntailmentTransitivity}
  Given two entailments $F_1 \entails \exists \vec{x}. F_2$ and $F_2
  \entails F_3$, where $\vec{x}$ is a list of variables. If both of them
  are valid, then the entailment $F_1 \entails \exists \vec{x}. F_3$ is
  also valid.
\end{theorem}

\vspace{-1.2em}
\begin{proof}
  Consider an arbitrary model $s,h$ such that $s,h \satisfies F_1$. Since
  $F_1 \entails \exists \vec{x}. F_2$ is valid, it follows that $s,h
  \satisfies \exists \vec{x}. F_2$. By the semantics of $\SLUID$'s
  formulas, $s$ can be extended with values $\vec{v}$ of $\vec{x}$ to
  obtain $s' = [s\,|\,\vec{x}{:}\vec{v}]$ such that $s',h \satisfies F_2$.
  Since $F_2 \entails F_3$ is valid, it follows that $s',h \satisfies F_3$.
  Since $s' = [s\,|\,\vec{x}{:}\vec{v}]$, it is implied by the semantics of
  $\SLUID$'s formulas again that $s,h \satisfies \exists \vec{x}. F_3$. We
  have shown that for an arbitrary model $s,h$, if $s,h \satisfies F_1$,
  then $s,h \satisfies \exists \vec{x}. F_3$. Therefore, $F_1 \entails
  \exists \vec{x}. F_3$ is valid.
\end{proof}

\vspace{-0.5em}
\textbf{Substitution.} We write $[e_1/v_1,\ldots,e_n/v_n]$, or
$\theta$ for short, to denote a simultaneous substitution; and
$F[e_1/v_1,\ldots,$ $e_n/v_n]$ denotes a formula that is obtained from $F$
by \emph{simultaneously replacing} all occurrences of the variables $v_1,
\ldots, v_n$ by the expressions $e_1, \ldots, e_n$. The simultaneous
substitution, or substitution for short, has the following properties,
given that \freevars{F} and \freevars{e} are lists of all free variables
respectively occurring in the formula $F$ and the expression $e$.

\begin{proposition}[Substitution law for formulas~\cite{Reynolds08}]
  \label{prop:SubstitutionLawFormulas}
  Given a separation logic formula $F$ and a substitution $\theta \equiv
  [e_1/v_1,\ldots,e_n/v_n]$. Let $s,h$ be a separation logic model, where
  $\funcDom{s}$ contains $(\freevars{F} \setminus \setenum{v_1,\ldots,v_n})
  \cup \freevars{e_1} \cup \ldots \cup \freevars{e_n}$, and let $\hat{s} =
  [s \,|\, v_1 : \evalForm{e_1}{s} \,|\, \ldots \,|\, v_n : \evalForm{e_n}{s}]$.
  Then $s,h \satisfies F\theta$ if and only if\, $\hat{s},h \satisfies F$
\end{proposition}

\vspace{-0.3em}
\begin{theorem}[Substitution law for entailments]
  \label{thm:SubstitutionLawEntailments}
  Given a separation logic entailment $F_1 \entails F_2$ and a substitution
  $\theta$. If $F_1 \entails F_2$ is valid, then $F_1\theta \entails
  F_2\theta$ is also valid.
\end{theorem}

\vspace{-1.2em}
\begin{proof}
  Suppose that $\theta \equiv [e_1/v_1,\ldots,e_n/v_n]$. Consider an
  arbitrary model $s,h$ such that $s,h \satisfies F_1\theta$. Let $\hat{s}
  = [s ~|~ v_1 : \evalForm{e_1}{s} ~|~ \ldots ~|~ v_n :
    \evalForm{e_n}{s}]$. By Proposition \ref{prop:SubstitutionLawFormulas},
  $s,h \satisfies F_1\theta$ implies that $\hat{s},h \satisfies F_1$. Since
  $F_1 \entails F_2$ is valid, it follows that $\hat{s},h \satisfies F_2$.
  By Proposition \ref{prop:SubstitutionLawFormulas} again, $s,h \satisfies
  F_2\theta$. We have shown that for an arbitrary model $s,h$, if $s,h
  \satisfies F_1\theta$, then $s,h \satisfies F_2\theta$. Therefore,
  $F_1\theta \entails F_2\theta$ is valid.
\end{proof}

\vspace{-0.5em} \textbf{Unknown entailments and unknown assumptions}.
Recall that we propose to synthesize lemmas by firstly discovering
essential lemma templates and then refining them. The template refinement
is conducted in 3 steps: (i) creating \emph{unknown entailments}, which
contain \emph{unknown relations} representing the lemmas' desired pure
constraints, (ii) proving the entailments by induction and collecting
\emph{unknown assumptions} about the relations, and (iii) solving the
assumptions to discover the lemmas. We formally define the \emph{unknown
  entailments} and the \emph{unknown assumptions} as follows.

\vspace{-0.3em}
\begin{definition}[Unknown entailment]
  An entailment $F_1 \entails F_2$ is called an {\em unknown entailment} if
  the antecedent $F_1$ or the consequent $F_2$ contains an unknown relation
  $\formp{\rUnk}{\vec{x}}$.
\end{definition}

\vspace{-0.8em}
\begin{definition}[Unknown assumption]
  A pure implication $\Pi_1 \pureImply \Pi_2$ is called an {\em unknown
    assumption} of the unknown relation $\formp{\rUnk}{\vec{x}}$ if
  $\formp{\rUnk}{\vec{x}}$ occurs in at least one of the two pure formulas
  $\Pi_1$ and $\Pi_2$.
\end{definition}

\vspace{-0.8em} \textbf{Syntactic equivalence.} An entailment induction
proof often contains a step that finds a substitution to unify the
antecedents of an induction hypothesis and of the goal entailment. In this
work, we syntactically check the unification between two spatial formulas
by using a \emph{syntactic equivalence} relation defined earlier in
\cite{TaLKC16}. We formally restate this relation in the below.

\vspace{-0.1em}
\begin{definition}[Syntactic equivalence~\cite{TaLKC16}]
  \label{def:Background:SyntacticEquiv}
  The syntactical equivalence relation
  of two spatial formulas $\Sigma_1$ and $\Sigma_2$,
  denoted as $\Sigma_1 \synequiv \Sigma_2$,
  is inductively defined as follows:\\[-1.6em]

  \begin{small}
  \hspace*{1em}(1)\hspace{0.5em} $\emp ~\,\synequiv~\, \emp$ \qquad
  (2)\hspace{0.5em} $\formst{\varSort}{u}{v_1,\ldots,v_n}
  ~\,\synequiv~\,
  \formst{\varSort}{u}{v_1,\ldots,v_n}$ \qquad
  (3)\hspace{0.5em} $\formp{\P}{u_1,\ldots,u_n} ~\,\synequiv~\,
  \formp{\P}{u_1,\ldots,u_n}$\\
  \hspace*{1em}(4)\hspace{0.5em} $(\Sigma_1 ~\synequiv~ \Sigma'_1)
  ~\wedge~  (\Sigma_2 ~\synequiv~ \Sigma'_2) \quad
  ~\,\,\mtimply~\,\,
  (\Sigma_1 * \Sigma_2 ~\synequiv~ \Sigma'_1 * \Sigma'_2)
  ~\wedge~
  (\Sigma_1 * \Sigma_2 ~\synequiv~ \Sigma'_2 * \Sigma'_1)$
  \end{small}
\end{definition}
\vspace{-0.6em}

%%%%%%%%%%%%%%%%%%%%%%%%%%%%%%%%%%%%%%%%%%%%%%%%%%%%%
%%%%%%%%%%%%%%%%%%%%%%%%%%%%%%%%%%%%%%%%%%%%%%%%%%%%%
\vspace{-0.6em}
\subsection{Structural induction for entailment proofs and lemma
  syntheses}
\label{sec:StructuralInduction}
\vspace{-0.3em}

We develop the lemma synthesis framework based on a standard structural
induction proof. This proof technique is an instance of {\em Noetherian
  induction}, a.k.a. well-founded induction \cite{Bundy01}. We will briefly
explain both Noetherian induction and structural induction here.
Nonetheless, \emph{our lemma synthesis idea can also be integrated into
  other induction-based proof techniques}.

\vspace{-0.1em} {\bf Noetherian induction}. Given $\mathcal{P}$ is a
conjecture on structures of type $\tau$ where $\lttau$ is a {\em
  well-founded relation} among these structures, i.e., there is no infinite
chain like $\ldots \,{\lttau}\, \alpha_n \,{\lttau}\, \ldots \,{\lttau}\,
\alpha_1$. Then the \emph{Noetherian induction principle} states that:
$\mathcal{P}$ is said to hold for all these structures, if for any
structure $\alpha$, the fact that $\mathcal{P}(\beta)$ holds for all
structures $\beta \lttau \alpha$ implies that $\mathcal{P}(\alpha)$ also
holds. Formally:

\vspace*{-0.8em}
\begin{figure}[H]
\small
\begin{prooftree}
  \Hypo{
    \forall \alpha\,{:}\,\tau.~ (
    \forall \beta\,{:}\,\tau
    \prec_{\tau}
    \alpha.~\, \mathcal{P}(\beta))
    \,\mtimply\, \mathcal{P}(\alpha)}
  \Infer1{\forall \alpha\,{:}\,\tau.~\, \mathcal{P}(\alpha)}
\end{prooftree}
\end{figure}
\vspace*{-1.8em}

{\bf Substructural relation.} We prove entailments by applying Noetherian
induction on the structure of inductive heap predicates. The {\em
  substructural relation} is formally defined in Definition
\ref{def:PredSubStructure}.

\vspace{-0.3em}
\begin{definition}[Substructural relation]
  \label{def:PredSubStructure}
  A heap predicate $\formp{\POne}{\vec{v}}$ is said to be a {\em
    substructure} of $\formp{\PTwo}{\vec{u}}$, denoted by
  $\formp{\POne}{\vec{v}} \ltpred \formp{\PTwo}{\vec{u}}$, if
  % $\formp{\PTwo}{\vec{u}}$ is satisfiable, i.e., $\formp{\PTwo}{\vec{u}}
  % \not\equiv \false$, and
  $\formp{\POne}{\vec{v}}$ occurs in a formula obtained from \emph{directly
    unfolding} $\formp{\PTwo}{\vec{u}}$ or from \emph{unfolding any
    substructure} of $\formp{\PTwo}{\vec{u}}$. These conditions are
  formally stated as follows: \vspace{-0.5em}

  \begin{small}
  \begin{enumerate}[noitemsep, leftmargin=2.6em]
  \item[(1)] $\exists \vec{w}, \Sigma, \Pi, \formF(\vec{u}).~
    (\formF(\vec{u}) \synequiv \exists \vec{w}. (\formp{\POne}{\vec{v}}
    \,{*}\, \Sigma \,{\wedge}\, \Pi) ~~\,\,\wedge~~\,\,
    \formF(\vec{u}) \defrightarrow
    \formp{\PTwo}{\vec{u}})$

  \item[(2)] $\exists \vec{w}, \Sigma, \Pi, \formF(\vec{t}),
    \formp{\P'}{\vec{t}}.~ (\formF(\vec{t}) \synequiv \exists \vec{w}.
    (\formp{\POne}{\vec{v}} \,{*}\, \Sigma \,{\wedge}\, \Pi)
    ~~\,\,\wedge~~\,\,
    \formF(\vec{t}) \defrightarrow \formp{\P'}{\vec{t}}
    ~~\,\,\wedge~~\,\,
    \formp{\P'}{\vec{t}} \,{\ltpred}\, \formp{\PTwo}{\vec{u}})$
  \end{enumerate}
  \end{small}
\end{definition}

\vspace*{-0.8em}
In the above definition, $\POne$ and $\PTwo$ can be the same or different
predicate symbols. The latter happens when they are mutually recursively
defined, such as $\ListE$ and $\ListO$ in Example \ref{example:ListEListO}.

%% ~~~~~~~~~~~~~~~~~~~~~~~~~~~~~~~~~~~~~~
\vspace*{-0.2em}
\begin{theorem}[Well-foundedness]
  \label{thm:WellFoundedness}
  Given an inductive heap predicate $\formp{\P_1}{\vec{u}_1}$. If it is
  satisfiable, i.e., $\formp{\P_1}{\vec{u}_1} \not\equiv \false$, then
  under the least fixed point semantics of inductive heap predicates, there
  is no infinite chain like $\ldots \ltpred \formp{\P_n}{\vec{u}_n} \ltpred
  \ldots \ltpred \formp{\P_1}{\vec{u}_1}$.
\end{theorem}

\vspace{-1.2em}
\begin{proof}
  Suppose that there exists an infinite chain $\ldots \ltpred
  \formp{\P_n}{\vec{u}_n} \ltpred \ldots \ltpred \formp{\P_1}{\vec{u}_1}$.
  For all $i \,{\geq}\, 1$, we can always insert all predicates derived
  when unfolding $\formp{\P_{i}}{\vec{u}_{i}}$ to obtain
  $\formp{\P_{i{+}1}}{\vec{u}_{i{+}1}}$ into the current chain. Therefore,
  w.l.o.g., we can assume that $\formp{\P_{i{+}1}}{\vec{u}_{i{+}1}}$ is
  obtained from directly unfolding $\formp{\P_{i}}{\vec{u}_{i}}$, for all
  $i\,{\geq}\,1$. Hence, $\ldots \ltpred \formp{\P_n}{\vec{u}_n} \ltpred
  \ldots \ltpred \formp{\P_1}{\vec{u}_1}$ is also the infinite unfolding
  chain of $\formp{\P_1}{\vec{u}_1}$. In the \emph{least fixed point
    semantics}, if a predicate is unfolded infinitely, it can be evaluated
  to $\false$. Therefore, $\formp{\P_1}{\vec{u}_1} \equiv \false$. This
  contradicts with the theorem's hypothesis that $\formp{\P_1}{\vec{u}_1}$
  is satisfiable.
  % Therefore, the infinite chain $\ldots \ltpred \formp{\P_n}{\vec{u}_n}
  % \ltpred \ldots \ltpred \formp{\P_1}{\vec{u}_1}$ does not exist.
\end{proof}

\vspace{-0.6em} {\bf Structural induction.} We apply the {\em substructural
  relation} $\ltpred$ to propose a {\em structural induction principle} for
the entailment proof. Consider an entailment $E$ whose antecedent contains
an inductive predicate $\formp{\P}{\vec{u}}$, that is, $E \triangleq F_1 *
\formp{\P}{\vec{u}} \entails F_2$, for some formulas $F_1, F_2$. We write
$\formr{E}{\formp{\P}{\vec{u}}}$ to parameterize $E$ by
$\formp{\P}{\vec{u}}$; and $\formr{E}{\formp{\P}{\vec{v}}}$ is an
entailment obtained from $\formr{E}{\formp{\P}{\vec{u}}}$ by replacing
$\formp{\P}{\vec{u}}$ by $\formp{\P}{\vec{v}}$ and respectively replacing
variables in $\vec{u}$ by variables in $\vec{v}$. Moreover, we write
$\satisfies \formr{E}{\formp{\P}{\vec{u}}}$ to denote that $E$ holds for
$\formp{\P}{\vec{u}}$. The structural induction principle is formally
stated as follows:

\vspace*{-0.1em}
\begin{theorem}[Structural induction]
  \label{thm:StructuralInduction}
  The entailment $\formr{E}{\formp{\P}{\vec{u}}} \,{\triangleq}\, F_1
  \,{*}\, \formp{\P}{\vec{u}} \,{\entails}\, F_2$ is valid, if for all
  predicate $\formp{\P}{\vec{u}}$, the fact that $E$ holds for all
  sub-structure predicates $\formp{\P}{\vec{v}}$ of~$\formp{\P}{\vec{u}}$
  implies that $E$ also holds for $\formp{\P}{\vec{u}}$. Formally:

  \vspace*{-0.1em}
  \begin{center}
  \small
  \begin{prooftree}
    \Hypo{
      \forall \formp{\P}{\vec{u}}.~
      (\forall \formp{\P}{\vec{v}} \ltpred \formp{\P}{\vec{u}}.~
       ~\satisfies \formr{E}{\formp{\P}{\vec{v}}})
      \,\mtimply~\,\, \satisfies \formr{E}{\formp{\P}{\vec{u}}}}
    \Infer1{\forall \formp{\P}{\vec{u}}.~
      \satisfies \formr{E}{\formp{\P}{\vec{u}}}}
  \end{prooftree}
  \end{center}
  \vspace*{-0.8em}
\end{theorem}

\vspace*{-0.2em}
\begin{proof}
  We consider two scenarios w.r.t. $\formp{\P}{\vec{u}}$ in the entailment
  $\formr{E}{\formp{\P}{\vec{u}}} \,{\triangleq}\, F_1 \,{*}\,
  \formp{\P}{\vec{u}} \,{\entails}\, F_2$ as follows. (1) If
  $\formp{\P}{\vec{u}} \,{\equiv}\, \false$. Then $F_1 \,{*}\,
  \formp{\P}{\vec{u}} \,{\equiv}\, \false$, therefore
  $\formr{E}{\formp{\P}{\vec{u}}}$ is valid. (2) If $\formp{\P}{\vec{u}}
  \not\equiv \false$. Then by Theorem \ref{thm:WellFoundedness}, the
  substructural relation $\ltpred$, which applies to $\formp{\P}{\vec{u}}$,
  is \emph{well-founded}. In this scenario, the above induction principle
  is an instance of the Noetherian induction \cite{Bundy01} where the
  \emph{substructural relation} $\ltpred$ is chosen as the
  \emph{well-founded relation}. Therefore, the correctness of this theorem
  is automatically implied by the soundness of the Noetherian induction
  principle.
\end{proof}

%%% Local Variables:
%%% mode: latex
%%% TeX-master: "main"
%%% End:

%%%%%%%%%%%%%%%%%%%%%%%%%%%%%%%%%%%%%%%%%%%%%%%%%%%%%%%%%%%%
%%%%%%%%%%%%%%%%%%%%%%%%%%%%%%%%%%%%%%%%%%%%%%%%%%%%%%%%%%%%
\vspace{-0.8em}
\section{The structural induction proof system}
\label{sec:ProofSystem}
\vspace{-0.3em}

In this section, we present a structural induction proof system for
separation logic entailments. Initially, this system aims to prove
\emph{normal entailments} using a set of inference rules (Section
\ref{sec:InferenceRules}). These rules include the two lemma application
rules which apply synthesized lemmas to assist in proving entailments
(Figure \ref{fig:LemmaRules}). We use the same proof system to reason about
\emph{unknown entailments}, introduced in the lemma synthesis, using a set
of synthesis rules (Section \ref{sec:SynthesisRules}). The proof system
also includes a proof search procedure which selectively applies the
aforementioned rules to prove a goal entailment (Section
\ref{sec:ProofSearchProcedure}). We will explain the details in the
following sections.

\vspace{-0.6em}
\subsection{Inference rules for standard entailments}
\label{sec:InferenceRules}
\vspace{-0.3em}

Each inference rule of our proof system contains zero or more premises, a
conclusion, and possibly a side condition. The premises and the conclusion
are in the form of $\Hypos,~\Lemmas,~ F_1 \entails F_2$, where $\Hypos$
and $\Lemmas$ are respectively sets of induction hypotheses and valid
lemmas, and $F_1 \entails F_2$ is the (sub-)goal entailment. An inference
rule can be interpreted as follows: {\em if all entailments in its premises
  are valid, and its side condition (if present) is satisfied, then its goal
  entailment is also valid}.

\vspace{-0.1em}
For brevity, we write $F$ to denote a symbolic-heap formula $\exists
\vec{x}. (\Sigma \wedge \Pi)$, where $\vec{x}$ is a list of quantified
variables (possibly empty). Furthermore, we define $F * \Sigma' \triangleq
\exists \vec{x}. (\Sigma * \Sigma'\!\wedge \Pi)$ and $F \wedge \Pi'
\triangleq \exists \vec{x}. (\Sigma \wedge \Pi \wedge \Pi')$, given that
$\freevars{\Sigma'} \,{\cap}\, \vec{x} = \setempty$ and $\freevars{\Pi'}
\,{\cap}\, \vec{x} = \setempty$. We also write $\vec{u} = \vec{v}$ to
denote $(u_1 {=}\, v_1) \wedge \ldots \wedge (u_n {=}\, v_n)$, given that
$\vec{u} \triangleq u_1, \ldots ,u_n$ and $\vec{v} \triangleq v_1, \ldots
,v_n$ are two variable lists of the same size. Finally, $\vec{u} \disjoins
\vec{v}$ indicates that the two lists $\vec{u}$ and $\vec{v}$ are disjoint,
i.e., $\nexists w. (w \,{\in}\, \vec{u} \wedge w \,{\in}\, \vec{v})$.

\vspace{-0.1em}
The set of inference rules comprises logical rules (Figure
\ref{fig:LogicalRules}) reasoning about the logical structure
of the entailments, induction rules (Figure \ref{fig:InductionRules})
proving the entailments by structural induction,
and lemma application rules (Figure \ref{fig:LemmaRules})
applying supporting lemmas to assist in proving the entailments.

\begin{figure}[H]
  \begin{small}
  \begin{tabular}{ll}
    \multicolumn{2}{l}{
      \minipageFalseLeftOne\hspace{1em}
      \minipageFalseLeftTwo\hspace{1em}
      \minipagePureEntail}
    \\[1.15em]

    \multicolumn{2}{l}{
      \minipageEqualLeft\hspace{2.4em}
      \minipageEmpLeft\hspace{4.5em}
      \minipageEmpRight}
    \\[1.4em]

    \minipageExistsLeft
    & \minipageCaseAnalysis
    \\[1.3em]

    \minipageExistsRight
    & \minipageStarData
    \\[1.3em]

    \minipagePredIntroRight\hspace{2em}
    & \minipageStarPred
  \end{tabular}
  \end{small}
  \vspace{-0.6em}
  \caption{Logical rules}
  \label{fig:LogicalRules}
\end{figure}

\vspace{-1.2em}
\textbf{Logical rules.} The set of logical rules in Figure
\ref{fig:LogicalRules} consists of:
\vspace{-0.1em}

\begin{itemize}[noitemsep, leftmargin=1em]
\setlength\itemsep{0.2em}
\item {\em Axiom rules} $\rulePureEntail, \ruleFalseLeftOne,
  \ruleFalseLeftTwo$. These rules prove pure entailments by invoking an
  off-the-shelf prover such as Z3~\cite{MouraB08} (as in the rule
  $\rulePureEntail$), or prove entailments whose antecedents are
  inconsistent, i.e., they contain overlaid singleton heaps
  ($\formst{\varSort_1}{u}{\vec{v}}$ and $\formst{\varSort_2}{u}{\vec{t}}$
  in the rule $\ruleFalseLeftOne$) or contradictions ($\Pi_1 \rightarrow
  \false$ in the rule $\ruleFalseLeftTwo$).

\item {\em Normalization rules} $\ruleExistsLeft, \ruleExistsRight,
  \ruleEqualLeft, \ruleEmpLeft, \ruleEmpRight$. These rules simplify the
  goal entailment by either eliminating existentially quantified variables
  ($\ruleExistsLeft, \ruleExistsRight$), or removing equalities
  ($\ruleEqualLeft$) or empty heap predicates ($\ruleEmpLeft,
  \ruleEmpRight$) from the entailment.

\item {\em Case analysis rule} $\ruleCaseAnalysis$. This rule performs a
  case analysis on a pure condition $\Pi$ by deriving two sub-goal
  entailments whose antecedents respectively contain $\Pi$ and $\neg\Pi$.
  The underlying principle of this rule is known as the \emph{law of
    excluded middle} \cite{WhiteheadR12}.

\item \emph{Unfolding rule} $\rulePredRight$. This rule derives a new
  entailment by unfolding a heap predicate in the goal entailment’s
  consequent by its inductive definition.

\item {\em Matching rules} $\ruleStarData,\ruleStarPred$. These rules
  remove {\em identical} singleton heap predicates ($\ruleStarData$) or
  inductive heap predicates ($\ruleStarPred$) from two sides of the goal
  entailments. Here, we ensure that these predicates are identical by
  adding equality constraints about their parameters into the derived
  entailments' consequents.
\end{itemize}

\vspace{-0.1em}
\begin{figure}[H]
  \begin{small}
    \begin{tabular}{l}
      \hspace{-1.5em}
      \minipageInduction
      \\[2.2em]

      \hspace{-1.5em}
      \minipageHypothesis
    \end{tabular}
  \end{small}

  \vspace{-0.6em}
  \caption{Induction rules}
  \label{fig:InductionRules}
\end{figure}

\vspace{-1.3em} \textbf{Induction rules.} The structural induction
principle is integrated into our proof system via the two induction rules
$\ruleInduction$ and $\ruleHypo$ (Figure \ref{fig:InductionRules}). These
rules are explained in details as follows:

\begin{itemize}[leftmargin=1em]
\item \emph{Rule $\ruleInduction$}. This rule performs the structural
  induction proof on a chosen heap predicate $\formp{\P}{\vec{u}}$ in the
  antecedent of the goal entailment $\Sigma_1 * \formp{\P}{\vec{u}} \wedge
  \Pi_1 \entails F_2$. In particular, the predicate $\formp{\P}{\vec{u}}$
  is unfolded by its inductive definition $\formp{\P}{\vec{u}} \definedas
  \formp{\formF^{\P}_1}{\vec{u}} \,{\vee} ... {\vee}\,
  \formp{\formF^{\P}_m}{\vec{u}}$ to derive new sub-goal entailments as
  shown in this rule's premises. Moreover, the goal entailment is also
  recorded as an induction hypothesis in the set $\Hypos'$ so that it can
  be utilized later to prove the sub-goal entailments.

\item \emph{Rule $\ruleHypo$}. This rule applies an appropriate recorded
  induction hypothesis $H \triangleq \Sigma_3 {*} \formp{\P}{\vec{v}}
  {\wedge} \Pi_3 \,{\entails}\, F_4$ to prove the goal entailment $\Sigma_1
  \,{*}\, \formp{\P}{\vec{u}} \,{\wedge}\, \Pi_1 \,{\entails}\, F_2$. It
  first checks the well-founded condition $\formp{\P}{\vec{u}} \ltpred
  \formp{\P}{\vec{v}}$, i.e. $\formp{\P}{\vec{u}}$ is a substructure of
  $\formp{\P}{\vec{v}}$, which is required by the structural induction
  principle. In practice, this condition can be easily examined by labeling
  each inductive heap predicate $\formp{\Q}{\vec{x}}$ in a proof tree with
  a set of its ancestor predicates, which are consecutively unfolded to
  derive $\formp{\Q}{\vec{x}}$. By that mean, $\formp{\P}{\vec{u}} \ltpred
  \formp{\P}{\vec{v}}$ iff $\formp{\P}{\vec{v}}$ appears in the label of
  $\formp{\P}{\vec{u}}$. Afterwards, the induction hypothesis application
  is performed in two steps:

  -- \emph{Unification step:} unify the antecedents of both the goal
  entailment and the induction hypothesis by syntactically finding a
  substitution $\theta$ and a spatial formula $\Sigma$ such that
  $\Sigma_1 \,{*}\, \formp{\P}{\vec{u}} \synequiv
  \Sigma_3\theta \,{*}\, \formp{\P}{\vec{v}}\theta \,{*}\, \Sigma$
  and $\Pi_1 {\rightarrow} \Pi_3\theta$. If these conditions hold, then it is
  certain that the entailment
  $E \triangleq \Sigma_1 * \formp{\P}{\vec{u}} \wedge \Pi_1 \entails
  \Sigma_3\theta * \formp{\P}{\vec{v}}\theta * \Sigma \wedge
  \Pi_3\theta \wedge \Pi_1$ is valid.

  -- \emph{Proving step:} if such $\theta$ and $\Sigma$ exist, then derive
  a new sub-goal entailment $F_4\theta * \Sigma \wedge \Pi_1 \entails F_2$.
  We will explain why this sub-goal entailment is derived. The induction
  hypothesis $\Sigma_3 * \formp{\P}{\vec{v}} \wedge \Pi_3 \entails F_4$
  implies that $H\theta \triangleq \Sigma_3\theta *
  \formp{\P}{\vec{v}}\theta \wedge \Pi_3\theta \entails F_4\theta$ is also
  valid, by Theorem \ref{thm:SubstitutionLawEntailments}. From $E$ and
  $H\theta$, we then have a derivation chain: $\Sigma_1 *
  \formp{\P}{\vec{u}} \wedge \Pi_1 \entails \Sigma_3\theta *
  \formp{\P}{\vec{v}}\theta * \Sigma \wedge \Pi_3\theta \wedge \Pi_1
  \entails F_4\theta * \Sigma \wedge \Pi_1$. Therefore, if the sub-goal
  entailment $F_4\theta * \Sigma \wedge \Pi_1 \entails F_2$ can be proved,
  then the goal entailment holds. Here, we propagate the pure condition
  $\Pi_1$ through the chain as we want the antecedent of the sub-goal
  entailment to be the strongest in order to prove $F_2$.
\end{itemize}

\vspace{0.8em}
\setlength{\intextsep}{0.3em}
\begin{figure}[H]
  \begin{small}
    \begin{tabular}{l}

      \hspace{-5em}
      \minipageLemmaLeft
      \\[1.5em]

      \hspace{-5em}
      \minipageLemmaRight
    \end{tabular}
  \end{small}

  \vspace{-0.3em}
  \caption{Lemma application rules}
  \label{fig:LemmaRules}
\end{figure}

\vspace{-0.3em}

\textbf{Lemma application rules.} The two rules
$\ruleLemmaLeft, \ruleLemmaRight$ in Figure \ref{fig:LemmaRules} derive a
new sub-goal entailment by applying a lemma on the goal entailment's
antecedent or consequent. In particular:

\begin{itemize}[leftmargin=1em]
\item The rule $\ruleLemmaLeft$ applies a lemma on the goal entailment's
  antecedent. It is similar to the induction application rule $\ruleHypo$,
  except that we do not need to check the well-founded condition of the
  structural induction proof, since the applied lemma is already proved
  valid.

\item The rule $\ruleLemmaRight$ applies a lemma on the goal entailment's
  consequent. It also needs to perform an unification step: finding a
  substitution $\theta$ and a formula $\Sigma$ so that the heap parts of
  the goal entailment's and the lemma's consequents are unified, i.e.,
  $\Sigma_4\theta * \Sigma \synequiv \Sigma_2$. If this step succeeds, then
  $E_1 \triangleq \Sigma_4\theta * \Sigma \wedge \Pi_2 \entails \Sigma_2
  \wedge \Pi_2$ is valid. The lemma $F_3 \entails \exists \vec{w}.(\Sigma_4
  \wedge \Pi_4)$ implies that $F_3 \entails \exists \vec{w}.\Sigma_4$ is
  valid. Hence, $F_3\theta \entails \exists \vec{w}\theta.\Sigma_4\theta$
  is valid by Theorem \ref{thm:SubstitutionLawEntailments}, following that
  $F_3\theta * \Sigma \wedge \Pi_2 \entails \exists
  \vec{w}\theta.(\Sigma_4\theta * \Sigma \wedge \Pi_2)$ is also valid. By
  the rule's side condition $\vec{w}\theta \subseteq \vec{x}$, we have
  another valid entailment $E_2 \triangleq F_3\theta * \Sigma \wedge \Pi_2
  \entails \exists \vec{x}.(\Sigma_4\theta * \Sigma \wedge \Pi_2)$.
  Therefore, by proving the entailment $F_1 \entails \exists
  \vec{x}.(F_3\theta * \Sigma \wedge \Pi_2)$ in this rule's premise and
  applying Theorem \ref{thm:EntailmentTransitivity} twice sequentially on
  $E_2$ and then $E_1$, we can conclude that the goal entailment $F_1
  \entails \exists \vec{x}.(\Sigma_2 \wedge \Pi_2)$ is also valid.

\end{itemize}

%% ~~~~~~~~~~~~~~~~~~~~~~~~~~~~~~~~~~~~~~~~~
\vspace{-1em}
\subsection{Synthesis rules for unknown entailments}
\label{sec:SynthesisRules}
\vspace{-0.3em}

Figure \ref{fig:SynthesisRules} presents synthesis rules, which deal with
unknown entailments introduced in the lemma synthesis. These rules share
the similar structure to the inference rules, except that each of them also
contains a special premise indicating an assumption set $\Assumpts$ of the
unknown relations. We will describe in details the synthesis rules as
follows:

\begin{itemize}[leftmargin=1.2em]
\vspace{-0.2em}
\item {\em Axiom synthesis rules} $\ruleSynPiOne, \ruleSynPiTwo,
  \ruleSynSigmaOne, \ruleSynSigmaTwo$. These rules conclude the validity of
  their unknown goal entailments under the assumption sets $\Assumpts$ in
  the rules' premises. The rule $\ruleSynPiOne$ and $\ruleSynPiTwo$ handle
  pure entailments with unknown relations either in the antecedents or the
  consequents. The rule $\ruleSynSigmaOne$ and $\ruleSynSigmaTwo$ deal with
  unknown entailments whose antecedents contain non-empty spatial formulas
  while the consequents are pure formulas or vice versa. In both cases, the
  antecedents must be inconsistent to make the goal entailments valid.
  Here, we only create assumptions on the pure parts, i.e., $\Pi_1 \wedge
  \formr{\rUnk} {\vec{x}} \pureImply \false$, since inconsistency in their
  heap parts, if any, can be detected earlier by the rule
  $\ruleFalseLeftOne$. Also, we do not consider the case that the unknown
  relations only appear in the consequents since these relations cannot
  make the antecedents inconsistent.

\vspace{-0.2em}
\item {\em Induction hypothesis synthesis rule} $\ruleSynHypo$. This rule
  applies an induction hypothesis to prove a derived unknown entailment.
  The induction hypothesis also contains unknown relations since it is
  recorded earlier from an unknown goal entailment. The rule $\ruleSynHypo$
  is similar to the normal induction hypothesis application rule
  $\ruleHypo$, except that it does not contain a side condition like $\Pi_1
  \wedge \formr{\rUnk}{\vec{x}} \pureImply (\Pi_3 \wedge
  \formr{\rUnk}{\vec{y}})\theta$, due to the appearance of the unknown
  relation $\rUnk$. Instead, this condition will be registered in the
  unknown assumption set $\Assumpts$ in the premises of $\ruleSynHypo$.

\end{itemize}

\vspace{0.5em}
\begin{figure}[H]
  \begin{small}
    \begin{tabular}{ll}
      \minipagePiOne\hspace{10em}
      & \minipageSigmaOne
      \\[1.4em]

      \minipagePiTwo
      & \minipageSigmaTwo
      \\[1.4em]

      \multicolumn{2}{l}{\hspace*{-0.8em}\minipageSynHypo}
      \\[1em]
    \end{tabular}
  \end{small}

  \vspace{-0.8em}
  \caption{Synthesis rules}
  \label{fig:SynthesisRules}
\end{figure}

%%%%%%%%%%%%%%%%%%%%%%%%%%%%%%%%%%%%%%%%%%%%%%%%%%%%%%%
\vspace{-1.8em}
\subsection{The entailment proving procedure}
\label{sec:ProofSearchProcedure}
\vspace{-0.3em}

Figure \ref{fig:ProofSearch} presents the core procedure $\Prove$ of our
proof system. In particular, the input of $\Prove$ includes an induction
hypothesis set $\Hypos$, a valid lemma set $\Lemmas$, and a goal entailment
$F_1 \entails F_2$. These three inputs correlate to an inference step of
the proof system. We also use an additional parameter $\ProofMode$ to
control when new lemmas can be synthesized (if $\ProofMode = \SynLemma$) or
strictly not (if $\ProofMode = \NoSynLemma$). The procedure $\Prove$
returns the validity of the goal entailment, a set of new lemmas
synthesized during the proof, and a set of assumptions to make the
entailment valid.

There are two main contexts where this procedure is initially invoked:

\vspace{-0.2em}
\begin{itemize}[label=--, leftmargin=1.2em]

\item In the {\em entailment proving} phase, $\Prove$ is invoked to prove a
  {\em normal} entailment $F_1 \entails F_2$, which does not contain any
  unknown relation, with the initial setting $\Hypos \,{=}\, \setempty,
  \Lemmas \,{=}\, \setempty, \ProofMode \,{=}\, \SynLemma$.

\vspace{-0.3em}
\item In the {\em lemma synthesis} phase, $\Prove$ is invoked either to
  prove an {\em unknown} entailment $F_1 \entails F_2$ related to a lemma
  template, or to verify whether a discovered lemma is an inductive lemma.
  In the first case, $\Prove$ will establish sufficient assumptions about
  unknown relations appearing in the entailment so that the entailment can
  become valid. Moreover, $\Prove$ is invoked with $\ProofMode =
  \NoSynLemma$ in both the two cases to avoid entering into nested lemma
  synthesis phases.
\end{itemize}
\vspace{-0.2em}

In Figure \ref{fig:ProofSearch}, we also present formal specifications of
$\Prove$ in pairs of pre- and post-conditions
($\Summary{Requires/Ensures}$). These specifications relate to three cases
when $\Prove$ is invoked to prove an \emph{unknown entailment} with the
lemma synthesis always being disabled ($\ProofMode = \NoSynLemma$), or to
prove a \emph{normal entailment} with the lemma synthesis being disabled
($\ProofMode = \NoSynLemma$) or enabled ($\ProofMode = \SynLemma$). We
write $\res$ to represent the returned result of $\Prove$. In addition,
$\hasUnk{E}$ indicates that the entailment $E$ has an unknown relation,
$\validlem{\Lemmas}$ and $\validlem{\Assumpts}$ mean that all lemmas in
$\Lemmas$ and all assumptions in $\Assumpts$ are {\em semantically valid}.
Moreover, $\validindt{\Hypos, \Lemmas, E}$ specifies that the entailment
$E$ is \emph{semantically valid} under the induction hypothesis set
$\Hypos$ and the lemma set $\Lemmas$. We will refer to these specifications
when proving the proof system's soundness. The formal verification of
$\Prove$ w.r.t. these specifications is illustrated in
  {\iftechreport
  	Appendix \ref{append:HoareProof}.
   \else
   	the technical report \cite{TaLKC17}.
   \fi}

\vspace{-0.1em}
\vspace{0.5em}
\begin{figure}[H]
  \algrenewcommand\algorithmicindent{2em}
  \begin{algorithm}[H]
    \small
    \caption*{\small{\bf Procedure} \Call{\Prove}{
        \Hypos, \Lemmas, F_1 \entails F_2, \ProofMode}}
    \begin{flushleft}
      \Summary{Input:}
      $F_1 \entails F_2$ is the goal entailment,
      $\Hypos$ is a set of induction hypotheses,
      $\Lemmas$ is a set of valid lemmas, and\\
      \hspace{2.4em} $\ProofMode$ controls whether new lemmas will be
      synthesized ($\SynLemma$) or strictly not ($\NoSynLemma$)\\
      \Summary{Output:} The goal entailment's validity
      ($\ValidWitness{\xi}$ or $\Unknown$,
      where $\xi$ is the witness valid proof tree),\\
      \hspace{3.3em}a set of new synthesized lemmas,
      and a set of unknown assumptions\\[0.3em]
      \Summary{Requires:}
      $(\ProofMode \,{=}\, \NoSynLemma)
      \wedge \hasUnk{F_1 {\entails} F_2}
      \wedge \validlem{\Lemmas}$\\
      \Summary{Ensures:} $(\res \,{=}\, (\Unknown, \setempty, \setempty))
      ~\vee~
      \exists\xi,\Assumpts.
      ((\res \,{=}\, (\ValidWitness{\xi}, \setempty, \Assumpts))
      \wedge (\validlem{\Assumpts} \rightarrow
      \validindt{\Hypos, \Lemmas, F_1 {\entails} F_2}))$\\[0.3em]
      \Summary{Requires:}
      $(\ProofMode \,{=}\, \NoSynLemma)
      \wedge \neg\hasUnk{F_1 {\entails} F_2}
      \wedge \validlem{\Lemmas}$\\
      \Summary{Ensures:} $(\res \,{=}\, (\Unknown, \setempty, \setempty))
      ~\vee~
      \exists\xi.
      ((\res \,{=}\, (\ValidWitness{\xi}, \setempty, \setempty))
      \wedge \validindt{\Hypos, \Lemmas, F_1 {\entails} F_2})$\\[0.3em]
      \Summary{Requires:}
      $(\ProofMode \,{=}\, \SynLemma)
      \wedge \neg\hasUnk{F_1 {\entails} F_2}
      \wedge \validlem{\Lemmas}$\\
      \Summary{Ensures:} $(\res \,{=}\, (\Unknown, \setempty, \setempty))
      \,{\vee}\,
      \exists\xi,\!\LemmasSyn.
      ((\res \,{=}\, (\ValidWitness{\xi}, \LemmasSyn, \setempty))
      \,{\wedge}\, \validlem{\!\LemmasSyn}
      \,{\wedge}\, \validindt{\Hypos, \Lemmas, F_1 {\entails} F_2})$\\[0.3em]
    \end{flushleft}
    \begin{algorithmic}[1]
      \def\indent{\hspace{1.2em}}
      \State \Assign{\sRuleSelected}{\setenum{
          \Call{\Unify}{\vRule, (\Hypos, \Lemmas, F_1 \entails F_2)}
          ~|~ \vRule \in \sRule}}
      \label{line:UnifyRules}
      \Comment{Find applicable inference and synthesis rules from $\sRule$}
      \State \Assign{\LemmasSyn}{\setempty}
      \Comment{Initialize the new synthesized lemma set}
      \If{(\ProofMode = \SynLemma) \Keyword{and}
          \Call{\NeedLemmas}{F_1 \entails F_2, \sRuleSelected}}
        \label{line:NeedLemmas}
        % \Comment{if it is needed,}
        \State \Assign{\LemmasSyn}{
          \Call{\SynthesizeLemma}{\Lemmas,\,F_1 \entails F_2}}
        \label{line:SynNewLemma}
        \Comment{Synthesize new lemmas}
        \State \Assign{\sRuleSelected}{
          \sRuleSelected \cup \setenum{\Call{\Unify}{
              \vRule, (\Hypos, \LemmasSyn, F_1 \entails F_2)}
            ~|~ \vRule \in \{\ruleLemmaLeft, \ruleLemmaRight\}}}
        \label{line:FindNewLemmaRule}
        \Comment{Update lemma application rules}
      \EndIf
      \For{\Keyword{each} \vRuleInst \Keyword{in} \sRuleSelected}
        % \Comment{rules $\ruleLemmaLeft, \ruleLemmaRight$ have the highest
        %   priority in $\sRuleSelected$, if existing}
        \label{line:ProveSubGoalsBegin} \label{line:FindRule}
        \State \Assign{\Assumpts}{\setempty}
        \Comment{Initialize the assumption set}
        \If{\Call{IsSynthesisRule}{\vRuleInst}}
          \Assign{\Assumpts}{\Call{\Assumptions}{\vRuleInst}}
          \Comment{$\vRuleInst \in \setenum{
              \ruleSynPiOne, \ruleSynPiTwo,
              \ruleSynSigmaOne, \ruleSynSigmaTwo, \ruleSynHypo}$}
        \EndIf
        \label{line:CollectAssumptions}
        \If{\Call{IsAxiomRule}{\vRuleInst}} %
          \label{line:AxiomRuleBegin}
          \Comment{$\vRuleInst \in \setenum{
              \rulePureEntail, \ruleFalseLeftOne, \ruleFalseLeftTwo,
              \ruleSynPiOne, \ruleSynPiTwo, \ruleSynSigmaOne,
              \ruleSynSigmaTwo}$}
          \State \Assign{\xi}{\Call{\CreateWitnessProofTree}{
              F_1 \entails F_2, \vRuleInst, \setempty}}
          \label{line:CreateProofTreeAxiom}
          \State \ReturnTuple{\ValidWitness{\xi}, \LemmasSyn, \Assumpts}
        \EndIf{}
        \label{line:AxiomRuleEnd}
        \label{line:SynRuleEnd}
        \label{line:PremiseBegin}
        \State \Assign{
          (\Hypos_i, \Lemmas_i, F_{1i} \entails F_{2i})_{i\,=\,1 \ldots n}}{
          \Call{\Premises}{\vRuleInst}}
        \label{line:OneRuleProveBegin}
        \State \Assign{(\vR_i,
          \Lemmas^i_{\mathrm{syn}},\Assumpts_i)_{i\,=\,1 \ldots n}}{
          \Call{\Prove}{
            \Hypos_i,\, \Lemmas \cup \LemmasSyn \cup \Lemmas_i,\,F_{1i} \entails
            F_{2i}, \ProofMode}_{i\,=\,1 \ldots n}}
        \Comment{Prove all sub-goals}
        \If{\vR_i = \ValidWitness{\xi_i}\Keyword{for~all~}i = 1 \ldots n}
          \State \Assign{\xi}{\Call{\CreateWitnessProofTree}{
              F_1 \entails F_2, \vRuleInst, \setenum{\xi_1, \ldots,
                \xi_n}}}
          \label{line:CreateProofTreeSubtrees}
          \State \ReturnTuple{\ValidWitness{\xi},~
            \LemmasSyn \cup \Lemmas^1_{\mathrm{syn}} \cup \ldots
            \cup \Lemmas^n_{\mathrm{syn}},~
            \Assumpts \cup \Assumpts_1 \cup \ldots \cup \Assumpts_n}
          \label{line:OneRuleProveEnd}
        \EndIf
        \label{line:ProveSubGoalsEnd}
      \EndFor{}
      \label{line:PremiseEnd}
      \State{\ReturnTuple{\Unknown, \setempty, \setempty}}
      \Comment{All rules fail to prove $F_1 \entails F_2$}
      \label{line:AllRulesFail}
      \algstore{myalg}
    \end{algorithmic}
  \end{algorithm}
  \vspace{-0.6em}
  \caption{The main proof search procedure with description
    (Input/Output) and specification (Requires/Ensures)}
  \label{fig:ProofSearch}
\end{figure}

%%% Local Variables:
%%% mode: latex
%%% TeX-master: "main"
%%% End:

\vspace{-0.2em}

% \wnsay{Is it possible for Prove to go into a infinite reucrsion?
% Need to say we have placed some bound?}
Given the goal entailment $F_1 \entails F_2$, the procedure $\Prove$ first
finds from all rules $\sRule$ (Figures \ref{fig:LogicalRules},
\ref{fig:InductionRules}, \ref{fig:LemmaRules}, and
\ref{fig:SynthesisRules}) a set of potential inference and synthesis rules
$\sRuleSelected$, which can be unified with $F_1 \entails F_2$ (line
\ref{line:UnifyRules}). When the lemma synthesis mode is enabled
($\ProofMode = \SynLemma$), it invokes a subroutine $\NeedLemmas$ to
examine the selected rules $\sRuleSelected$ and the goal entailment $F_1
\entails F_2$ to decide whether it really needs to synthesize new lemmas
(line \ref{line:NeedLemmas}). Note that the input valid lemma set $\Lemmas$
is also exploited to make the lemma synthesis more effective (line
\ref{line:SynNewLemma}). The new synthesized lemmas, if any, will be
utilized to discover new lemma application rules (lines
\ref{line:FindNewLemmaRule}). Thereafter, $\Prove$ successively applies
each rule $\vRuleInst \in \sRuleSelected$ to derive new sub-goal
entailments, as in the premises of $\vRuleInst$, and recursively searches
for their proofs (line \ref{line:ProveSubGoalsBegin} {--}
\ref{line:ProveSubGoalsEnd}). It returns the {\em valid} result
($\ValidWitness{\xi}$, where $\xi$ is the witness proof tree) if the
selected rule $\vRuleInst$ does not introduce any new sub-goals (lines
\ref{line:AxiomRuleBegin} -- \ref{line:AxiomRuleEnd}), or all derived
sub-goals are successfully proved (line \ref{line:OneRuleProveBegin} {--}
\ref{line:OneRuleProveEnd}). In essence, the proof tree $\xi$ is composed
of a root (the goal entailment $F_1 \entails F_2$), a label (the applied
rule $\vRuleInst$), and sub-trees, if any, corresponding to the sub-goal
entailments' proofs (lines \ref{line:CreateProofTreeAxiom} and
\ref{line:CreateProofTreeSubtrees}). Its form is intuitively similar to the
proof trees depicted in Figures \ref{fig:ProofTreeMotiv},
\ref{fig:ProofTreeLemmaL1}, and \ref{fig:ProofTree:TrivialLemma}. On the
other hand, $\Prove$ announces the {\em unknown} result $(\Unknown)$ when
all selected rules in $\sRuleSelected$ fail to prove the goal entailment
(line \ref{line:AllRulesFail}). Besides, $\Prove$ also returns a set of new
synthesized lemmas and a set of unknown assumptions. These lemmas are
discovered when $\Prove$ is invoked in the {\em entailment proving} phase.
The unknown assumptions are collected by the synthesis rules (line
\ref{line:CollectAssumptions}), when $\Prove$ is executed in the {\em lemma
  synthesis} phase.

Details of the lemma synthesis will be presented in
Section~\ref{sec:LemmaSynthesis}. In the following, we will explain when
$\Prove$ decides to synthesize new lemmas (line \ref{line:NeedLemmas}). The
procedure $\NeedLemmas$ returns {\em true} when all the following
conditions are satisfied:
\vspace{-0.1em}

\begin{enumerate}[label=--, noitemsep]
\setlength\itemsep{0.2em}
\item $F_1 \,{\entails}\, F_2$ \emph{is not} an unknown entailment, which
  implies that lemmas are possibly needed.
\item $\sRuleSelected$ \emph{does not contain} any axiom or normalization
  ($\rulePureEntail, \ruleFalseLeftOne,\ruleFalseLeftTwo, \ruleExistsLeft,
  \ruleExistsRight, \ruleEqualLeft, \ruleEmpLeft,\ruleEmpRight$), matching
  rules of identical heap predicates ($\ruleStarData, \ruleStarPred$), or
  unfolding rules that introduce identical predicates ($\ruleInduction,
  \rulePredRight$). This condition indicates that all rules in
  $\sRuleSelected$ cannot make any immediate proof progress.
\item $\sRuleSelected$ \emph{does not have} any induction hypothesis or
  lemma application rules ($\ruleHypo, \ruleLemmaLeft, \ruleLemmaRight$),
  or any case analysis rule ($\ruleCaseAnalysis$) that potentially leads to
  the application of an induction hypothesis or a lemma. This condition
  conveys that existing induction hypotheses and lemmas are inapplicable
\item $F_1 \,{\entails}\, F_2$ \emph{is not} a good induction hypothesis
  candidate, which indicates that the induction hypothesis recorded from
  $F_1 \entails F_2$ cannot be used to prove other derived entailments.
\end{enumerate}
\vspace{-0.1em}

While the first three conditions can be checked syntactically on $F_1
\entails F_2$ and $\sRuleSelected$, the last condition can be tested by a
trial and error method. Specifically, $F_1 \entails F_2$ will firstly be
recorded as a temporary induction hypothesis candidate, and then each
inductive heap predicate in $F_1$ will be consecutively unfolded to search
for an application of the recorded induction hypothesis via the rule
$\ruleHypo$. If no induction hypothesis application can be found, then $F_1
\entails F_2$ is evidently not a good candidate.

\vspace{-0.7em}
\subsection{Soundness of the proof system}
\vspace{-0.3em}

Recall that our proof search procedure $\Prove$ is implemented in a
recursive manner. When it is invoked to prove a normal goal entailment $E$,
which does not contain any unknown relation, the initial setting $\Hypos
\,{=}\, \setempty, \Lemmas \,{=}\, \setempty, \ProofMode \,{=}\, \SynLemma$
indicates that no induction hypothesis or lemma is provided beforehand, and
the proof system can synthesize new lemmas to assist in proving $E$. When
synthesizing the new supporting lemmas, the proof system can be utilized to
prove an \emph{unknown entailment} related to a lemma template or to verify
a discovered lemma, which is a \emph{normal entailment} not containing any
unknown relation. In addition, the proof system is also invoked to prove
sub-goal entailments, which are \emph{normal entailments} derived from $E$.
All of these scenarios are summarized by the three specifications in Figure
\ref{fig:ProofSearch}.

In the following, we present Propositions
\ref{thm:SoundnessProofSystemUnknown},
\ref{thm:SoundnessProofSystemClassic}, and
\ref{thm:SoundnessProofSystemLemma}, which respectively specify the
soundness of our proof system in the three typical scenarios: (i) proving
an unknown entailment with the lemma-synthesis-disabled mode (the first
pre/post specification of $\Prove$), (ii) verifying a discovered lemma,
i.e., proving a normal entailment with the lemma-synthesis-disabled mode
(the second pre/post specification), or (iii) proving a normal entailment
with the lemma-synthesis-enabled mode (the third pre/post specification).
Finally, we describe the overall soundness of the proof system in Theorem
\ref{thm:SoundnessProofSystem} when $\Prove$ is invoked with the initial
setting $\Hypos \,{=}\, \setempty, \Lemmas \,{=}\, \setempty, \ProofMode
\,{=}\, \SynLemma$. Note that Propositions
\ref{thm:SoundnessProofSystemUnknown} and
\ref{thm:SoundnessProofSystemClassic} are relevant to the lemma synthesis
in Section \ref{sec:LemmaSynthesis}. Proposition
\ref{thm:SoundnessProofSystemLemma} directly relates to the overall
soundness in Theorem \ref{thm:SoundnessProofSystem}.

\vspace{-0.1em}
\begin{restatable}[Proof of an unknown
    entailment]{proposition}{TheoremSoundnessProofSystemUnknown}
  \label{thm:SoundnessProofSystemUnknown}
  \!\! Given an unknown entailment $E$. If the procedure $\Prove$ returns
  $\ValidWitness{\_}$ and generates an assumption set $\A$ when proving $E$
  in the lemma-synthesis-disabled mode ($\NoSynLemma$), using an empty
  induction hypothesis set $(\Hypos {=} \setempty)$ and a valid lemma set
  $\Lemmas$ as its inputs, then $E$ is semantically valid, given that all
  assumptions in $\A$ are valid.
\end{restatable}

\vspace{-0.6em}
\begin{restatable}[Proof of a normal entailment when the lemma synthesis is
    disabled]{proposition}{TheoremSoundnessProofSystemClassic}
  \label{thm:SoundnessProofSystemClassic}
  Given a normal entailment $E$ which does not contain any unknown
  relation. If the procedure $\Prove$ returns $\ValidWitness{\_}$ when
  proving $E$ in the lemma-synthesis-disabled mode ($\NoSynLemma$), using
  an empty induction hypothesis set $(\Hypos {=} \setempty)$ and a valid
  lemma set $\Lemmas$ as its inputs, then $E$ is semantically valid.
\end{restatable}

\vspace{-0.6em}
\begin{restatable}[Proof of a normal entailment when the lemma synthesis is
    enabled]{proposition}{TheoremSoundnessProofSystemLemma}
  \label{thm:SoundnessProofSystemLemma}
  Given a normal entailment $E$ which does not contain any unknown
  relation. If the procedure $\Prove$ returns $\ValidWitness{\_}$ and
  synthesizes a set of lemmas $\LemmasSyn$ when proving $E$ in the
  lemma-synthesis-enabled mode ($\SynLemma$), using an empty induction
  hypothesis set $(\Hypos {=} \setempty)$ and a valid lemma set $\Lemmas$
  as its inputs, then the entailment $E$ and all lemmas in $\LemmasSyn$ are
  semantically valid.
\end{restatable}

\vspace{-1.1em}
\begin{proof}[Proofs of Propositions \ref{thm:SoundnessProofSystemUnknown},
    \ref{thm:SoundnessProofSystemClassic},
    \ref{thm:SoundnessProofSystemLemma}]
  We first show that all inference and synthesis rules are sound, and the
  proof system in the $\NoSynLemma$ mode is sound. Based on that, we can
  prove Propositions \ref{thm:SoundnessProofSystemUnknown} and
  \ref{thm:SoundnessProofSystemClassic}. The proof of Proposition
  \ref{thm:SoundnessProofSystemLemma} will be argued based on the lemma
  synthesis's soundness in Section \ref{sec:LemmaSynthesis}. Details of all
  these proofs are presented in
  {\iftechreport
  	Appendix \ref{append:SoundnessProof}.
   \else
   	the technical report \cite{TaLKC17}.
   \fi}
\end{proof}

\vspace{-0.5em}
\begin{restatable}[The overall soundness of the proof
    system]{theorem}{TheoremSoundnessProofSystem}
  \label{thm:SoundnessProofSystem}
  Given a normal entailment $E$ which does not contain any unknown
  relation, if the procedure $\Prove$ returns $\ValidWitness{\_}$ when
  proving $E$ in the initial context that there is no induction hypothesis
  or lemma provided beforehand and the lemma synthesis is enabled ($\Hypos
  \,{=}\, \setempty, \Lemmas \,{=}\, \setempty, \ProofMode \,{=}\,
  \SynLemma$), then $E$ is semantically valid.
\end{restatable}

\vspace{-1.1em}
\begin{proof}
  Since $E$ does not contain any unknown relation, both the input induction
  hypothesis set and lemma set are empty ($\Hypos {=} \setempty, \Lemmas
  {=} \setempty$), and $\Prove$ is invoked in the $\SynLemma$ mode, it
  follows from Proposition \ref{thm:SoundnessProofSystemLemma} that if
  $\Prove$ returns $\ValidWitness{\_}$ when proving $E$, then $E$ is
  semantically valid.
\end{proof}

\vspace{-0.8em}
\section{The Lemma Synthesis Framework}
\vspace{-0.3em}
\label{sec:LemmaSynthesis}

We are now ready to describe our lemma synthesis framework. It consists of
a main procedure, presented in Subsection~\ref{sec:LemmaSynthesisMain}, and
auxiliary subroutines, described in Subsections~\ref{sec:LemmaTemplate},
\ref{sec:FineTuneAnte}, and \ref{sec:FineTuneConseq}. Similar to the proof
system in Section \ref{sec:ProofSystem}, we also provide the input/output
description and the formal specification for each of these synthesis
procedures. We use the same keyword $\res$ to represent the returned result
and $\validlem{\Lemmas}$ indicates that all lemmas in $\Lemmas$ are
semantically valid.

\vspace{-0.6em}
\subsection{The main synthesis procedure}
\label{sec:LemmaSynthesisMain}

\vspace{-0.6em}

\begin{wrapfigure}{r}{0.555\textwidth}
\vspace{-0.8em}
\begin{minipage}{0.555\textwidth}
  \begin{figure}[H]
    \begin{algorithm}[H]
      \small
      \caption*{\small{\bf Procedure} \Call{\SynthesizeLemma}{
          \Lemmas, F_1 \entails F_2}}
      \begin{flushleft}
        \Summary{Input:} $F_1 {\entails} F_2$ is the goal entailment,
        $\Lemmas$ is a valid lemma set\\
        \Summary{Output:} A set of new synthesized lemmas\\[0.2em]
        \Summary{Requires:} $\validlem{\Lemmas}$\\
        \Summary{Ensures:}\phantom{i} $ %% (res \,{=}\, \setempty)
        \validlem{res}$
        % $\forall L \in res.\, \setempty, \setempty, L$
      \end{flushleft}
      \begin{algorithmic}[1]
        \def\indent{\hspace{1.2em}}
        \algrestore{myalg}
%        \label{line:CreateLemmaTemplate}
        \For{\Keyword{each}
            \Sigma_1 \,{\entails}\, \exists \vec{v}. \Sigma_2
            \Keyword{in}
            \Call{\CreateLemmaTemplate}{\Lemmas, F_1 \,{\entails}\, F_2}}
          \label{line:CreateLemmaTemplate}
          % \State \Assign{\Sigma_1 \wedge \Pi_1 \entails
          % \exists \vec{v}. \Sigma_2}{
          % \Call{\Preprocess}{
          % \Sigma_1 \entails \exists \vec{v}. \Sigma_2}}
          % \label{line:Preprocess}
          \State \Assign{\Lemmas_1}{
            \Call{\RefineAnte}{\Lemmas,~ \Sigma_1 \wedge \true
              \entails \Sigma_2}}
          \label{line:FineTuneAnte}
          \label{line:FineTuneBegin}
          \State \Assign{\Lemmas_2}{
            \Call{\RefineConseq}{\Lemmas,~ \Sigma_1
              \entails \exists \vec{v}. \Sigma_2}}
          \label{line:FineTuneConseq}
          \label{line:FineTuneEnd}
          \label{line:SynFoundLemmaBegin}
          \If{(\Lemmas_1 \cup \Lemmas_2) \neq \setempty}
            \Return{(\Lemmas_1 \cup \Lemmas_2)}
            \label{line:SynFoundLemmaEnd}
          \EndIf
        \EndFor{}
        \State \Return{\setempty}
        \label{line:SynNoLemma}
        \algstore{myalg}
      \end{algorithmic}
    \end{algorithm}
    \vspace{-0.8em}
    \caption{The main lemma synthesis procedure}
    \label{fig:SynthesizeLemmas}
  \end{figure}
\end{minipage}
\end{wrapfigure}

Figure \ref{fig:SynthesizeLemmas} presents the main lemma synthesis
procedure $(\SynthesizeLemma)$. Its inputs include a goal entailment $F_1
\entails F_2$ which needs to be proved by new lemmas, and a set of
previously synthesized lemmas $\Lemmas$ which will be exploited to make the
current synthesis more effective. The procedure $\SynthesizeLemma$ first
identifies a set of desired lemma templates based on the entailment's heap
structure, via the invocation of the procedure $\CreateLemmaTemplate$ (line
\ref{line:CreateLemmaTemplate}). These lemma templates are of the form
$\Sigma_1 \entails \exists \vec{v}. \Sigma_2$, in which $\Sigma_1$ and
$\Sigma_2$ are spatial formulas constructed from heap predicate symbols
appearing in $F_1$ and $F_2$, respectively, and $\vec{v}$ are all free
variables in $\Sigma_2$. By this construction, each synthesized lemma will
share a similar heap structure with the goal entailment, hence they can be
unified by the lemma application rules $\ruleLemmaLeft$ and
$\ruleLemmaRight$. We will formally define the lemma templates in
Subsection \ref{sec:LemmaTemplate}. In our implementation,
$\CreateLemmaTemplate$ returns a list of possible lemma templates, which
are sorted in the ascending order of their simplicity, i.e., templates
containing less spatial atoms are on the top of the list. Moreover, any
template sharing the same heap structure with a previously synthesized
lemma in $\Lemmas$ will not be considered.

The framework then successively refines each potential lemma template by
continuously discovering and adding in \emph{pure constraints} of its
variables until valid \emph{inductive lemmas} are found. In essence, for a
given lemma template, the refinement is performed by a 3-step recipe:

\begin{enumerate}[noitemsep, leftmargin = 2em]
\setlength\itemsep{0.1em}
\item Establishing an unknown relation representing a desired constraint
  inside the template and creating an unknown entailment.
\item Proving the unknown entailment by structural induction and collecting
  assumptions about the unknown relation.
\item Solving the assumptions to find out the actual definition of the
  unknown relation, thus discovering the desired inductive lemma.
\end{enumerate}

There are two possible places to refine a lemma template: on its antecedent
(line \ref{line:FineTuneBegin}) and its consequent (line
\ref{line:FineTuneEnd}). We aim to synthesize a lemma which has an {\em as
  weak as possible} antecedent or an {\em as strong as possible}
consequent. We will elaborate the details in Subsections
\ref{sec:FineTuneAnte} and \ref{sec:FineTuneConseq}. Our framework
\emph{immediately} returns a non-empty set of synthesized lemmas (lines
\ref{line:SynFoundLemmaEnd}) once it successfully refines a template.
Otherwise, it returns an empty set ($\setempty$) indicating that no lemma
can be synthesized (line \ref{line:SynNoLemma}).

\vspace{-0.5em}
\subsection{Discovering lemma templates}
\label{sec:LemmaTemplate}
\vspace{-0.3em}

The lemma templates for a given goal entailment can be discovered from the
entailment's heap structure. In the followings, we present the formal
definition of the lemma templates and also illustrate by examples how to
create them.

\begin{definition}[Lemma template]
  \label{def:LemmaTemplate}
  A lemma template for a goal entailment $F_1 \entails F_2$ is an
  entailment of the form $ \Sigma_1 \entails \exists \vec{v}. \Sigma_2 $,
  where:

  \vspace{-0.1em}

  %\begin{itemize}[label=-,noitemsep,nosep,nolistsep]
  \begin{enumerate}[noitemsep, leftmargin = 2em]
  \setlength\itemsep{0.1em}
  \item $\Sigma_1$ and $\Sigma_2$ are spatial formulas containing at least one
    inductive heap predicate.

  \item Heap predicate symbols in $\Sigma_1$ and $\Sigma_2$ are
    sub-multisets of those in $F_1$ and $F_2$.

  \item Variables in $\Sigma_1$ and $\Sigma_2$ are separately named (no
    variables appear twice) and $\vec{v} \equiv \freevars{\Sigma_2}$.
  \end{enumerate}
  %\end{itemize}
\end{definition}

\vspace{-0.3em}

The condition (1) limits the templates for only inductive lemmas. The
condition (2) guarantees that the synthesized lemmas are {\em unifiable}
with the goal entailment via applications of the rules $\ruleLemmaLeft$ and
$\ruleLemmaRight$. Moreover, the condition (3) ensures that each template
is as general as possible so that desirable pure constraints can be
subsequently discovered in the next phases.

For instance, the following lemma templates can be created given the
motivating entailment in Section \ref{sec:MotivatingExample}: $E_1
\triangleq \formp{\DllRev}{x,y,u,v,n} * \formp{\Dll}{v,u,z,t,200} \wedge
n {\ge} 100 \entails \exists r. (\formp{\Dll}{x,y,r,z,n{+}199} *
\forms{z}{r,t})$. Note that the template $T_1$ is used to synthesize the
lemma $L_1 \triangleq \formp{\DllRev}{a,b,c,d,m} \entails
\formp{\Dll}{a,b,c,d,m}$.

\begin{center}
\begin{small}
\begin{tabular}{l }
  $T_1 \triangleq \formp{\DllRev}{x_1,x_2,x_3,x_4,n_1}
  \entails \exists x_5,x_6,x_7,x_8,n_2.
  \formp{\Dll}{x_5,x_6,x_7,x_8,n_2}$\\

  $\formp{\Dll}{x_1,x_2,x_3,x_4,n_1}
  \entails \exists x_5,x_6,x_7,x_8,x_9,x_{10},x_{11},n_2.
  (\formp{\Dll}{x_5,x_6,x_7,x_8,n_2} * \forms{x_9}{x_{10},x_{11}} )$\\

  $\formp{\DllRev}{x_1,x_2,x_3,x_4,n_1}
  \entails \exists x_5,x_6,x_7,x_8,x_9,x_{10},x_{11},n_2.
  (\formp{\Dll}{x_5,x_6,x_7,x_8,n_2} * \forms{x_9}{x_{10},x_{11}} )$\\

  $\formp{\DllRev}{x_1,x_2,x_3,x_4,n_1} *
  \formp{\Dll}{x_5,x_6,x_7,x_8,n_2}
  \entails \exists x_9,x_{10},x_{11},x_{12},n_3.
  \formp{\Dll}{x_9,x_{10},x_{11},x_{12},n_3}$
\end{tabular}
\end{small}
\end{center}

Similarly, there can be several possible lemma templates relating to the
entailments $E_2$ and $E_3$ in Figure \ref{fig:ProofTreeMotiv}. Among them,
the templates $T_2$ and $T_3$ below are used to synthesize the lemmas $L_2$
and $L_3$:

\vspace{-0.3em}
\begin{center}
\begin{small}
\begin{tabular}{l}
  $T_2 \triangleq \formp{\Dll}{x_1,x_2,x_3,x_4,n_1} *
  \formp{\Dll}{x_5,x_6,x_7,x_8,n_2}
  \entails \exists x_9,x_{10},x_{11},x_{12},n_3.
  \formp{\Dll}{x_9,x_{10},x_{11},x_{12},n_3}$\\

  $T_3 \triangleq \formp{\Dll}{x_1,x_2,x_3,x_4,n_1}
  \entails \exists x_5,x_6,x_7,x_8,x_9,x_{10},x_{11},n_2.
  (\formp{\Dll}{x_5,x_6,x_7,x_8,n_2} * \forms{x_9}{x_{10},x_{11}})$
\end{tabular}
\end{small}
\end{center}

\vspace{-0.7em}
\subsection{Refining the lemma template's antecedent}
\label{sec:FineTuneAnte}
\vspace{-0.5em}

Figure \ref{fig:FineTuneAnte} presents the antecedent refinement procedure
($\RefineAnte$), which aims to strengthen the antecedent of a lemma
template $\Sigma_1 \wedge \Pi_1 \entails \Sigma_2$ with pure constraints of
all its variables. Note that when $\RefineAnte$ is invoked in the first
time by $\SynthesizeLemma$ (Figure \ref{fig:SynthesizeLemmas}, line
\ref{line:FineTuneAnte}), $\Pi_1$ is set to $\true$ and the existential
quantification $\exists \vec{v}$ in the original template $\Sigma_1
\entails \exists \vec{v}. \Sigma_2$ is removed since the antecedent will be
strengthened with constraints of all variables in the template.

\begin{wrapfigure}{r}{0.59\textwidth}
\vspace{-0.6em}
\begin{minipage}{0.59\textwidth}
\begin{figure}[H]
  \begin{algorithm}[H]
    \small
    \caption*{\small{\bf Procedure} \Call{\RefineAnte}{
        \Lemmas,~ \Sigma_1
        \,{\wedge}\, \Pi_1 \entails \Sigma_2}}
    \begin{flushleft}
      \Summary{Input:} $\Sigma_1 {\wedge} \Pi_1 {\entails} \Sigma_2$
      is a lemma template, $\Lemmas$ is a valid lemma set\\
      \Summary{Output:} a set of at most one synthesized lemma\\[0.2em]
      \Summary{Requires:} $\validlem{\Lemmas}$ \\
      \Summary{Ensures:} $(res \,{=}\, \setempty)
      \vee \exists\,\Pi'_1.(res \,{=}\, \setenum{
        \Sigma_1 {\wedge} \Pi_1 {\wedge} \Pi'_1 {\entails} \Sigma_2}
        \wedge \validlem{res})$
    \end{flushleft}
    \begin{algorithmic}[1]
      \algrestore{myalg}
      \State \Assign{\vec{u}}{\Call{\FreeVars}{\Sigma_1, \Sigma_2, \Pi_1}}
      \label{line:TuneAnteCreateUnknownBegin}
      \State \Assign{\formr{\rUnk}{\vec{u}}}{
        \Call{\CreateUnknownRelation}{\vec{u}}}
      \label{line:TuneAnteCreateUnknownEnd}
      \State \Assign{\vR, \_, \Assumpts}{\Call{\Prove}{
          \setempty,~\Lemmas,~
          \Sigma_1 {\wedge}\, \Pi_1 {\wedge}\, \formr{\rUnk}{\vec{u}}
          \,{\entails}\, \Sigma_2,~ \NoSynLemma}}
      \label{line:TuneAnteProve}
      % \Comment{search for a proof}
      \If{\vR = \ValidWitness{\_}}
        \label{line:LemmaUnivSolve}
        \label{line:LemmaUnivSolveBegin}
        \label{line:LemmaUnivFullEnd}
        \For{\Keyword{each} \Assumpts'
            \Keyword{in} \Call{SuperSet}{\Assumpts}}
          \label{line:TuneAnteSolveSuperSet}
          \label{line:TuneAnteSolveBegin}
          \State \Assign{\formr{\rUnkSol}{\vec{u}}}{
            \Call{\Solve}{\Assumpts', \formr{\rUnk}{\vec{u}}}}
          \label{line:TuneAnteSolveEnd}
          \If{\Sigma_1 {\wedge}\, \Pi_1 {\wedge}\, \formr{\rUnkSol}{\vec{u}}
              \not\equiv \false}
            \label{line:TuneAnteSpuriousSolution}
            \State \Assign{L}{(\Sigma_1 {\wedge}\, \Pi_1 {\wedge}\,
              \formr{\rUnkSol}{\vec{u}} \,{\entails}\, \Sigma_2)}
            \State \Assign{\vR',\,\_,\,\_}{
              \Call{\Prove}{\setempty, \,\Lemmas,\, L, \,\NoSynLemma}}
            \Comment{Verify ...}
            \label{line:TuneAnteVerify}
            \label{line:TuneAnteTestLemmaBegin}
            \If{\vR'\,{=}\,\ValidWitness{\xi} \Keyword{and}
                \ruleHypo \,{\in}\, \xi}
              \label{line:TuneAnteTestLemmaEnd}
              \Comment{and return ...}
              \State \Return{\setenum{\vL}}
              \Comment{the first inductive lemma,}
              \label{line:TuneAnteReturnLemma}
              \label{line:TuneAnteIncrBegin}
            \Else~
              \Return{\Call{\RefineAnte}{\Lemmas,\,\vL}}
              \Comment{or continue refining.}
            \EndIf
            \label{line:TuneAnteIncrEnd}
          \EndIf
        \EndFor
      \EndIf
      \State \Return{\setempty}
      \label{line:TuneAnteNoLemma}
      \Comment{No lemmas are found.}
    \algstore{myalg}
    \end{algorithmic}
  \end{algorithm}
  \vspace{-0.8em}
  \caption{Refining a lemma template's antecedent}
  \label{fig:FineTuneAnte}
\end{figure}
\end{minipage}
\end{wrapfigure}

Initially, $\RefineAnte$ creates an unknown entailment $\Sigma_1 \wedge
\Pi_1 \wedge \formr{\rUnk}{\vec{u}} \entails \Sigma_2$, where
$\formr{\rUnk}{\vec{u}}$ is an unknown relation of all variables $\vec{u}$
in $\Sigma_1,\Sigma_2,\Pi_1$ (lines \ref{line:TuneAnteCreateUnknownBegin},
\ref{line:TuneAnteCreateUnknownEnd}). Then, it proves the entailment by
induction and collects a set $\Assumpts$ of unknown assumptions, if any,
about $\formr{\rUnk}{\vec{u}}$ (line \ref{line:TuneAnteProve}). In this
proof derivation (also in other invocations of $\Prove$ in the lemma
synthesis), we prevent the proof system from synthesizing new lemmas (by
passing $\NoSynLemma$) to avoid initiating a new lemma synthesis phase
inside the current synthesis, i.e., to prohibit the nested lemma synthesis.
In addition, the assumption set $\Assumpts$ will be examined, via the
procedure $\Solve$, to discover a solution of $\formr{\rUnk}{\vec{u}}$
(lines \ref{line:TuneAnteSolveBegin}, \ref{line:TuneAnteSolveEnd}). We
implement in $\Solve$ a constraint solving technique using Farkas' lemma
\cite{Schrijver1986, ColonSS03}. This technique assigns
$\formr{\rUnk}{\vec{u}}$ to a predefined linear arithmetic formula with
unknown coefficients, and applies Farkas' lemma to transform $\Assumpts$
into a set of constraints involving only the unknown coefficients. Then, it
utilizes an off-the-shelf prover such as Z3 \cite{MouraB08} to find a
concrete model of the coefficients, thus obtains the actual definition
$\formr{\rUnkSol}{\vec{u}}$ of $\formr{\rUnk}{\vec{u}}$. We will describe
this technique in Subsection \ref{sec:SolveConstraint}.

Furthermore, we aim to find a \emph{non-spurious} solution
$\formr{\rUnkSol}{\vec{u}}$ which does not refute the antecedent, i.e.,
$\Sigma_1 \wedge \Pi_1 \wedge \formr{\rUnkSol}{\vec{u}} \not\equiv \false$,
to avoid creating an \emph{useless} lemma: $\false \,{\entails}\, \Sigma_2$
(line \ref{line:TuneAnteSpuriousSolution}). To examine this refutation, we
follow the literature \cite{BrotherstonFPG14, LeSC16} to implement an
unsatisfiability checking algorithm, which over-approximates a
symbolic-heap formula to a pure formula and invokes the off-the-shelf
prover (such as Z3) to check the pure formula's unsatisfiability, thus
concludes about the unsatisfiability of the original formula.

\vspace{-0.1em} In general, discovering a \emph{non-spurious} solution
$\formr{\rUnkSol}{\vec{u}}$ is challenging, because: \vspace{-0.2em}

\begin{itemize}[noitemsep, label=--, leftmargin=1em]
\setlength\itemsep{0.3em}
\item The assumption set $\Assumpts$ can be complicated since the
  parameters $\vec{u}$ of the unknown relation $\formr{\rUnk}{\vec{u}}$ are
  all variables in the lemma template. This complexity can easily overwhelm
  the underlying prover when finding the model of the corresponding unknown
  coefficients.\vspace{-0.2em}

\item The discovered proof tree of the unknown entailment might not be
  similar to the actual proof tree of the desired lemma, due to the
  occurrence of the unknown relation. Therefore, the set $\Assumpts$ might
  contain {\em noise} assumptions of $\formr{\rUnk}{\vec{u}}$, which
  results in a spurious solution. Nonetheless, a part of the expected
  solution can still be discovered from a subset of $\Assumpts$, which
  corresponds to the common part of the discovered and the desired proof
  trees.
\end{itemize}

The above challenges inspire us to design an {\em exhaustive approach} to
solve $\Assumpts$ (line \ref{line:TuneAnteSolveSuperSet}). In particular,
$\RefineAnte$ first solves the entire set $\Assumpts$ (the first element in
$\Call{SuperSet}{\Assumpts}$) to find a {\em complete} solution, which
satisfies {\em all} assumptions in $\Assumpts$. If such solution is
\emph{not} available, $\RefineAnte$ iteratively examines each subset of
$\Assumpts$ (among the remaining elements in $\Call{SuperSet}{\Assumpts}$)
to discover a {\em partial} solution, which satisfies {\em some}
assumptions in $\Assumpts$.

The discovered solution (complete or partial) will be verified whether it
can form a valid inductive lemma. In particular, the proof system will be
invoked to prove the candidate lemma $L$ (line
\ref{line:TuneAnteTestLemmaBegin}). The verification is successful when $L$
is proved valid and its witness proof tree $\xi$ contains an induction
hypothesis application (labeled by $\ruleHypo$) (line
\ref{line:TuneAnteTestLemmaEnd}). The latter condition ensures that the
returned lemma is actually an inductive lemma.

We also follow an \emph{incremental approach} to refine the lemma
template's antecedent. That is, the antecedent will be strengthened with
the discovered solution $\formr{\rUnkSol}{\vec{u}}$ to derive a new
template. The new template will be refined again until the first valid
lemma is discovered (lines \ref{line:TuneAnteTestLemmaEnd} --
\ref{line:TuneAnteIncrEnd}). Note that this refinement stops at the first
solution to ensure that the discovered antecedent is as weak as possible.
Finally, $\RefineAnte$ returns $\setempty$ if no valid inductive lemma can
be discovered (line \ref{line:TuneAnteNoLemma}).

For example, given the template $T_1 {\triangleq}\,
\formp{\DllRev}{x_1,x_2,x_3,x_4,n_1} \,{\entails}\, \exists
x_5,x_6,x_7,x_8,n_2. \formp{\Dll}{x_5,x_6,x_7,x_8,n_2}$, $\RefineAnte$
creates an unknown relation $\formr{\rUnk}{x_1,...,x_8,n_1,n_2}$ and
introduces the entailment $E_{u_1}$:

\vspace{-0.1em}
\begin{center}
\begin{small}
$E_{u_1} \triangleq \formp{\DllRev}{x_1,x_2,x_3,x_4,n_1}
\wedge \formr{\rUnk}{x_1,...,x_8,n_1,n_2} \entails
\formp{\Dll}{x_5,x_6,x_7,x_8,n_2}$
\end{small}
\end{center}
\vspace{0.35em}

\begingroup
\let\oldmapsto\mapsto
\renewcommand\mapsto{\mathord{\scalebox{0.75}[1.0]{$\oldmapsto$}}}
\relax
\begin{figure}[H]
  \begin{small}
    \begin{prooftree}[label separation = 0.16em,
      separation = 0.6em, rule margin = 0.26em]
    \def\defaultHypSeparation{\hskip 1em}
    \def\ScoreOverhang{0.2em}
    \Hypo{\hboxauto{
        $\A_1 \triangleq \{x_1{\sequal}x_3 \,{\wedge}\, n_1{\sequal}1
        \,{\wedge}\, \formr{\rUnk}{\vec{x},\vec{n}} \,{\rightarrow}\,$\\
        $\hspace*{0em}
        x_1{\sequal}x_5 {\wedge}
        x_2{\sequal}x_6 {\wedge}
        x_4{\sequal}x_8 {\wedge}
        x_5{\sequal}x_7 {\wedge} n_2{\sequal}1\}$
      }}
    % \rewrite{\color{red}\box\treebox}
    \Infer1[{\footnotesize $\ruleSynPiOne$}]{\hboxauto{
        $ x_1{\sequal}x_3 \,{\wedge}\, n_1{\sequal}1
        \,{\wedge}\, \formr{\rUnk}{\vec{x},\vec{n}} \,{\entails}\,
        x_1{\sequal}x_5 \,{\wedge}\,$\\
        \hspace*{1.5em}$
        x_2{\sequal}x_6 \,{\wedge}\,
        x_4{\sequal}x_8 \,{\wedge}\,
        x_5{\sequal}x_7 \,{\wedge}\, n_2{\sequal}1$
      }}
    \Infer1[{\footnotesize $\ruleStarData$}]{\hboxauto{
        $\forms{x_1}{x_2,x_4} \,{\wedge}\,
        x_1{\sequal}x_3 \,{\wedge}\, n_1{\sequal}1
        \,{\wedge}\, \formr{\rUnk}{\vec{x},\vec{n}}$\\
        \hspace*{2.6em}$ \,{\entails}\,\forms{x_5}{x_6,x_8} \,{\wedge}\,
        x_5{\sequal}x_7 \,{\wedge}\, n_2{\sequal}1$
      }}
    \Infer1[{\footnotesize $\rulePredRight$}]{\hboxauto{
        $\forms{x_1}{x_2,x_4} \,{\wedge}\, x_1{\sequal}x_3
        \,{\wedge}\, n_1{\sequal}1
        \,{\wedge}\, \formr{\rUnk}{\vec{x},\vec{n}}$\\
        \hspace*{5.2em}$\,{\entails}\,\formp{\Dll}{x_5,x_6,x_7,x_8,n_2}$
      }}
    %===================================
    \Hypo{\A_2 \triangleq
      \{\formr{\rUnk}{\vec{x},\vec{n}} \,{\rightarrow}\, \false\}}
    % \rewrite{\color{red}\box\treebox}
    \Infer1[{\footnotesize $\ruleSynSigmaOne$}]{\hboxauto{
        $\forms{x_3}{u,x_4} \,{\wedge}\, \formr{\rUnk}{\vec{x},\vec{n}}
        \,{\entails}\, a_5{\sequal}x_5 \,{\wedge}\,  $ \\
        $a_6{\sequal}x_6 \,{\wedge}\,
        a_7{\sequal}x_7 \,{\wedge}\, a_8{\sequal}x_8 \,{\wedge}\,
        b_2{\sequal}n_2$}}
    \Infer1[{\footnotesize $\ruleStarPred$}]{\hboxauto{
        $\formp{\Dll}{a_5,a_6,a_7,a_8,b_2}
        * \forms{x_3}{u,x_4} \,{\wedge}\, $\\
        \hspace*{1.2em}
        $\formr{\rUnk}{\vec{x},\vec{n}} \,{\entails}\,
        \formp{\Dll}{x_5,x_6,x_7,x_8,n_2}$
      }}
    \Hypo{
      \A_3 \,{\triangleq}\, \{\formr{\rUnk}{\vec{x},\vec{n}}
      \,{\rightarrow}\, \formr{\rUnk}{\vec{a},\vec{b}}\theta\}}
    % \rewrite{\color{red}\box\treebox}
    \Infer[separation = 0.2em]2
    % [{\color{red}\footnotesize $\ruleSynHypo$}]
    [\footnotesize $\ruleSynHypo$]{\hboxauto{
        $\formp{\DllRev}{x_1,x_2,u,x_3,n_1{\sminus}1} * \forms{x_3}{u,x_4}
        \,{\wedge}\, \formr{\rUnk}{\vec{x},\vec{n}}$\\
        \hspace*{16em}$ \,{\entails}\,
        \formp{\Dll}{x_5,x_6,x_7,x_8,n_2}$
      }}
    \Infer2
    % [{\color{red}\footnotesize $\ruleInduction$}]
    [\footnotesize $\ruleInduction$]{
      E_{u_1} \triangleq \formp{\DllRev}{x_1,x_2,x_3,x_4,n_1} \,{\wedge}\,
      \formr{\rUnk}{x_1,x_2,x_3,x_4,x_5,x_6,x_7,x_8,n_1,n_2}
      \,{\entails}\,
      \formp{\Dll}{x_5,x_6,x_7,x_8,n_2}
      }

  \end{prooftree}
  \vspace{-0.6em}
  \caption{A possible proof tree of the unknown entailment $E_{u_1}$}

  \vspace{0.2em}
  where $\formr{\rUnk}{\vec{x},\vec{n}} \equiv
  \formr{\rUnk}{x_1,x_2,x_3,x_4,x_5,x_6,x_7,x_8,n_1,n_2}$;
  $\formr{\rUnk}{\vec{a},\vec{b}} \equiv
  \formr{\rUnk}{a_1,a_2,a_3,a_4,a_5,a_6,a_7,a_8,b_1,b_2}$;\\
  the rule $\ruleInduction$ is performed to record
  the IH $H \triangleq \formp{\DllRev}{a_1,a_2,a_3,a_4,b_1} \wedge
  \formr{\rUnk}{\vec{a}, \vec{b}} \entails
  \formp{\Dll}{a_5,a_6,a_7,a_8,b_2}$;\\
  the rule $\ruleSynHypo$ applies $H$
  with $\theta = [x_1/a_1,x_2/a_2,u/a_3,x_3/a_4,n_1{-}1/b_1]$;

  \vspace{-0.3em}
  \label{fig:ProofTreeUnknEntail}
  \end{small}
\end{figure}
\endgroup

Figure \ref{fig:ProofTreeUnknEntail} presents a possible proof tree of
$E_{u_1}$. From this proof, we obtain a set $\A = \A_1 \cup \A_2 \cup \A_3$
of three unknown assumptions about the relation
$\formr{\rUnk}{x_1,x_2,x_3,x_4,x_5,x_6,x_7,x_8,n_1,n_2}$:

\begin{small}
  \begin{enumerate}[noitemsep, leftmargin=3em]
  \item $x_1{=}x_3 \,{\wedge}\, n_1{=}1 \,{\wedge}\,
    \formr{\rUnk}{x_1,x_2,x_3,x_4,x_5,x_6,x_7,x_8,n_1,n_2} ~\rightarrow~
    x_1{=}x_5 \,{\wedge}\, x_2{=}x_6 \,{\wedge}\, x_4{=}x_8 \,{\wedge}\,
    x_5{=}x_7 \,{\wedge}\, n_2{=}1$
  \item $\formr{\rUnk}{x_1,x_2,x_3,x_4,x_5,x_6,x_7,x_8,n_1,n_2} ~\rightarrow~
    \false$
  \item $\formr{\rUnk}{x_1,x_2,x_3,x_4,x_5,x_6,x_7,x_8,n_1,n_2}
    ~\rightarrow~
    \formr{\rUnk}{x_1,x_2,u,x_3,a_5,a_6,a_7,a_8,n_1{-}1,b_2}$
  \end{enumerate}
\end{small}

Our framework first attempts to solve the full assumption set $\A$.
Unfortunately, there is only a {\em spurious solution}
$\formr{\rUnkSol}{x_1,x_2,x_3,x_4,x_5,x_6,x_7,x_8,n_1,n_2} \equiv \false$,
since the assumption (2) is too strong. It then tries to find another
solution by {\em partially solving} the set $\A$. In this case, it can
discover the following partial solution $\rUnkSol$ when solving a subset
containing only the assumption (1):

\vspace{-0.2em}
\begin{center}
\begin{small}
  $\formr{\rUnkSol}{x_1,x_2,x_3,x_4,x_5,x_6,x_7,x_8,n_1,n_2}
  \equiv x_1{=}x_5 \wedge x_2{=}x_6 \wedge x_3{=}x_7 \wedge x_4{=}x_8 \wedge
  n_1{=}n_2$
\end{small}
\end{center}
\vspace{-0.2em}

Since the above is only a partial solution, the framework needs to verify
whether it can construct a {\em valid inductive lemma}. Indeed, $\rUnkSol$
can be used to derive the following entailment $\overbar{E}_{u_1}$, which
can be simplified to become $\formp{\DllRev}{x_1,x_2,x_3,x_4,n_1} \entails
\formp{\Dll}{x_1,x_2,x_3,x_4,n_1}$. The latter entailment can be
equivalently transformed into the motivating lemma $L_1$ by a renaming on
its variables.

\vspace{-0.1em}
\begin{center}
\begin{small}
  $\overbar{E}_{u_1} \triangleq
  \formp{\DllRev}{x_1,x_2,x_3,x_4,n_1} \wedge x_1{=}x_5 \wedge x_2{=}x_6
  \wedge x_3{=}x_7 \wedge x_4{=}x_8 \wedge n_1{=}n_2 \entails
  \formp{\Dll}{x_5,x_6,x_7,x_8,n_2}$
\end{small}
\end{center}
\vspace{0.3em}

\vspace{-1em}
\subsection{Refining the lemma template's consequent}
\vspace{-0.3em}
\label{sec:FineTuneConseq}

The consequent refinement is presented in the procedure $\RefineConseq$
(Figure \ref{fig:RefineConseq}). Unlike the antecedent refinement, this
refinement is not straightforward by simply adding pure constraints into
the template's consequent, since the existing template's antecedent might
not be strong enough to prove any formulas derived from the consequent. For
example, all entailments derived from adding pure constraints to only the
consequent of the lemma template $T_3$ are invalid, because when $n_1{=}1$,
the list $\formp{\Dll}{x_1,x_2,x_3,x_4,n_1}$ has the length of $1$, thus
cannot be split into a singleton heap $\forms{x_9}{x_{10},x_{11}}$ and a
list $\formp{\Dll}{x_5,x_6,x_7,x_8,n_2}$, whose length is at least 1, by
the definition of $\Dll$.

\begin{center}
\begin{small}
$T_3 \triangleq \formp{\Dll}{x_1,x_2,x_3,x_4,n_1}
\entails \exists x_5,x_6,x_7,x_8,x_9,x_{10},x_{11},n_2.
(\formp{\Dll}{x_5,x_6,x_7,x_8,n_2} * \forms{x_9}{x_{10},x_{11}})$
\end{small}
\end{center}

\begin{wrapfigure}[12]{r}{0.666\textwidth}
\vspace{-0.5em}
\begin{minipage}{0.666\textwidth}
\begin{figure}[H]
  \begin{algorithm}[H]
    \small
    \caption*{\small{\bf Procedure} \Call{\RefineConseq}{
        \Lemmas,~
        \Sigma_1 \,{\entails}\, \exists \vec{v}. \Sigma_2}}
    \begin{flushleft}
      \Summary{Input:} $\Sigma_1 \entails \exists \vec{v}. \Sigma_2$
      is a lemma template, $\Lemmas$ is a set of valid lemmas\\
      \Summary{Output:} a set of at most one synthesized lemma\\[0.1em]
      \Summary{Requires:} $\validlem{\Lemmas}$\\
      \Summary{Ensures:} $(res \,{=}\, \setempty)
      \,{\vee}\,
      \exists\,\Pi_1,\!\Pi_2.(res \,{=}\,
      \setenum{\Sigma_1 {\wedge} \Pi_1 {\entails}
        \exists \vec{v}.(\Sigma_2 {\wedge} \Pi_2)}
      \,{\wedge}\, \validlem{res})$
    \end{flushleft}
    \begin{algorithmic}[1]
      \algrestore{myalg}
      \def\indent{\hspace{1.2em}}
      \State \Assign{\Templates}{
        \Call{\Preprocess}{\Lemmas,~
          \Sigma_1 \,{\entails}\, \exists \vec{v}. \Sigma_2}}
      \label{line:RefineConseqPreprocess}
      \If{\Templates = \setenum{
            \Sigma_1 \,{\wedge}\, \Pi_1 \,{\entails}\,
            \exists \vec{v}. \Sigma_2}}
      \label{line:RefineConseqPreprocessEnd}
      \State \Return{\Call{\FineTuneConseq}{
          \Lemmas,~ \Sigma_1 \,{\wedge}\, \Pi_1 \,{\entails}\,
            \exists \vec{v}. (\Sigma_2 \,{\wedge}\, \true)}}
        \label{line:RefineConseqFineTune}
      \EndIf
      \State \Return{\setempty}
      % \Comment{no lemmas are found}
      \algstore{myalg}
    \end{algorithmic}
  \end{algorithm}
  \vspace{-0.6em}
  \caption{Refining a lemma template's consequent}
  \label{fig:RefineConseq}
\end{figure}
\end{minipage}
\end{wrapfigure}

\vspace{-0.2em} To overcome this problem, we decompose the consequent
refinement into two phases, \emph{preprocessing} and \emph{fine-tuning}.
For an input lemma template $\Sigma_1 \,{\entails}\, \exists \vec{v}.
\Sigma_2$, in the first phase, the framework infers a pure constraint
$\Pi_1$ so that the entailment $\Sigma_1 \,{\wedge}\, \Pi_1 \,{\entails}\,
\exists \vec{v}. \Sigma_2$ is valid (lines
\ref{line:RefineConseqPreprocess}, \ref{line:RefineConseqPreprocessEnd}).
In the second phase, it incrementally strengthens the consequent of the
derived entailment until discovers a valid inductive lemma (line
\ref{line:RefineConseqFineTune}). Note that, we retain the existential
quantification in this consequent, and any constraints added into the
consequent will be bound by this quantification. These two phases will be
elaborated in details as follows.

\textbf{Preprocessing a lemma template.}
\vspace{-0.2em}

Figure \ref{fig:PreprocessTemplate} presents our antecedent preprocessing
procedure ($\Preprocess$). Similar to the antecedent refinement in Section
\ref{sec:FineTuneAnte}, this procedure strengthens the antecedent of the
template $\Sigma_1 \entails \exists \vec{v}. \Sigma_2$ with a {\em
  non-spurious} condition $\Pi_1$ to make the lemma template valid. We also
prevent the framework from entering nested lemma synthesis phases by
invoking $\Prove$ in the $\NoSynLemma$ mode (line
\ref{line:BasicProofSearch}).

However, this preprocessing step differs from the antecedent refinement
(Section \ref{sec:FineTuneAnte}) when it creates an unknown entailment by
introducing the unknown relation $\formr{\rUnk}{\vec{u}}$ on only the
antecedent's free variables $\vec{u}$ (lines
\ref{line:PreprocessCreateUnknownBegin},
\ref{line:PreprocessCreateUnknownEnd}). We also equip the consequent
$\exists \vec{v}. \Sigma_2$ of the unknown entailment with a conjunction
$\Pi_{inv}$ of its inductive heap predicates' pure invariants (lines
\ref{line:PreprocessInvariantBegin}, \ref{line:PreprocessInvariantEnd}).
These invariants are pure formulas representing Boolean constraints of the
predicates' variables. We will briefly describe the invariant construction
in Section \ref{sec:HeapPredicateInvariant}. This construction is also
well-studied in separation logic literature \cite{ChinDNQ12,
  BrotherstonFPG14, LeSC16}.

\begin{wrapfigure}[15]{r}{0.58\textwidth}
\vspace{-0.35em}
\begin{minipage}{0.58\textwidth}
\begin{figure}[H]
  \begin{algorithm}[H]
    \small
    \caption*{\small{\bf Procedure} \Call{\Preprocess}{\Lemmas,\,
        \Sigma_1 \entails{} \exists \vec{v}. \Sigma_2}}
    \begin{flushleft}
      \Summary{Input:} $\Sigma_1 {\entails} \exists \vec{v}. \Sigma_2$
      is a lemma template, $\Lemmas$ is a valid lemma set\\
      \Summary{Output:} a set of at most one refined template\\[0.3em]
      \Summary{Requires:} $\validlem{\Lemmas}$\\
      \Summary{Ensures:} $(res \,{=}\, \setempty)
      \vee \exists\,\Pi_1.(res \,{=}\,
      \setenum{\Sigma_1 {\wedge} \Pi_1 {\entails} \exists \vec{v}.\Sigma_2}
      \,{\wedge}\, \validlem{res})$
    \end{flushleft}
    \begin{algorithmic}[1]
      \algrestore{myalg}
      \def\indent{\hspace{1.2em}}
      % \State \Assign{\vec{u}}{\Call{\pFV}{\Sigma_1}};~
      \State \Assign{\vec{u}}{\Call{\FreeVars}{\Sigma_1}}
      \label{line:PreprocessCreateUnknownBegin}
      \State \Assign{\formr{\rUnk}{\vec{u}}}{
        \Call{\CreateUnknownRelation}{\vec{u}}}
      \label{line:PreprocessCreateUnknownEnd}
      \State \Assign{\Pi_{inv}}{
        \midwedge_{\,\formp{\P}{\vec{x}} \,\in\, \Sigma_2}
        \Call{\Invariant}{\formp{{\P}}{\vec{x}}}}
      \label{line:PreprocessInvariantBegin}
      \State \Assign{\vR, \_, \Assumpts}{\Call{\Prove}{
          \setempty,\, \Lemmas,\, \Sigma_1 {\wedge} \formr{\rUnk}{\vec{u}}
          {\entails} \exists \vec{v}. (\Sigma_2 {\wedge} \Pi_{inv}),\,
          \NoSynLemma}}
      \label{line:PreprocessInvariantEnd}
      \label{line:BasicProofSearch}
      % \Comment{about vars of $\Sigma_1$}
      \If{\vR = \ValidWitness{\_}}
        \label{line:PreprocessSolveBegin}
        \State \Assign{\formr{\rUnkSol}{\vec{u}}}{
          \Call{\Solve}{\Assumpts, \formr{\rUnk}{\vec{u}}}}
        \label{line:PreprocessSolve}
        \label{line:PreprocessSolveEnd}
        \If{\Sigma_1 {\wedge}\, \formr{\rUnkSol}{\vec{u}} \not\equiv \false}
          % \Comment{found a productive solution}
          \Return{\setenum{\Sigma_1 {\wedge}\,
              \formr{\rUnkSol}{\vec{u}}
            \,{\entails}\, \exists \vec{v}. \Sigma_2}}
        \EndIf
        \label{line:PreprocessReturnTemplate}
      \EndIf
      \State \Return{\setempty}
      \Comment{No refined templates are found.}
      \label{line:PreprocessNoRefine}
      \algstore{myalg}
    \end{algorithmic}
  \end{algorithm}
  \vspace{-0.8em}
  \caption{Preprocess a lemma template's antecedent}
  \label{fig:PreprocessTemplate}
\end{figure}
\end{minipage}
\end{wrapfigure}

In theory, equipping additional pure invariants of inductive heap
predicates \emph{does not} weaken or strengthen the lemma template's
consequent, i.e., $\Sigma_2 \equiv \Sigma_2 \wedge \Pi_{inv}$. In our
approach, $\Preprocess$ solves the entire assumption constraint set $\A$ at
once (line \ref{line:PreprocessSolve}), and not incrementally as in the
antecedent refinement (Section \ref{sec:FineTuneAnte}). Therefore, the
additional pure invariant $\Pi_{inv}$ is useful for $\Preprocess$ to solve
the entire set $\A$ more precisely and effectively.

For example, given the template $T_3$, $\Preprocess$ sets up an unknown
relation $\formr{\rUnk}{x_1,x_2,x_3,x_4,n_1}$ in the template's antecedent,
and introduces the invariant $n_2 {\geq} 1$ of
$\formp{\Dll}{x_5,x_6,x_7,x_8,n_2}$ in the template's consequent to create
the following unknown entailment $E_{u_2}$.

\vspace{-0.35em}
\begin{center}
\begin{small}
  $E_{u_2} \triangleq
  \formp{\Dll}{x_1,x_2,x_3,x_4,n_1} \wedge
  \formr{\rUnk}{x_1,x_2,x_3,x_4,n_1} \entails \exists x_{5,\ldots,11},n_2.
  (\formp{\Dll}{x_5,x_6,x_7,x_8,n_2} * \forms{x_9}{x_{10}, x_{11}} \wedge
  n_2{\geq}1)$
\end{small}
\end{center}
\vspace{0.1em}

\begingroup
\let\oldmapsto\mapsto
\renewcommand\mapsto{\mathord{\scalebox{0.75}[1.0]{$\oldmapsto$}}}
\relax
\begin{figure}[H]
  \begin{footnotesize}
    \begin{prooftree}[label separation = 0.16em,
        separation = 0.8em, rule margin = 0.3em]
      \def\defaultHypSeparation{\hskip 1em}
      \def\ScoreOverhang{0.2em}
      \Hypo{\A_1 \triangleq \{x_1{\sequal}x_3 \,{\wedge}\, n_1{\sequal}1
        \,{\wedge}\, \formr{\rUnk}{x_{1,2,3,4},n_1}\,{\rightarrow}\,
        \false\}}
      % \rewrite{\color{red}\box\treebox}
      \Infer1[{\footnotesize $\ruleSynSigmaTwo$}]{\hboxauto{
          $x_1{\sequal}x_3 \,{\wedge}\, n_1{\sequal}1
          \,{\wedge}\, \formr{\rUnk}{x_{1,2,3,4},n_1}\,{\entails}\,$\\
          \hspace*{1em}$\exists x_{5,...,11},n_2.(
          \formp{\Dll}{x_5,x_6,x_7,x_8,n_2} \,{\wedge}\,$\\
          \hspace*{5em}$x_1{\sequal}x_9 \,{\wedge}\, x_2{\sequal}x_{10}
          \,{\wedge}\, x_4{\sequal}x_{11} \,{\wedge}\, n_2{\geq}1)$
        }}
      \Infer1[{\footnotesize $\ruleStarData$}]{\hboxauto{
          $\forms{x_1}{x_2,x_4} \,{\wedge}\, x_1{\sequal}x_3
          \,{\wedge}\, n_1{\sequal}1
          \,{\wedge}\, \formr{\rUnk}{x_{1,2,3,4},n_1}\,{\entails}\,$\\
          \hspace*{1em}$\exists x_{5,...,11},n_2.
          (\formp{\Dll}{x_5,x_6,x_7,x_8,n_2} \,{*}\, $\\
          \hspace*{10em}$ \forms{x_9}{x_{10}, x_{11}}
          \,{\wedge}\, n_2{\geq}1)$
        }}
      \Infer[rule style = no rule, rule margin = 0em]1{
        \hboxauto{\hspace*{16em}}}
      % ===================================
      \Hypo{\hboxauto{
          $\A_2 \triangleq \{
          \formr{\rUnk}{x_{1,2,3,4},n_1}\,{\rightarrow}\,
          \exists x_{5,...,11},n_2.(
          x_1{\sequal}x_9 \,{\wedge}\, x_2{\sequal}x_{10} \,{\wedge}\,$\\
          \hspace*{0.5em}$u{\sequal}x_{11} \,{\wedge}\, u{\sequal}x_5
          \,{\wedge}\,
          x_1{\sequal}x_6 \,{\wedge}\, x_3{\sequal}x_7 \,{\wedge}\,
          x_4{\sequal}x_8 \,{\wedge}\, n_1{\sminus}1{\sequal}n_2
          \,{\wedge}\, n_2{\geq}1)\}$
        }}
      % \rewrite{\color{red}\box\treebox}
      \Infer1
      [{\footnotesize $\ruleSynPiOne$}]{\hboxauto{
          $\formr{\rUnk}{x_{1,2,3,4},n_1}\,{\entails}\,
          \exists x_{5,...,11},n_2.(
          x_1{\sequal}x_9 \,{\wedge}\, x_2{\sequal}x_{10} \,{\wedge}\,$\\
          \hspace*{0.5em}$u{\sequal}x_{11}\,{\wedge}\, u{\sequal}x_5
          \,{\wedge}\,
          x_1{\sequal}x_6 \,{\wedge}\, x_3{\sequal}x_7 \,{\wedge}\,
          x_4{\sequal}x_8 \,{\wedge}\, n_1{\sminus}1{\sequal}n_2
          \,{\wedge}\, n_2{\geq}1)$
        }}
      \Infer1
      [{\footnotesize $\ruleStarPred$}]{\hboxauto{
          $\formp{\Dll}{u,x_1,x_3,x_4,n_1{\sminus}1}
          \,{\wedge}\, \formr{\rUnk}{x_{1,2,3,4},n_1}\,{\entails}\,
          \exists x_{5,...,11},n_2.$\\
          \hspace*{0.5em}$(\formp{\Dll}{x_5,x_6,x_7,x_8,n_2} \,{\wedge}\,
          x_1{\sequal}x_9 \,{\wedge}\, x_2{\sequal}x_{10} \,{\wedge}\,
          u{\sequal}x_{11} \,{\wedge}\, n_2{\geq}1)$
        }}
      \Infer1
      [{\footnotesize $\ruleStarData$}]{\hboxauto{
          $\forms{x_1}{x_2,u} *
          \formp{\Dll}{u,x_1,x_3,x_4,n_1{\sminus}1}
          \,{\wedge}\, \formr{\rUnk}{x_{1,2,3,4},n_1}\,{\entails}\,$\\
          \hspace*{0.5em}$\exists x_{5,...,11},n_2.
          (\formp{\Dll}{x_5,x_6,x_7,x_8,n_2} * \forms{x_9}{x_{10}, x_{11}}
          \wedge n_2{\geq}1)$
        }}
      \Infer[rule style = no rule, rule margin = 0em]1{
        \hboxauto{\hspace*{22em}}}
      \Infer2[{\footnotesize $\ruleInduction$}]{
        E_{u_2} \triangleq
        \formp{\Dll}{x_1,x_2,x_3,x_4,n_1} \,{\wedge}\,
        \formr{\rUnk}{x_1,x_2,x_3,x_4,n_1}
        \,{\entails}\,
        \exists x_{5,...,11},n_2.
        (\formp{\Dll}{x_5,x_6,x_7,x_8,n_2} * \forms{x_9}{x_{10}, x_{11}}
        \wedge n_2{\geq}1)}
    \end{prooftree}
    \vspace{-0.8em}
    \caption{A possible proof tree of the unknown entailment $E_{u_2}$}
    \vspace{-0.65em}
    \label{fig:ProofTreeUnknEntailPreprocess}
  \end{footnotesize}
\end{figure}
\endgroup

This entailment will be proved by induction to collect constraints about
the unknown relation $\rUnk$. We present its detailed proof in Figure
\ref{fig:ProofTreeUnknEntailPreprocess}. Observe that the assumption
constraint $\formr{\rUnk}{x_1,x_2,x_3,x_4,n_1} \rightarrow \exists
x_5,\ldots,x_{11},n_2.( x_1{=}x_9 \wedge x_2{=}x_{10} \wedge u{=}x_{11}
\wedge u{=}x_5 \wedge x_1{=}x_6 \wedge x_3{=}x_7 \wedge x_4{=}x_8 \wedge
n_1{\sminus}1{=}n_2 \wedge n_2{\geq}1)$ in $\A_2$ can be simplified by
eliminating existentially quantified variables to become
$\formr{\rUnk}{x_1,x_2,x_3,x_4,n_1} \rightarrow n_1{\geq}2$. Therefore, we
can obtain a set $\A = \A_1 \cup \A_2$ containing the following unknown
assumptions:

\vspace{-0.2em}
\begin{center}
\begin{small}
  (1) $x_1{=}x_3 \wedge n_1{=}1 \wedge \formr{\rUnk}{x_1,x_2,x_3,x_4,n_1}
  \rightarrow \false$ \hspace{8em}
  (2) $\formr{\rUnk}{x_1,x_2,x_3,x_4,n_1} \rightarrow n_1{\geq}2$
\end{small}
\end{center}
\vspace{-0.2em}

The procedure $\Preprocess$ can solve the \emph{full assumption constraint
  set} $\A$ to discover a potential solution
$\formr{\rUnkSol}{x_1,x_2,x_3,x_4,n_1} \equiv n_1{\geq}2$, which allows
$T_3$ to be refined to a new lemma template $T'_3$:

\vspace{-0.2em}
\begin{center}
\begin{small}
  $T'_3 \triangleq \formp{\Dll}{x_1,x_2,x_3,x_4,n_1} \wedge
  n_1 {\geq} 2 \entails
  \exists x_5,\ldots,x_{11},n_2.
  (\formp{\Dll}{x_5,x_6,x_7,x_8,n_2} * \forms{x_9}{x_{10},x_{11}})$
\end{small}
\end{center}

\vspace{0.2em}
{\bf Fine-tuning a lemma template's consequent.}

This step aims to refine further the consequent of the lemma template
discovered in the preprocessing phase. The refinement is performed by the
recursive procedure $\FineTuneConseq$ (Figure \ref{fig:FineTuneConseq}).
Its input is a (refined) lemma template $\Sigma_1 \wedge \Pi_1 \entails
\exists \vec{v}. (\Sigma_2 \wedge \Pi_2)$. When $\FineTuneConseq$ is
invoked in the first time by $\RefineConseq$, $\Pi_2$ is set to $\true$
(Figure \ref{fig:RefineConseq}, line \ref{line:RefineConseqFineTune}).

Initially, $\FineTuneConseq$ establishes an unknown
relation $\formr{\rUnk}{\vec{u}}$ on all variables in the templates and
creates an unknown entailment $\Sigma_1 \wedge \Pi_1 \entails \exists
\vec{v}. (\Sigma_2 \wedge \Pi_2 \wedge \formr{\rUnk}{\vec{u}})$ (lines
\ref{line:TuneConseqCreateUnknownBegin},
\ref{line:TuneConseqCreateUnknownEnd}). Then, it proves the unknown
entailment by induction to collect a set $\Assumpts$ of unknown assumptions
(line \ref{line:TuneConseqProve}). Similarly to the antecedent refinement
(Section \ref{sec:FineTuneAnte}), $\Assumpts$ will be {\em exhaustively
  solved}: the entire set and its subsets will be examined until a feasible
solution is found to obtain a valid inductive lemma (lines
\ref{line:TuneConseqSolveBegin} -- \ref{line:TuneConseqRefineEnd}).

\begin{wrapfigure}{r}{0.671\textwidth}
\vspace{-0.6em}
\begin{minipage}{0.671\textwidth}
\begin{figure}[H]
  \begin{algorithm}[H]
    \small
    \caption*{\small{\bf Procedure} \Call{\FineTuneConseq}{
        \Lemmas,\,
        \Sigma_1 \wedge \Pi_1 \entails \exists \vec{v}. (\Sigma_2 \wedge \Pi_2)}}
    \begin{flushleft}
      \Summary{Input:} $\Sigma_1 {\wedge} \Pi_1
      {\entails} \exists \vec{v}.
      (\Sigma_2 {\wedge} \Pi_2)$ is a lemma template,
      $\Lemmas$ is a valid lemma set\\
      \Summary{Output:} a set of at most one synthesized lemma\\
      \Summary{Requires:} $\validlem{\Lemmas}$\\
      %% $\satisfies \{\Prove, \Preprocess, \Solve\}$\\
      \Summary{Ensures:} $(res \,{=}\, \setempty)
      \vee
      \exists\,\Pi'_2.(res\,{=}\,
      \setenum{\Sigma_1 {\wedge} \Pi_1 {\entails}
        \exists \vec{v}.(\Sigma_2 {\wedge} \Pi_2 {\wedge} \Pi'_2)}
      \,{\wedge}\, \validlem{res})$
    \end{flushleft}
    \begin{algorithmic}[1]
      \algrestore{myalg}
      \State \Assign{\vec{u}}{\Call{\FreeVars}{
          \Sigma_1, \Sigma_2, \Pi_1, \Pi_2}}
      \label{line:TuneConseqCreateUnknownBegin}
      \State \Assign{\formr{\rUnk}{\vec{u}}}{
        \Call{\CreateUnknownRelation}{\vec{u}}}
      \label{line:TuneConseqCreateUnknownEnd}
      \State \Assign{\vR,\, \_,\, \Assumpts}{\Call{\Prove}{
          \setempty,\, \Lemmas,\,
          \Sigma_1 {\wedge}\, \Pi_1 {\entails}\, \exists \vec{v}.
          (\Sigma_2 {\wedge}\, \Pi_2 {\wedge}
          \formr{\rUnk}{\vec{u}}),\, \NoSynLemma}}
      \label{line:TuneConseqProve}
      % \Comment{search for a proof}
      \If{\vR  = \ValidWitness{\_}}
        \For{\Keyword{each} \Assumpts'
            \Keyword{in} \Call{SuperSet}{\Assumpts}}
          \label{line:TuneConseqSolveBegin}
          % \Comment{exhaustively solve constraint set}
          \State \Assign{\formr{\rUnkSol}{\vec{u}}}{
            \Call{\Solve}{\Assumpts', \formr{\rUnk}{\vec{u}}}}
          \label{line:TuneConseqSolve}
          \State \Assign{\vL}{(\Sigma_1 \,{\wedge}\, \Pi_1
            \,{\entails}\,
            \exists \vec{v}. (\Sigma_2 \,{\wedge}\, \Pi_2
            \,{\wedge}\, \formr{\rUnkSol}{\vec{u}})}
          \State \Assign{\vR',\,\_,\,\_}{
            \Call{\Prove}{\setempty,\, \Lemmas,\, L, \, \NoSynLemma}}
          \Comment{Verify ...}
          \label{line:TuneConseqVerify}
          \If{\vR' {=}\, \ValidWitness{\xi} \Keyword{and}
                \ruleHypo \,{\in}\, \xi}
            \label{line:TuneConseqVerifyIH}
            \label{line:TuneConseqRefineBegin}
            \Comment{the inductive lemma, and ...}
            \State \Assign{\LemmasSyn}{\Call{\FineTuneConseq}{\Lemmas,\,\vL}}
            \Comment{strengthen its consequent.}
            \If{\LemmasSyn = \setempty} \Return{\setenum{\vL}}
            \Else~ \Return{\LemmasSyn}
            \EndIf
            \label{line:TuneConseqRefineEnd}
          \EndIf
          \label{line:TuneConseqSolveEnd}
        \EndFor
      \EndIf
      \State \Return{\setempty}
      \Comment{No lemmas are found.}
    \end{algorithmic}
  \end{algorithm}
\vspace{-0.6em}
  \caption{Fine-tuning the consequent of a lemma template}
  \label{fig:FineTuneConseq}
\end{figure}
\end{minipage}
\end{wrapfigure}

We also prevent the proof system from entering nested lemma synthesis
phases, controlled by the argument $\NoSynLemma$, when $\Prove$ is invoked
to collect the unknown assumption set $\Assumpts$ (line
\ref{line:TuneConseqProve}) or to verify the inductive lemma (line
\ref{line:TuneConseqVerify}). Witness proof tree of a candidate lemma, if
having, is examined for the occurrence of the induction hypothesis
application rule $(\ruleHypo)$, to determine if this candidate is a valid
inductive lemma (line \ref{line:TuneConseqVerifyIH}).

However, unlike the antecedent refinement, $\FineTuneConseq$ keeps refining
the template until its consequent cannot be refined further (line
\ref{line:TuneConseqRefineBegin}, \ref{line:TuneConseqRefineEnd}). This
repetition ensures that the returned lemma's consequent, if discovered, is
as strong as possible.

For example, given the preprocessed template $T'_3$, $\FineTuneConseq$
creates an unknown relation $\formr{\rUnk}{x_1,\ldots,x_{11},n_1,n_2}$, or
$\formr{\rUnk}{\vec{x}, \vec{n}}$ for short, involving all variables in
$T'_3$ and constructs the unknown entailment $E_{u_3}$. This entailment
will be proved again to collect constraints about $\formr{\rUnk}{\vec{x},
  \vec{n}}$.

\begin{center}
\begin{small}
  $E_{u_3} \triangleq \formp{\Dll}{x_1,x_2,x_3,x_4,n_1}
  \wedge n_1{\geq}2 \entails
  \exists x_5,\ldots,x_{11},n_2.(
  \formp{\Dll}{x_5,x_6,x_7,x_8,n_2} *
  \forms{x_9}{x_{10},x_{11}} \wedge
  \formr{\rUnk}{\vec{x}, \vec{n}})$
\end{small}
\end{center}

\vspace{0.6em}
\begingroup
\let\oldmapsto\mapsto
\renewcommand\mapsto{\mathord{\scalebox{0.75}[1.0]{$\oldmapsto$}}}
\relax
\begin{figure}[H]
  \begin{footnotesize}
    \begin{prooftree}[label separation = 0.16em,
        separation = 0.2em, rule margin = 0.33em]
      \def\defaultHypSeparation{\hskip 1em}
      \def\ScoreOverhang{0.2em}
      \Hypo{\A_1}
      % \rewrite{\color{red}\box\treebox}
      \Infer[rule margin = 1.1em]1{$\ldots$}
      % ===================================
      \Hypo{\A_2}
      % \rewrite{\color{red}\box\treebox}
      \Infer[rule margin = 1.2em]1{$\ldots$}
      \Infer[rule style = no rule]1{}
      \Hypo{\hboxauto{
          $\A_3 \triangleq \{n_1\!{\geq}2
          \wedge r{\sequal}x_3
          \wedge n_1{\sminus}2{\sequal}1
          \rightarrow
          \exists x_{5,...,11}{,}n_2{,}v.(
          \formr{\rUnk}{\vec{x}{,}\vec{n}}
          \wedge x_1{\sequal}x_5~\wedge $\\
          \hspace*{3em}$x_2{\sequal}x_6
          \wedge u{\sequal}v
          \wedge v{\sequal}x_7
          \wedge r{\sequal}x_8
          \wedge r{\sequal}x_9
          \wedge u{\sequal}x_{10}
          \wedge x_4{\sequal}x_{11}
          \wedge n_2{\sminus}1{\sequal}1)\}$
        }}
      % \rewrite{\color{red}\box\treebox}
      \Infer1
      [{\footnotesize $\ruleSynPiOne$}]{\hboxauto{
          $ n_1{\geq}2
          \wedge r{\sequal}x_3
          \wedge n_1{\sminus}2{\sequal}1
          \entails
          \exists x_{5,...,11}{,}n_2{,}v.(
          \formr{\rUnk}{\vec{x}{,}\vec{n}}
          \wedge x_1{\sequal}x_5 ~\wedge$\\
          \hspace*{2em}$x_2{\sequal}x_6
          \wedge u{\sequal}v
          \wedge v{\sequal}x_7
          \wedge r{\sequal}x_8
          \wedge r{\sequal}x_9
          \wedge u{\sequal}x_{10}
          \wedge x_4{\sequal}x_{11}
          \wedge n_2{\sminus}1{\sequal}1)$
        }}
      \Infer1
      [{\footnotesize $\ruleStarData$}]{\hboxauto{
          $\forms{r}{u{,}x_4}
          \wedge n_1{\geq}2
          \wedge r{\sequal}x_3
          \wedge n_1{\sminus}2{\sequal}1
          \entails
          \exists x_{5,...,11}{,}n_2{,}v.(\forms{x_9}{x_{10}{,} x_{11}}
          ~\wedge$\\
          \hspace*{2em}$\formr{\rUnk}{\vec{x}{,}\vec{n}}
          \wedge x_1{\sequal}x_5
          \wedge x_2{\sequal}x_6
          \wedge u{\sequal}v
          \wedge v{\sequal}x_7
          \wedge r{\sequal}x_8
          \wedge n_2{\sminus}1{\sequal}1)$
        }}
      \Hypo{\A_4}
      % \rewrite{\color{red}\box\treebox}
      \Infer[rule margin = 1.15em]1{$\ldots$}
      \Infer[rule style = no rule]1{\hspace*{3em}}
      \Infer[separation = 1em]2
      [{\footnotesize $\ruleInduction$}]{\hboxauto{
          $\formp{\Dll}{r{{,}}u{{,}}x_3{{,}}x_4{{,}}n_1{\sminus}2}
          \wedge n_1{\geq}2 \entails
          \exists x_{5,...,11}{,}n_2{,}v.(\forms{x_9}{x_{10}{{,}} x_{11}}
          ~\wedge$\\
          \hspace*{2em}$ \formr{\rUnk}{\vec{x}{{,}}\vec{n}}
          \wedge x_1{\sequal}x_5
          \wedge x_2{\sequal}x_6
          \wedge u{\sequal}v
          \wedge v{\sequal}x_7
          \wedge r{\sequal}x_8
          \wedge n_2{\sminus}1{\sequal}1)$
        }}
      \Infer1
      [{\footnotesize $\ruleStarData$}]{\hboxauto{
          $\forms{u}{x_1{{,}}r} {*}
          \formp{\Dll}{r{{,}}u{{,}}x_3{{,}}x_4{{,}}n_1{\sminus}2}
          \wedge n_1{\geq}2 \entails
          \exists x_{5,...,11}{,}n_2{,}v.(\forms{v}{x_5{{,}}x_8}
          ~*$\\
          \hspace*{5em}$\forms{x_9}{x_{10}{{,}} x_{11}}
          \wedge \formr{\rUnk}{\vec{x}{{,}}\vec{n}}
          \wedge x_1{\sequal}x_5
          \wedge x_2{\sequal}x_6
          \wedge u{\sequal}v
          \wedge v{\sequal}x_7
          \wedge n_2{\sminus}1{\sequal}1)$
        }}
      \Infer[separation = 2em]2
      [{\footnotesize $\ruleInduction$}]{\hboxauto{
          $\formp{\Dll}{u{,}x_1{,}x_3{,}x_4{,}n_1{\sminus}1}
          {\wedge} n_1\!{\geq}2 {\entails}
          \exists x_{5,...,11}{,}n_2{,}v.(\forms{v}{x_5{,}x_8} {*}
          \forms{x_9}{x_{10}{,} x_{11}}
          {\wedge} \formr{\rUnk}{\vec{x}{,}\vec{n}}
          {\wedge} x_1{\sequal}x_5
          {\wedge} x_2{\sequal}x_6
          {\wedge} u{\sequal}v
          {\wedge} v{\sequal}x_7
          {\wedge} n_2{\sminus}1{\sequal}1)$
        }}
      \Infer1
      [{\footnotesize $\rulePredRight$}]{\hboxauto{
          $\formp{\Dll}{u{,}x_1{,}x_3{,}x_4{,}n_1{\sminus}1}
          {\wedge} n_1\!{\geq}2 {\entails}
          \exists x_{5,...,11}{,}n_2{,}v.(
          \formp{\Dll}{v{,}x_5{,}x_7{,}x_8{,}n_2{\sminus}1}
          {*} \forms{x_9}{x_{10}{,} x_{11}}
          {\wedge} \formr{\rUnk}{\vec{x}{,}\vec{n}}
          {\wedge} x_1{\sequal}x_5
          {\wedge} x_2{\sequal}x_6
          {\wedge} u{\sequal}v)$
        }}
      \Infer1
      [{\footnotesize $\ruleStarData$}]{\hboxauto{
          $\forms{x_1}{x_2{,}u} *
          \formp{\Dll}{u{,}x_1{,}x_3{,}x_4{,}n_1{\sminus}1}
          {\wedge} n_1\!{\geq}2 {\entails}
          \exists x_{5,...,11}{,}n_2{,}v.(
          \forms{x_5}{x_6{{,}}v}{*}
          \formp{\Dll}{v{,}x_5{,}x_7{,}x_8{,}n_2{\sminus}1}
          {*} \forms{x_9}{x_{10}{,} x_{11}}
          {\wedge} \formr{\rUnk}{\vec{x}{,}\vec{n}})$
        }}
      \Infer1
      [{\footnotesize $\rulePredRight$}]{\hboxauto{
          $\forms{x_1}{x_2{,}u} *
          \formp{\Dll}{u{,}x_1{,}x_3{,}x_4{,}n_1{\sminus}1}
          {\wedge} n_1\!{\geq}2 {\entails}
          \exists x_{5,...,11}{,}n_2.(
          \formp{\Dll}{x_5{,}x_6{,}x_7{,}x_8{,}n_2} {*}
          \forms{x_9}{x_{10}{,} x_{11}}
          {\wedge} \formr{\rUnk}{\vec{x}{,}\vec{n}})$
        }}
      % \Infer[rule style = no rule{,} rule margin = 0em]1{
      %   \hboxauto{\hspace*{28.5em}}}
      \Infer2[{\footnotesize $\ruleInduction$}]{
        E_{u_3} \triangleq
        \formp{\Dll}{x_1{,}x_2{,}x_3{,}x_4{,}n_1} \,{\wedge}\,
        n_1\!{\geq}2
        \,{\entails}\,
        \exists x_{5,...,11}{,}n_2.
        (\formp{\Dll}{x_5{,}x_6{,}x_7{,}x_8{,}n_2} * \forms{x_9}{x_{10}{,} x_{11}}
        \,{\wedge}\, \formr{\rUnk}{\vec{x}{,}\vec{n}})}
    \end{prooftree}
    \vspace{-0.6em}
    \caption{A partial proof tree of the unknown entailment $E_{u_3}$}
    \label{fig:ProofTreeUnknEntailRefineConseq}
  \end{footnotesize}
\end{figure}
\endgroup

We present a partial proof tree of $E_{u_3}$ in Figure
\ref{fig:ProofTreeUnknEntailRefineConseq}, where $\FineTuneConseq$ is able
to collect a set of unknown assumptions $\A ~{=}~ \A_1 \,{\cup}\, \A_2
\,{\cup}\, \A_3 \,{\cup}\, \A_4$. More specifically, $\A_3 \triangleq
\{n_1{\geq}2 \wedge r{\sequal}x_3 \wedge n_1{\sminus}2{\sequal}1 \entails
\exists x_{5,...,11},n_2,v.\,(\formr{\rUnk}{\vec{x},\vec{n}} \wedge
x_1{\sequal}x_5 \wedge x_2{\sequal}x_6 \wedge u{\sequal}v \wedge
v{\sequal}x_7 \wedge r{\sequal}x_8 \wedge r{\sequal}x_9 \wedge
u{\sequal}x_{10} \wedge x_4{\sequal}x_{11} \wedge
n_2{\sminus}1{\sequal}1)\}$ is an assumption subset derived from a
potential proof path of an inductive lemma, and $\A_1, \A_2, \A_4$ relate
to other proof paths. The set $\A$ can be partially solved by considering
only $\A_3$. Here, a possible solution of $\A_3$ is
$\formr{\rUnkSol_p}{x_1,...,x_{11},n_1,n_2} \equiv x_1{=}x_5 \wedge
x_2{=}x_6 \wedge x_3{=}x_8 \wedge x_4{=}x_{11} \wedge x_7{=}x_{10} \wedge
x_8{=}x_9$. This solution can be substituted back to $E_{u_3}$ to derive
the refined lemma template $T''_3$.

\vspace{-0.2em}
\begin{center}
\begin{small}
  $T''_3 \triangleq \formp{\Dll}{x_1,x_2,x_3,x_4,n_1}
  \wedge n_1{\geq}2 \entails
  \exists x_7,n_2.(
  \formp{\Dll}{x_1,x_2,x_7,x_3,n_2} *
  \forms{x_3}{x_7,x_4})$
\end{small}
\end{center}
\vspace{-0.2em}

Then, $\FineTuneConseq$ constructs another unknown entailment
$\formr{\rUnk'}{x_1,x_2,x_3,x_4,x_7,n_1,n_2}$ on all variables of $T''_3$
and creates a new unknown entailment, which will be proved again to find a
solution $\formr{\rUnkSol'_p}{x_1,x_2,x_3,x_4,x_7,n_1,n_2} \equiv
n_1{=}n_2{+}1$. This solution helps to refine the template $T''_3$ to
obtain an inductive lemma $\overbar{T}_3 \triangleq
\formp{\Dll}{x_1,x_2,x_3,x_4,n_1} \wedge n_1{\geq}2 \entails \exists
x_7.(\formp{\Dll}{x_1,x_2,x_7,x_3,n_1{-}1} *\forms{x_3}{x_7,x_4})$, which
can be equivalently transformed to the motivating lemma $L_3$ in Section
\ref{sec:MotivatingExample}.

\vspace{-0.6em}
\subsection{Inferring pure invariants of inductive heap predicates}
\label{sec:HeapPredicateInvariant}
\vspace{-0.5em}

We present in this subsection the construction of inductive heap
predicates' invariants, which are mainly utilized to preprocess the lemma
templates' antecedents (Section \ref{sec:FineTuneConseq}). Our construction
is inspired from separation logic literature \cite{ChinDNQ12,
  BrotherstonFPG14, LeSC16}.

\vspace{-0.1em}
\begin{definition}[Invariant of inductive heap predicates]
  Given an inductive heap predicate $\formp{\P}{\vec{x}}$, a pure formula
  $\Call{\Invariant}{\formp{\P}{\vec{x}}}$ is called an invariant of
  $\formp{\P}{\vec{x}}$ iff $s,h \satisfies \formp{\P}{\vec{x}}$ implies
  that $s \satisfies \Call{\Invariant}{\formp{\P}{\vec{x}}}$, for all model
  $s,h$. Formally, $\forall s, h.\,( s,h \satisfies \formp{\P}{\vec{x}}
  \rightarrow s \satisfies \Call{\Invariant}{\formp{\P}{\vec{x}}})$.
\end{definition}

\vspace{-0.6em}
\textbf{Constructing pure invariants}. We also exploit a
template-based approach to discover the pure invariants. For a system of
$k$ (mutually) inductive heap predicates $\formp{\P_1}{\vec{x}_1}, \ldots,
\formp{\P_k}{\vec{x}_k}$, we first create $k$ unknown relations
$\formr{U_{\P_1}}{\vec{x}_1}, \ldots, \formr{U_{\P_k}}{\vec{x}_k} $ to
respectively represent their desired invariants. Then, we follow the
predicates' definitions to establish constraints about the unknown
relations.

\vspace{-0.2em}
In particular, consider each definition case $\formr{F^i_j}{\vec{x_i}}$
of a predicate $\formp{\P_i}{\vec{x}_i}$, which is of the form
$\formr{F^i_j}{\vec{x_i}} \triangleq \exists
\vec{u}.(\Sigma^i_j \wedge \Pi^i_j)$. Suppose that
$\formp{\Q_1}{\vec{u}_1}, ..., \formp{\Q_n}{\vec{u}_n}$ are {\em all} inductive
heap predicates in $\Sigma^i_j$, where $\Q_1, ..., \Q_n \,{\in}\,
\setenum{\P_1, ..., \P_k}$, we create the following unknown assumption:

\vspace{-0.3em}
\begin{center}
\begin{small}
  $\formr{U_{\Q_1}}{\vec{u}_1} \wedge \ldots \wedge
  \formr{U_{\Q_n}}{\vec{u}_n} \wedge \Pi^i_j
  \quad \rightarrow \quad \formr{U_{\P_i}}{\vec{x}_i}$
\end{small}
\end{center}
\vspace{-0.2em}

Thereafter, the system of all unknown assumptions can be solved by the
constraint solving technique based on Farkas' lemma (Section
\ref{sec:SolveConstraint}) to discover the actual invariants of
$\formp{\P_1}{\vec{x_1}}, \ldots, \formp{\P_k}{\vec{x_k}}$.

For example, given the predicate $\formp{\Dll}{hd,pr,tl,nt,len}$
in Section \ref{sec:MotivatingExample}, whose definition is

\vspace{-0.5em}
\begin{center}
\begin{small}
\begin{tabular}{rll}

  $\formp{\Dll}{hd,pr,tl,nt,len}$
  & $\defeq$
  & $(\forms{hd}{pr,nt} \wedge hd{=}tl \wedge len{=}1)
  ~~\vee~~ \exists u. (\forms{hd}{pr,u} * \formp{\Dll}{u,hd,tl,nt,len{-}1})$

\end{tabular}
\end{small}
\end{center}
\vspace{-0.2em}

We create an unknown relation $\formr{\rUnk_{\Dll}}{hd,pr,tl,nt,len}$
representing its invariant and follow the definition of
$\formp{\Dll}{hd,pr,tl,nt,len}$ to establish the following constraints:

\vspace{-0.3em}
\begin{center}
\begin{small}
  (1)~~ $hd{=}tl \wedge len{=}1
  \rightarrow
  \formr{\rUnk_{\Dll}}{hd,pr,tl,nt,len}$
  \hspace{2em}
  (2)~~ $\formr{\rUnk_{\Dll}}{u,hd,tl,nt,len{-}1}
  \rightarrow
  \formr{\rUnk_{\Dll}}{hd,pr,tl,nt,len}$
\end{small}
\end{center}
\vspace{-0.2em}

We then solve these constraints to obtain the solution
$\formr{\rUnkSol_{\Dll}}{hd,pr,tl,nt,len} \equiv len\,{\geq}\,1$, which is
also the invariant of $\formp{\Dll}{hd,pr,tl,nt,len}$.

% \textbf{Soundness of the invariant construction}. We claim that our
% invariant constructions are sound. This can be proved by induction on the
% inductive heap predicates' structures. Due to page limit, we present this
% soundness proof in the Appendix \ref{append:SoundnessProof}.

\vspace{-0.6em}
\subsection{Solving assumption constraints with Farkas' lemma}
\label{sec:SolveConstraint}
\vspace{-0.5em}

In this subsection, we describe the underlying constraint solving technique
based on Farkas' lemma. This technique is implemented in the procedure
$\Solve$, which is frequently invoked in the lemma synthesis to solve an
unknown assumption set (Sections \ref{sec:FineTuneAnte},
\ref{sec:FineTuneConseq}, and \ref{sec:HeapPredicateInvariant}). We will
formally restate Farkas' lemma and explain how it is applied in constraint
solving.

{\bf Farkas' lemma} \cite{Schrijver1986}. Given a conjunction of linear
constraints $\midwedge^m_{j=1}\, \midsum^n_{i=1}\!\! {a_{ij}x_i} \,{+}\,
b_j \,{\geq}\, 0$, which is satisfiable, and a linear constraint $\midsum^n_{i=1}\!\! {c_{i}x_i}
\,{+}\, \gamma \,{\geq}\, 0$, we have:

\vspace*{-1.5em}
\begin{small}
\[
\forall x_1 \ldots x_n .~
\big(
\underset{{j=1}}{\overset{m}{\midwedge}}\,
\underset{i=1}{\overset{n}{\midsum}}\!\! {a_{ij}x_i} \,{+}\, b_j \,{\geq}\, 0
\big)
\pureImply
\underset{i=1}{\overset{n}{\midsum}}\!\! {c_{i}x_i} \,{+}\, \gamma \,{\geq}\, 0
\quad \text{iff} \quad
\exists \lambda_1 \ldots \lambda_m \,{\geq}\, 0 .~
\big(
\underset{{i=1}}{\overset{n}{\midwedge}}\!\! c_i \,{=}
\underset{j=1}{\overset{m}{\midsum}}\!\! \lambda_j a_{ij} ~ \wedge
\underset{j=1}{\overset{m}{\midsum}}\!\! \lambda_j b_j \,{\leq}\, \gamma
\big)
\]
\end{small}
\vspace*{-1.3em}

{\bf Solving constraints.} Given an assumption set $\Assumpts$ of an
unknown relation $\formr{\rUnk}{x_1,\ldots,x_m,u_1,\ldots,u_n}$, where
$x_1,\ldots,x_m$ are spatial variables and $u_1,\ldots,u_n$ are integer
variables, the procedure $\Solve$ aims to find an actual definition
$\rUnkSol$ of $\rUnk$ in the following template, where $c_{ij}, d_{ij}$ are
unknown coefficients, $M, N$ are pre-specified numbers of conjuncts.

\vspace*{-1.3em}
\begin{small}
\[
\formr{\rUnk}{x_1,\ldots,x_m,u_1,\ldots,u_n}
\quad\triangleq\quad
\big(\midwedge^M_{j=1}\, \midsum^m_{i=1}
\!\! {c_{ij}x_i} + c_{0j} \geq 0\big)
~\wedge~
\big(\midwedge^N_{j=1}\, \midsum^n_{i=1}
\!\! d_{ij}u_i + d_{0j} \geq 0\big)
\]
\end{small}
\vspace*{-1.3em}

Recall that our lemma synthesis framework can incrementally discover the
final solution $\rUnkSol$ in several passes. Hence, we can set $M,N$ small,
e.g., $1{\leq} M{,}N{\leq}3$, or $M {=} 0,N {\leq} 6$ (no spatial
contraints), or $M {\leq} 6,N {=} 0$ (no arithmetic constraints) to make
the constraint solving more effective. We restrict the coefficients
$c_{ij}$ to represent equality or disequality constraints of spatial
variables, e.g., $x_k{=}x_l$, $x_k{\neq}x_l$, $x_k{=}0$, and $x_k{\neq}0$,
where $0$ denotes $\nil$. The constraint $x_k{=}x_l$ can be encoded as $x_k
{-} x_l {\geq} 0 \wedge -x_k {+} x_l {\geq} 0$, or the encoding of
$x_k{\neq}x_l$ is $x_k {-} x_l {-} 1 {\geq} 0$. Therefore, it requires that
$-1 {\leq} c_{ij} {\leq} 1$ for $i{>}0$, and $c_{0j} {=} 0$ (for
equalities) or $c_{0j} {=} {-1}$ (for disequalities). We also add the
restrictions $-1\,{\leq}\,{\midsum^m_{i=1}\!\! c_{ij}}\,{\leq}\,1$ and
$1\,{\leq}\,{\midsum^m_{i=1}\!\! |c_{ij}|}\,{\leq}\,2$ to ensure that the
spatial constraints involve at most two variables.

In summary, the assumption set $\Assumpts$ of the unknown relation $\rUnk$
can be solved in three steps:

\vspace{-0.1em}

\begin{enumerate}[label=--,leftmargin=1em, noitemsep, nosep]
\item Normalize assumptions in the set $\Assumpts$ into the form of Horn
  clause, and substitute all occurrences of $U$ in the normalized
  assumptions by its template to obtain a set of normalized constraints.

\item Apply Farkas' lemma to eliminate universal quantification to obtain
  new constraints with only existential quantification over the unknown
  coefficients $c_{ij}, d_{ij}$ and the factors $\lambda_j$.

\item Use an off-the-shelf prover, such as Z3, to find a model of the
  unknown coefficients from their constraints, and replace them back in the
  template to discover the actual definition of $\rUnk$.
\end{enumerate}

\vspace{-0.6em}
\subsection{Soundness of the lemma synthesis framework}
\vspace{-0.3em}

We claim that our lemma synthesis framework is sound, that is, all lemmas
synthesized by the framework are semantically valid. We formally state the
soundness in the following Theorem \ref{thm:SoundnessLemmaSynthesis}.

\begin{restatable}[Soundness of the lemma
    synthesis]{theorem}{TheoremSoundnessLemmaSynthesis}
  \label{thm:SoundnessLemmaSynthesis}
  Given a normal goal entailment $E$ which does not contain any unknown
  relation and a set of valid input lemma $\Lemmas$, if the lemma synthesis
  procedure $\SynthesizeLemma$ returns a set of lemmas $\LemmasSyn$, then
  all lemmas in $\LemmasSyn$ are semantically valid.
\end{restatable}

\vspace{-1.2em}
\begin{proof}
  Figure \ref{fig:SynthesizeLemmas} shows that all lemmas returned by
  $\SynthesizeLemma$ are directly discovered by $\RefineAnte$ and
  $\RefineConseq$. In addition, all lemmas returned by $\RefineConseq$ are
  realized by $\FineTuneConseq$ (Figure \ref{fig:RefineConseq}, line
  \ref{line:RefineConseqFineTune}). Moreover, all lemmas discovered by
  $\RefineAnte$ and $\FineTuneConseq$ are verified by $\Prove$ in a setting
  that disables the lemma synthesis ($\NoSynLemma$) and utilizes the valid
  lemma set $\Lemmas$ (Figure \ref{fig:FineTuneAnte}, line
  \ref{line:TuneAnteVerify} and Figure \ref{fig:FineTuneConseq}, line
  \ref{line:TuneConseqVerify}). It follows from Proposition
  \ref{thm:SoundnessProofSystemClassic} that if $\Prove$ returns
  $\ValidWitness{\_}$ when verifying a lemma, then the lemma is
  semantically valid. Consequently, all lemmas returned by
  $\SynthesizeLemma$ are semantically valid.
\end{proof}

%%% Local Variables:
%%% mode: latex
%%% TeX-master: "main"
%%% End:

%%%%%%%%%%%%%%%%%%%%%%%%%%%%%%%%%%%%%%%%%%%%%%%%%%%%%
%%%%%%%%%%%%%%%%%%%%%%%%%%%%%%%%%%%%%%%%%%%%%%%%%%%%%
\vspace{-0.8em}
\section{Experiments}
\label{sec:Experiment}
\vspace{-0.3em}

We have implemented the lemma synthesis framework into a prototype prover,
named \songbirdLS\footnote{\,\songbirdLS is built on top of an existing
  prover \songbird \cite{TaLKC16}; its name stands for ``\textbf{S}ongbird
  + \textbf{L}emma \textbf{S}ynthesis''} and have conducted two experiments
to evaluate its ability in proving entailments. Both the prover and the
experiment details are available online at {\songbirdLSURL}.

The first experiment was done on literature benchmarks, which include
\textsf{sll\_entl, slrd\_entl} from the separation logic competition
SL-COMP 2014 \cite{SighireanuC14}, and \textsf{slrd\_indt} by
\citet{TaLKC16}. These benchmarks focus on representing the data
structures' shapes, hence, their pure formulas solely contain equality
constraints among spatial variables. Therefore, we decided to compose a
richer entailment benchmark \textsf{slrd\_lm}, which captures also
arithmetic constraints of the data structures' size and content, and then
performed the second experiment. Moreover, we only considered entailments
that are marked as {\em valid}. Our experiments were conducted on a Ubuntu
14.04 LTS machine with CPU Intel\textsuperscript{\textregistered}
Core\textsuperscript{{\tiny TM}} i7-6700 (3.4GHz) and RAM 16GB. We compared
$\songbirdLS$ against state-of-the-art separation logic provers, including
$\slide$ \cite{IosifRS13}, $\sleek$ \cite{ChinDNQ12}, $\spen$
\cite{EneaLSV14}, $\cyclist$ \cite{BrotherstonDP11} and $\songbirdMI$
\cite{TaLKC16}. These provers were configured to run with a timeout of 180
seconds for each entailment.

\vspace{0.3em}
\begin{table}[t]
  \caption{Evaluation on the existing entailment benchmarks, where
    participants are \slide~(\slideAbr), \sleek~(\sleekAbr),
    \spen~(\spenAbr), \cyclist~(\cyclistAbr),
    \songbird (\songbirdMiAbr)
    and our prototype prover \songbirdLS}
  \vspace{-0.65em}
  \label{fig:AllEntl}

  \setlength\tabcolsep{0.4pt}
  \def\arraystretch{1.1}%  1 is the default, change whatever you need
  \scriptsize
  \begin{tabular}{|g|Lc|gwgwgw|gwgwgw|gwgwgw|gwgw|gwg|}
    \hline
    \multicolumn{3}{|c|}{{\bf ~Benchmark}}
    & \multicolumn{6}{c|}{\bf Proved Entailments}
    & \multicolumn{6}{c|}{\bf Total Proving Time (s)}
    & \multicolumn{6}{c|}{\bf Average Time (s)}
    % & \multicolumn{7}{c|}{\bf Lemma statistics}
    & \multicolumn{4}{c|}{\bf Lemma Syn}
    & \multicolumn{3}{c|}{\bf Lemma App}
    \\

    \cline{1-28}

    % \rowcolor{LightCyan}
    \multicolumn{2}{|L}{~Category} & \#En
    & \slideAbr & \sleekAbr & \spenAbr
    & \cyclistAbr & \songbirdMiAbr & \songbirdLsAbr

    & \slideAbr & \sleekAbr & \spenAbr
    & \cyclistAbr & \songbirdMiAbr & \songbirdLsAbr

    & \slideAbr \,& \sleekAbr \,& \spenAbr
    \,& \cyclistAbr \,& \songbirdMiAbr \,& \songbirdLsAbr

    & \#Lm & T\,(s) & A\,(s) & O\,(\%)
    & \#Cv & \#Sp & \#Cb
    \\

    \hline
    & ~\textsf{bolognesa} & 57
    & 0 & 0 & {\bf 57}
    & 0 & {\bf 57} & {\bf 57}
    & 0.0 & 0.0 & 23.6
    & 0.0 & 140.3 &	18.5
    & -- & -- & 0.41
    & -- & 2.46 & 0.32

    & 0 & 0.0 & -- & 0.0
    & 0 & 0 & 0
    \\

    & ~\textsf{clones} & 60
    & 0 & {\bf 60} & {\bf 60}
    & {\bf 60} & {\bf 60} & {\bf 60}
    & 0.0 & 3.7 & 3.5
    & 0.5 & 3.9	& 0.7
    & -- & 0.06 & 0.06
    & 0.01 & 0.07 & 0.01

    & 0 & 0.0 & -- & 0.0
    & 0 & 0 & 0
    \\

    \multirow{-3}{*}{\,\rotatebox[origin=c]{90}{\textsf{sll\_entl}}\,}
    & ~\textsf{smallfoot} & 54
    & 0 & {\bf 54} & {\bf 54}
    & {\bf 54} & {\bf 54} & {\bf 54}
    & 0.0 & 2.7 & 2.5
    & 11.8 & 3.5 & 4.2
    & -- & 0.05 & 0.05
    & 0.22 & 0.06 & 0.08

    & 0 & 0.0 & -- & 0.0
    & 0 & 0 & 0
    \\
    \hline

    %%%%%%%%%%%%%%%%%%%%%%%%%%%%%%%%%%%%%%%%%%%%%%%%%%%%%%%%%%%%%%
    & ~\textsf{singly-ll} & 64
    & 12 & 48 & 3
    & 63 & {\bf 64} & {\bf 64}
    & 1.0 & 6.3 & 0.1
    & 2.1 & 8.3	& 1.7
    & 0.08 & 0.13 & 0.04
    & 0.03 & 0.13 & 0.03

    & 0 & 0.0 & -- & 0.0
    & 0 & 0 & 0
    \\

    & ~\textsf{doubly-ll} & 37
    & 14 & 17 & 9
    & 29 & 25 & {\bf 35}
    & 38.3 & 3.1 & 0.4
    & 112.5 & 11.3 & 91.9
    & 2.74 & 0.18 & 0.04
    & 3.88 & 0.45 & 2.63

    & 18[10]  & 51.2 & 2.8 & 55.7
    & 8\,(68) & 0 & 2\,(2)
    \\

    & ~\textsf{nested-ll} & 11
    & 0 & 5 & {\bf 11}
    & 7 & {\bf 11} & {\bf 11}
    & 0.0 & 2.2 & 0.5
    & 16.7 & 2.3 & 0.4
    & -- & 0.44 & 0.04
    & 2.38 & 0.21 & 0.04

    & 0 & 0.0 & -- & 0.0
    & 0 & 0 & 0
    \\

    & ~\textsf{skip-list} & 13
    & 0 & 4 & {\bf 13}
    & 5 & {\bf 13} & {\bf 13}
    & 0.0 & 1.1 & 1.1
    & 0.6 & 8.1 & 1.3
    & -- & 0.27 & 0.08
    & 0.11 & 0.63 & 0.10

    & 0 & 0.0 & -- & 0.0
    & 0 & 0 & 0
    \\

    \multirow{-5}{*}{\rotatebox[origin=c]{90}{\textsf{slrd\_entl}}}
    & ~\textsf{tree} & 26
    & 12 & 14 & 0
    & 23 & 23 & {\bf 24}
    & 110.0 & 2.9 & 0.0
    & 58.8 & 11.5 &	2.2
    & 9.16 & 0.21 & --
    & 2.55 & 0.50 & 0.09

    & 0 & 0.0 & -- & 0.0
    & 0 & 0 & 0
    \\
    \hline

    % %%%%%%%%%%%%%%%%%%%%%%%%%%%%%%%%%%%%%%%%%%%%%%%%%%%%%%%%%%%%%%
    & ~\textsf{ll/ll2} & 24
    & 0 & 0 & 0
    & {\bf 24} & {\bf 24} & {\bf 24}
    & 0.0 & 0.0 & 0.0
    & 60.2 & 11.9	& 78.5
    & -- & -- & --
    & 2.51 & 0.50 & 3.27

    & 14[10] & 12.6 & 0.9 & 16.1
    & 4\,(4) & 0 & 6\,(11)
    \\

    & ~\textsf{ll-even/odd} & 20
    & 0 & 0 & 0
    & {\bf 20} & {\bf 20} & {\bf 20}
    & 0.0 & 0.0 & 0.0
    & 34.6 & 50.3	& 1.3
    & -- & -- & --
    & 1.73 & 2.52 & 0.07

    & 0 & 0.0 & -- & 0.0
    & 0 & 0 & 0
    \\

    & ~\textsf{ll-left/right} & 20
    & 0 & 0 & 0
    & {\bf 20} & {\bf 20} & {\bf 20}
    & 0.0 & 0.0 & 0.0
    & 19.5 & 10.6 &	1.3
    & -- & -- & --
    & 0.97 & 0.53 & 0.06

    & 0 & 0.0 & -- & 0.0
    & 0 & 0 & 0
    \\

    \multirow{-4}{*}{\rotatebox[origin=c]{90}{\textsf{slrd\_indt}}}
    & ~\textsf{misc.} & 32
    & 0 & 0 & 0
    & 31 & {\bf 32} & {\bf 32}
    & 0.0 & 0.0 & 0.0
    & 254.2 & 55.0 & 19.5
    & -- & -- & --
    & 8.20 & 1.72 & 0.61

    & 2[2] & 1.3 & 0.7 & 6.7
    & 2\,(3) & 0 & 0
    \\

    \hline
    \multicolumn{2}{|L}{~Total} & 418
    & 38 & 202 & 207
    & 336 & 403 & {\bf 414}
    & 149.2 & 21.9 & 31.6
    & 571.5 & 317.2	& 221.6
    & 3.93 & 0.11 & 0.15
    & 1.70 & 0.79 & 0.54

    & 34[22] & \,65.2 & \,1.9 & \,29.4
    & 14\,(75) & 0 & 8\,(13)
    \\
    \hline
  \end{tabular}
  \vspace{-1.4em}
\end{table}

\vspace{-0.1em}
We present the first experiment in Table \ref{fig:AllEntl}. The benchmark
\textsf{sll\_entl} relates to only singly linked list and is categorized by
its original sub-benchmarks. The two benchmarks \textsf{slrd\_entl} and
\textsf{slrd\_indt} are classified based on the related data structures,
which are variants of linked lists and trees. We report in each category
the number of \emph{entailments successfully proved} by each prover, where
the best results are in {\bf bold} (the total number of entailments is
shown in the column \textbf{\#En}). We also report the \emph{total } and
the \emph{average time} in seconds of each prover for all \emph{proved
  entailments}. To be fair, time spent on \emph{unproved entailments} are
\emph{not considered} since a prover might spend up to the timeout of
180(s) for each such entailment. We also provide the statistics of how
lemmas are synthesized and applied by {\bf $\songbirdLS$}. The numbers of
lemmas synthesized and used are presented in the column {\bf \#Lm}, where
$x [y]$ means that there are totally $x$ lemmas synthesized, and only $y$
of them are successfully used. The \emph{total} and the \emph{average time}
spent on synthesizing all lemmas are displayed in the columns {\bf T} and
{\bf A}. The synthesis \emph{overhead} is shown in the column {\bf O},
which is the percentage (\%) of the total synthesizing time over the total
proving time. We also classify the synthesized lemmas into three groups of
{\em conversion}, {\em split}, and {\em combination} lemmas ({\bf \#Cv},
{\bf \#Sp}, and {\bf \#Cb}), in a spirit similar to the motivating lemmas
$L_1, L_2$ and $L_3$ (Section \ref{sec:MotivatingExample}). The number $x
(y)$ in each group indicates that there are $x$ lemmas applied, and they
are repeatedly applied $y$ times.

\vspace{-0.1em} Table \ref{fig:AllEntl} shows that $\songbirdLS$ can
prove {\em most} of the entailments in existing benchmarks (414/418
$\approx$ 99\%) on an average of 0.54 seconds per entailment, which is 1.46
times faster than the second best prover $\songbirdMI$. Moreover,
\songbirdLS outperforms the third best prover $\cyclist$ in both the number
of proved entailments (more than 78 entailments) and the average proving
time (3.15 times faster). Other provers like $\sleek$, $\spen$ and $\slide$
can prove no more than 50\% of the entailments that $\songbirdLS$ can
prove. In addition, {$\songbirdLS$} achieves the best results in {\em all}
categories, which demonstrates the effectiveness of our lemma synthesis
technique. Regarding the 4 entailments that $\songbirdLS$ cannot prove,
they are related to the doubly linked list or the tree data structures
(\textsf{doubly-ll}/\textsf{tree}), where the needed lemmas are too
complicated to be synthesized within a timeout of 180 seconds.

\vspace{-0.1em}
In the first experiment, the lemma synthesis is only triggered in some
categories with an overhead of 29.4\%. Although this overhead is
considerable, it does not overwhelm the overall proving process since much
of the proving time is saved via the lemma application. The lemma
synthesis's {\em efficacy\/}, determined by the ratio of the number of
lemmas applied to the number of lemmas synthesized, is about 65\% (22
lemmas used in the total of 34 lemmas synthesized). More interestingly,
these 22 lemmas were applied totally 88 times. This fact implies that a
lemma can be (re)used multiple times. In this experiment, the conversion
lemmas is applied more often than other lemmas (75/88 times).

%% ~~~~~~~~~~~~~~~~~~~~~~~~~~~~~~~~~~~~~~~~~~~~~~~~~~
\begin{table}[t]
  \caption{Evaluation on the benchmark \textsf{slrd\_lm}, where
    $\bullet$ marks categories with arithmetic constraints}
  \vspace{-0.65em}
  \label{fig:ExpLemmaBench}
  \setlength\tabcolsep{1.pt}
  \def\arraystretch{1.1}%  1 is the default, change whatever you need
  \scriptsize
  \begin{tabular}{|Lc|gwg|wgw|gwg|wgwg|wgw|}
    \hline
    \multicolumn{2}{|c|}{\bf Benchmark \textsf{slrd\_lm}}
    & \multicolumn{3}{c|}{\bf Proved Entails}
    & \multicolumn{3}{c|}{\bf Proving Time (s)}
    & \multicolumn{3}{c|}{\bf Average Time (s)}
    & \multicolumn{4}{c|}{\bf Lemma Synthesis}
    & \multicolumn{3}{c|}{\bf Lemma Application}
    \\
    \hline

    ~Category & \#En\,
    & \cyclistAbr & \songbirdMiAbr & \songbirdLsAbr
    & \cyclistAbr & \songbirdMiAbr & \songbirdLsAbr
    & \cyclistAbr & \songbirdMiAbr & \songbirdLsAbr
    & \#Lm & T\,(s) & A\,(s) & O\,(\%)
    & \#Cv & \#Sp & \#Cb
    \\
    \hline

    ~\textsf{ll/rev} & 20
    & 17 & 18 & {\bf 20}
    & 446.5 & 87.1 & 205.0
    & 26.27 & 4.84 & 10.25
    & 28\,[20] & 8.1 & 0.3 & 3.9
    & 18\,(25) & 0 & 2\,(2)
    \\

    ~\textsf{ll-even/odd} & 17
    & {\bf 17} & {\bf 17} & {\bf 17}
    & 3.5 & 0.6 & 0.4
    & 0.21 & 0.03 & 0.02
    & 0 & 0.0 & -- & 0.0
    & 0 & 0 & 0
    \\

    ~\textsf{ll/rev+arith}$^\bullet$ & 72
    & 0 & 10 & {\bf 72}
    & 0.0 & 0.4 & 203.2
    & -- & 0.04 & 2.82
    & 81\,[81] & 50.9 & 0.6 & 25.1
    & 0 & 81\,(143) & 0
    \\

    ~\textsf{ll-sorted}$^\bullet$ & 25
    & 0 & 3 & {\bf 19}
    & 0.0 & 0.1 & 8.7
    & -- & 0.03 & 0.46
    & 14\,[14] & 4.9 & 0.3 & 55.8
    & 0 & 0 & 14\,(30)
    \\

    ~\textsf{dll/rev/null} & 22
    & {\bf 22} & {\bf 22} & {\bf 22}
    & 8.4 & 17.6 & 149.8
    & 0.38 & 0.80 & 6.81
    & 12\,[7] & 59.1 & 4.9 & 39.5
    & 7\,(8) & 0 & 0
    \\

    ~\textsf{dll/rev/null+arith}$^\bullet$ & 94
    & 0 & 17 & {\bf 94}
    & 0.0 & 27.6 & 2480.6
    & -- & 1.63 & 26.39
    & 236\,[158] & 1902.7 & 8.1 & 76.7
    & 74\,(74) & 69\,(103) & 15\,(15)
    \\

    ~\textsf{ll/dll-mixed} & 5
    & {\bf 5} & {\bf 5} & {\bf 5}
    & 0.4 & 0.1 & 0.2
    & 0.07 & 0.03 & 0.03
    & 0 & 0.0 & -- & 0.0
    & 0 & 0 & 0
    \\

    ~\textsf{ll/dll-mixed+arith}$^\bullet$ & 15
    & 0 & 5 & {\bf 15}
    & 0.0 & 0.4 & 328.3
    & -- & 0.08 & 21.89
    & 12\,[10] & 227.2 & 18.9 & 69.2
    & 8\,(8) & 2\,(2) & 0
    \\

    ~\textsf{tree/tseg} & 18
    & 7 & 7 & {\bf 18}
    & 1.3 & 0.3 & 135.0
    & 0.18 & 0.04 & 7.50
    & 18\,[12] & 3.8 & 0.2 & 2.8
    & 8\,(11) & 0 & 4\,(5)
    \\

    ~\textsf{tree/tseg+arith}$^\bullet$ & 12
    & 0 & 3 & {\bf 12}
    & 0.0 & 3.5 & 513.7
    & -- & 1.15 & 42.81
    & 20\,[16] & 211.2 & 10.6 & 41.1
    & 8\,(8) & 4\,(5) & 4\,(4)
    \\

    \hline
    ~Total & 300
    & 68 & 107 & {\bf 294}
    & 460.0 & 137.7 & 4024.8
    & 6.76 & 1.29 & 13.69
    & 421\,[318] & 2467.7 & 5.9 & 61.3
    & 123\,(134) & 156\,(253) & 39\,(56)
    \\
    \hline
  \end{tabular}
  \vspace{-1.5em}
\end{table}

In the second experiment, we apply the best three provers in the first
experiment, i.e., $\cyclist$, $\songbirdMI$, and $\songbirdLS$, on the more
challenging benchmark \textsf{slrd\_lm}. This benchmark was constructed by
enhancing the existing inductive heap predicates with richer {\em numeric}
properties of the data structures' size and content. These new predicates
enable more challenging entailments to be designed, such as the motivating
entailment $E_1$ from Section~\ref{sec:MotivatingExample}.

\vspace{-0.1em}
\begin{center}
\begin{small}
  $E_1 \triangleq \formp{\DllRev}{x,y,u,v,n} * \formp{\Dll}{v,u,z,t,200}
  \wedge n{\ge}100
  \entails
  \exists r. (\formp{\Dll}{x,y,r,z,n{+}199} * \forms{z}{r,t})$
\end{small}
\end{center}
\vspace{-0.1em}

Details of the second experiment is presented in Table
\ref{fig:ExpLemmaBench}. Entailments in the benchmark \textsf{slrd\_lm} are
categorized based on the participating heap predicates and their numeric
properties. To minimize the table's size, we group various heap predicates
related to the same data structure in one category. For example, the
category \textsf{dll/rev/null} contains entailments about doubly linked
lists, including normal lists (\textsf{dll}), reversed lists
(\textsf{dllrev}), or null-terminated lists (\textsf{dllnull}). Entailments
in the category \textsf{ll/dll-mixed} involve both the singly and the
doubly linked lists.

Table \ref{fig:ExpLemmaBench} shows that {\bf $\songbirdLS$} outperforms
other provers at {\em all} categories. In total, {\bf $\songbirdLS$} can
prove 98\% of the benchmark problems (294/300 entailments), which is 2.7
times better than the second best prover $\songbirdMI$ (107/300
entailments). More impressively, in the 5 categories with arithmetic
constraints, marked by $\bullet$, $\songbirdLS$ can prove more than 5.5
times the number of entailments that $\songbirdMI$ can (212 vs. 38
entailments). On the other hand, $\cyclist$ performs poorly in this
experiment because it does not support numeric properties yet. However,
among 82 non-arithmetic entailments, $\cyclist$ can prove 68 entailments
(about 80\%), whereas {\bf $\songbirdLS$} can prove all of them.

In the second experiment, the lemma synthesis overhead is higher (61.3\%)
since there are many lemmas synthesized (421). However, the overall
efficacy is also improved (75.5\%) when 318 synthesized lemmas are actually
used to prove goal entailments. It is worth noticing that 100\% lemmas
synthesized in the two arithmetic categories \textsf{ll/rev+arith} and
\textsf{ll-sorted} are utilized; this shows the usefulness of our proposed
framework in proving sophisticated entailments. In this experiment, the
\emph{split} lemmas are synthesized (49.1\% of the total synthesized
lemmas) and prominently used (57\% of the total number of lemma
application). This interesting fact shows that the \textsf{slrd\_lm}
benchmark, though handcrafted, was well designed to complement the existing
benchmarks.

%%% Local Variables:
%%% mode: latex
%%% TeX-master: "main"
%%% End:

\vspace{-0.8em}
\section{Related work}
\label{sec:RelatedWorks}
\vspace{-0.3em}

There have been various approaches proposed to prove separation logic
entailments. A popular direction is to restrict the inductive heap
predicates to certain classes such as: predicates whose syntax and
semantics are defined beforehand \cite{BerdineCO04, BerdineCO05,
  PiskacWZ13, PiskacWZ14, BozgaIP10, PerezR11, PerezR13}, predicates
describing \emph{variants of linked lists} \cite{EneaLSV14}, or predicates
satisfying a particular {\em bounded tree width property} \cite{IosifRS13,
  IosifRV14}. These restrictions enable the invention of practical and
effective entailment proving techniques. However, these predicate classes
\emph{cannot model sophisticated constraints} of data structures, which
involve not only the shape but also the size or the content, like the
predicates $\Dll$ and $\DllRev$ in Section \ref{sec:MotivatingExample}. In
addition, the existing techniques are tied to fixed sets of inductive heap
predicates and \emph{could not be automatically extended} to handle new
predicates. This extension requires extra efforts.

Another research direction is to focus on a broader class
of \emph{user-defined inductive heap predicates}. In particular,
\citet{ChinDNQ12} proposed a proof system based on the
\emph{unfold-and-match} technique: heap predicates in a goal entailment can
be \emph{unfolded} by their definitions to produce possibly identical
predicates in the antecedent and consequent, which can be \emph{matched}
and {\em removed} to derive simpler sub-goal entailments. However, an
inductive heap predicate can be unfolded infinitely often, which leads to
the infinite derivation of an entailment proof. To deal with such
situation, \citet{BrotherstonDP11} and \citet{ChuJT15} proposed inspiring
techniques respectively based on cyclic and induction proofs where
\emph{the infinite unfolding sequences can be avoided by induction
  hypothesis applications}. In their works, the induction hypotheses are
discovered directly from the candidate entailments; they might not be
sufficiently general to prove sophisticated entailments. Therefore, these
entailments' proofs often require the supporting lemmas, such as $L_1, L_2,
L_3$ (Section \ref{sec:MotivatingExample}). These lemmas are also needed by
other non-induction based verification systems, such as \cite{ChinDNQ12,
  QiuGSM13}. At present, these systems require users to manually provide
the lemmas.

To the best of our knowledge, there have been two approaches aiming to
automatically discover the supporting lemmas. The first approach is the
\emph{mutual induction} proof presented in \cite{TaLKC16}. This work
speculates lemmas from \emph{all entailments} which are already derived in
an on-going induction proof to assist in proving future entailments
introduced within the same proof. This speculation provides more
lemma/induction hypothesis candidates than the cyclic-based
\cite{BrotherstonDP11} and induction-based \cite{ChuJT15} techniques.
Consequently, it can increase the chance of successfully proving the
entailments. However, the mutual induction proof \emph{cannot handle}
sophisticated entailments, such as $E_1$ in Section
\ref{sec:MotivatingExample}. All entailments derived from $E_1$ may contain
specific constraints and cannot be applied to prove other derived
entailments.

The second approach is the \emph{lemma generation} presented in
\cite{EneaSW15}. This work considers an interesting class of inductive heap
predicates satisfying the notions of \emph{compositionality} and
\emph{completion}. Under these properties, a class of lemmas can be
enumerated beforehand, either to \emph{convert} inductive predicates
\emph{of the same arity}, or to \emph{combine} two inductive predicates.
However, this technique \emph{cannot generate} lemmas which convert
predicates of different arities, or combine a singleton heap predicate with
an inductive heap predicate, or split an inductive heap predicate into
other predicates. In addition, an inductive heap predicate satisfying the
\emph{compositionality} property has exactly one base-case definition,
whose heap part is also empty. Moreover, each inductive-case definition
must contain a singleton heap predicate whose \emph{root} address is one of
the inductive heap predicate's arguments, like the \emph{compositional}
predicate $\formp{\Ls}{x,y} ~{\defeq}~ (x{=}y) \,{\vee}\, \exists u.
(\forms{x}{u} \,{*}\, \formp{\Ls}{u,y})$. This technique, therefore, cannot
generate lemmas for predicates with non-empty base cases, e.g., $\Dll$ and
$\DllRev$ in Section \ref{sec:MotivatingExample}, or lemmas for the
predicates defined in a \emph{reverse-fashion}, like the {\em
  non-compositional} predicate $\formp{\LsRev}{x,y} ~{\defeq}~ (x{=}y)
\,{\vee}\, \exists u. (\formp{\LsRev}{x,u} \,{*}\, \forms{u}{y})$. These
\emph{reverse-fashion} predicates are prevalent in SL-COMP 2014's
benchmarks, such as $\mathsf{RList}, \mathsf{ListO}, \mathsf{DLL\_plus},
\mathsf{DLL\_plus\_rev}, \mathsf{DLL\_plus\_mid}$.

In the \emph{synthesis} context, there is an appealing approach called
SyGuS, a.k.a., Syntax-Guided Synthesis \cite{AlurBDF0JKMMRSSSSTU15}. This
approach aims to \emph{infer computer programs} satisfying certain
restrictions on its syntax and semantics (respectively constrained by a
context free grammar and a SMT formula). Techniques following SyGuS often
operate on a \emph{learning phase} which proposes a candidate program, and
a \emph{verification phase} which checks the proposal against the semantic
restriction. To some extent, our lemma synthesis approach is similar to
SyGuS since we also discover potential lemmas and verify their validity.
However, we \emph{focus on synthesizing separation logic lemmas} but not
the computer programs. Our syntactic restriction is \emph{more
  goal-directed} since the lemmas is controlled by specific templates,
unlike the context-free-grammar restriction of SyGuS. Moreover, our
semantic restriction \emph{cannot be represented} by a SMT formula since we
require that if a lemma can be proved valid, its proof must contain an
induction hypothesis application. Therefore, we believe that the induction
proof is \emph{necessary} in both the lemma discovery and verification
phases. This proof technique is currently \emph{not supported} by any
SyGuS-based approaches.

\vspace{-0.8em}
\section{Conclusion}
\label{sec:Conclusion}
\vspace{-0.3em}

We have proposed a novel framework for synthesizing lemmas to assist in
proving separation logic entailments in the fragment of symbolic-heap
separation logic with inductive heap predicates and linear arithmetic. Our
framework is able to synthesize various kinds of \emph{inductive lemmas},
which help to modularize the proofs of sophisticated entailments. The
synthesis of inductive lemmas is non-trivial since induction proof is
required by both the lemma discovery and validation phases. In exchange,
these lemmas can significantly improve the completeness of induction proof
in separation logic. We have shown by experiment that our
lemma-synthesis-assisted prover $\songbirdLS$ is able to prove many
entailments that could not be proved by the state-of-the-art separation
logic provers.

We shall now discuss two limitations of our approach. Firstly, our current
implementation cannot simultaneously derive new constraints from both the
antecedent and the consequent of a lemma template. Theoretically, our
framework can handle a lemma template with different unknown relations on
both these two sides. However, the set of unknown assumptions which is
introduced corresponding to these relations is far too complicated to be
discharged by the current underlying prover. Secondly, we only support to
infer linear arithmetic constraints with Farkas' lemma. In future, we would
like to extend the lemma synthesis framework with suitable constraint
solving techniques to support more kinds of pure constraints, such as sets
or multisets of values.

\vspace{-0.8em}
\section*{Acknowledgments}
\vspace{-0.35em}
We would like to thank the reviewers of POPL'18 PC and AEC for the
constructive comments on the paper and the artifact. We wish to thank Dr.
Aleksandar Nanevski for his valuable suggestions on preparing the final
version of this paper, and Dr. Andrew C. Myers for his dedication as the
program chair of POPL'18. We are grateful for the encouraging feedback
from the reviewers of OOPSLA'17 on our previous submission. The first
author wish to thank Ms. Mirela Andreea Costea and Dr. Makoto Tatsuta for
the inspiring discussions about the entailment proof. This research is
partially supported by an NUS research grant R-252-000-553-112 and an MoE
Tier-2 grant MOE2013-T2-2-146.
\vspace*{0.8em}

\bibliography{main}

%%% -*-BibTeX-*-
%%% Do NOT edit. File created by BibTeX with style
%%% ACM-Reference-Format-Journals [18-Jan-2012].

\begin{thebibliography}{38}

%%% ====================================================================
%%% NOTE TO THE USER: you can override these defaults by providing
%%% customized versions of any of these macros before the \bibliography
%%% command.  Each of them MUST provide its own final punctuation,
%%% except for \shownote{}, \showDOI{}, and \showURL{}.  The latter two
%%% do not use final punctuation, in order to avoid confusing it with
%%% the Web address.
%%%
%%% To suppress output of a particular field, define its macro to expand
%%% to an empty string, or better, \unskip, like this:
%%%
%%% \newcommand{\showDOI}[1]{\unskip}   % LaTeX syntax
%%%
%%% \def \showDOI #1{\unskip}           % plain TeX syntax
%%%
%%% ====================================================================

\ifx \showCODEN    \undefined \def \showCODEN     #1{\unskip}     \fi
\ifx \showDOI      \undefined \def \showDOI       #1{#1}\fi
\ifx \showISBNx    \undefined \def \showISBNx     #1{\unskip}     \fi
\ifx \showISBNxiii \undefined \def \showISBNxiii  #1{\unskip}     \fi
\ifx \showISSN     \undefined \def \showISSN      #1{\unskip}     \fi
\ifx \showLCCN     \undefined \def \showLCCN      #1{\unskip}     \fi
\ifx \shownote     \undefined \def \shownote      #1{#1}          \fi
\ifx \showarticletitle \undefined \def \showarticletitle #1{#1}   \fi
\ifx \showURL      \undefined \def \showURL       {\relax}        \fi
% The following commands are used for tagged output and should be
% invisible to TeX
\providecommand\bibfield[2]{#2}
\providecommand\bibinfo[2]{#2}
\providecommand\natexlab[1]{#1}
\providecommand\showeprint[2][]{arXiv:#2}

\bibitem[\protect\citeauthoryear{Albarghouthi, Berdine, Cook, and
  Kincaid}{Albarghouthi et~al\mbox{.}}{2015}]%
        {AlbarghouthiBCK15}
\bibfield{author}{\bibinfo{person}{Aws Albarghouthi}, \bibinfo{person}{Josh
  Berdine}, \bibinfo{person}{Byron Cook}, {and} \bibinfo{person}{Zachary
  Kincaid}.} \bibinfo{year}{2015}\natexlab{}.
\newblock \showarticletitle{Spatial Interpolants}. In \bibinfo{booktitle}{{\em
  European Symposium on Programming (ESOP)}}. \bibinfo{pages}{634--660}.
\newblock


\bibitem[\protect\citeauthoryear{Alur, Bod{\'{\i}}k, Dallal, Fisman, Garg,
  Juniwal, Kress{-}Gazit, Madhusudan, Martin, Raghothaman, Saha, Seshia, Singh,
  Solar{-}Lezama, Torlak, and Udupa}{Alur et~al\mbox{.}}{2015}]%
        {AlurBDF0JKMMRSSSSTU15}
\bibfield{author}{\bibinfo{person}{Rajeev Alur}, \bibinfo{person}{Rastislav
  Bod{\'{\i}}k}, \bibinfo{person}{Eric Dallal}, \bibinfo{person}{Dana Fisman},
  \bibinfo{person}{Pranav Garg}, \bibinfo{person}{Garvit Juniwal},
  \bibinfo{person}{Hadas Kress{-}Gazit}, \bibinfo{person}{P. Madhusudan},
  \bibinfo{person}{Milo M.~K. Martin}, \bibinfo{person}{Mukund Raghothaman},
  \bibinfo{person}{Shambwaditya Saha}, \bibinfo{person}{Sanjit~A. Seshia},
  \bibinfo{person}{Rishabh Singh}, \bibinfo{person}{Armando Solar{-}Lezama},
  \bibinfo{person}{Emina Torlak}, {and} \bibinfo{person}{Abhishek Udupa}.}
  \bibinfo{year}{2015}\natexlab{}.
\newblock \showarticletitle{Syntax-Guided Synthesis}.
\newblock In \bibinfo{booktitle}{{\em Dependable Software Systems
  Engineering}}. \bibinfo{pages}{1--25}.
\newblock


\bibitem[\protect\citeauthoryear{Berdine, Calcagno, and O'Hearn}{Berdine
  et~al\mbox{.}}{2004}]%
        {BerdineCO04}
\bibfield{author}{\bibinfo{person}{Josh Berdine}, \bibinfo{person}{Cristiano
  Calcagno}, {and} \bibinfo{person}{Peter~W. O'Hearn}.}
  \bibinfo{year}{2004}\natexlab{}.
\newblock \showarticletitle{{A Decidable Fragment of Separation Logic}}. In
  \bibinfo{booktitle}{{\em International Conference on Foundations of Software
  Technology and Theoretical Computer Science (FSTTCS)}}.
  \bibinfo{pages}{97--109}.
\newblock


\bibitem[\protect\citeauthoryear{Berdine, Calcagno, and O'Hearn}{Berdine
  et~al\mbox{.}}{2005a}]%
        {Smallfoot05}
\bibfield{author}{\bibinfo{person}{Josh Berdine}, \bibinfo{person}{Cristiano
  Calcagno}, {and} \bibinfo{person}{Peter~W. O'Hearn}.}
  \bibinfo{year}{2005}\natexlab{a}.
\newblock \showarticletitle{{Smallfoot: Modular Automatic Assertion Checking
  with Separation Logic}}. In \bibinfo{booktitle}{{\em International Symposium
  on Formal Methods for Components and Objects}}. \bibinfo{pages}{115--137}.
\newblock


\bibitem[\protect\citeauthoryear{Berdine, Calcagno, and O'Hearn}{Berdine
  et~al\mbox{.}}{2005b}]%
        {BerdineCO05}
\bibfield{author}{\bibinfo{person}{Josh Berdine}, \bibinfo{person}{Cristiano
  Calcagno}, {and} \bibinfo{person}{Peter~W. O'Hearn}.}
  \bibinfo{year}{2005}\natexlab{b}.
\newblock \showarticletitle{{Symbolic Execution with Separation Logic}}. In
  \bibinfo{booktitle}{{\em Asian Symposium on Programming Languages and Systems
  (APLAS)}}. \bibinfo{pages}{52--68}.
\newblock


\bibitem[\protect\citeauthoryear{Berdine, Cook, and Ishtiaq}{Berdine
  et~al\mbox{.}}{2011}]%
        {BerdineCI11}
\bibfield{author}{\bibinfo{person}{Josh Berdine}, \bibinfo{person}{Byron Cook},
  {and} \bibinfo{person}{Samin Ishtiaq}.} \bibinfo{year}{2011}\natexlab{}.
\newblock \showarticletitle{SLAyer: Memory Safety for Systems-Level Code}. In
  \bibinfo{booktitle}{{\em International Conference on Computer Aided
  Verification (CAV)}}. \bibinfo{pages}{178--183}.
\newblock


\bibitem[\protect\citeauthoryear{Bozga, Iosif, and Perarnau}{Bozga
  et~al\mbox{.}}{2010}]%
        {BozgaIP10}
\bibfield{author}{\bibinfo{person}{Marius Bozga}, \bibinfo{person}{Radu Iosif},
  {and} \bibinfo{person}{Swann Perarnau}.} \bibinfo{year}{2010}\natexlab{}.
\newblock \showarticletitle{{Quantitative Separation Logic and Programs with
  Lists}}.
\newblock \bibinfo{journal}{{\em J. Autom. Reasoning\/}} \bibinfo{volume}{45},
  \bibinfo{number}{2} (\bibinfo{year}{2010}), \bibinfo{pages}{131--156}.
\newblock


\bibitem[\protect\citeauthoryear{Brotherston, Distefano, and
  Petersen}{Brotherston et~al\mbox{.}}{2011}]%
        {BrotherstonDP11}
\bibfield{author}{\bibinfo{person}{James Brotherston}, \bibinfo{person}{Dino
  Distefano}, {and} \bibinfo{person}{Rasmus~Lerchedahl Petersen}.}
  \bibinfo{year}{2011}\natexlab{}.
\newblock \showarticletitle{{Automated Cyclic Entailment Proofs in Separation
  Logic}}. In \bibinfo{booktitle}{{\em International Conference on Automated
  Deduction (CADE)}}. \bibinfo{pages}{131--146}.
\newblock


\bibitem[\protect\citeauthoryear{Brotherston, Fuhs, P{\'{e}}rez, and
  Gorogiannis}{Brotherston et~al\mbox{.}}{2014}]%
        {BrotherstonFPG14}
\bibfield{author}{\bibinfo{person}{James Brotherston}, \bibinfo{person}{Carsten
  Fuhs}, \bibinfo{person}{Juan A.~Navarro P{\'{e}}rez}, {and}
  \bibinfo{person}{Nikos Gorogiannis}.} \bibinfo{year}{2014}\natexlab{}.
\newblock \showarticletitle{{A decision procedure for satisfiability in
  Separation Logic with inductive predicates}}. In \bibinfo{booktitle}{{\em
  Joint Meeting of International Conference on Computer Science Logic and
  Symposium on Logic in Computer Science, {CSL-LICS}}}.
  \bibinfo{pages}{25:1--25:10}.
\newblock


\bibitem[\protect\citeauthoryear{Brotherston, Gorogiannis, Kanovich, and
  Rowe}{Brotherston et~al\mbox{.}}{2016}]%
        {BrotherstonGKR16}
\bibfield{author}{\bibinfo{person}{James Brotherston}, \bibinfo{person}{Nikos
  Gorogiannis}, \bibinfo{person}{Max~I. Kanovich}, {and}
  \bibinfo{person}{Reuben Rowe}.} \bibinfo{year}{2016}\natexlab{}.
\newblock \showarticletitle{{Model checking for Symbolic-Heap Separation Logic
  with inductive predicates}}. In \bibinfo{booktitle}{{\em Symposium on
  Principles of Programming Languages (POPL)}}. \bibinfo{pages}{84--96}.
\newblock


\bibitem[\protect\citeauthoryear{Brotherston and Simpson}{Brotherston and
  Simpson}{2011}]%
        {BrotherstonS11}
\bibfield{author}{\bibinfo{person}{James Brotherston} {and}
  \bibinfo{person}{Alex Simpson}.} \bibinfo{year}{2011}\natexlab{}.
\newblock \showarticletitle{Sequent calculi for induction and infinite
  descent}.
\newblock \bibinfo{journal}{{\em J. Log. Comput.\/}} \bibinfo{volume}{21},
  \bibinfo{number}{6} (\bibinfo{year}{2011}), \bibinfo{pages}{1177--1216}.
\newblock


\bibitem[\protect\citeauthoryear{Bundy}{Bundy}{2001}]%
        {Bundy01}
\bibfield{author}{\bibinfo{person}{Alan Bundy}.}
  \bibinfo{year}{2001}\natexlab{}.
\newblock \showarticletitle{The Automation of Proof by Mathematical Induction}.
  In \bibinfo{booktitle}{{\em Handbook of Automated Reasoning (in 2 volumes)}}.
  \bibinfo{pages}{845--911}.
\newblock


\bibitem[\protect\citeauthoryear{Calcagno, Distefano, Dubreil, Gabi,
  Hooimeijer, Luca, O'Hearn, Papakonstantinou, Purbrick, and
  Rodriguez}{Calcagno et~al\mbox{.}}{2015}]%
        {CalcagnoDDGHLOP15}
\bibfield{author}{\bibinfo{person}{Cristiano Calcagno}, \bibinfo{person}{Dino
  Distefano}, \bibinfo{person}{J{\'{e}}r{\'{e}}my Dubreil},
  \bibinfo{person}{Dominik Gabi}, \bibinfo{person}{Pieter Hooimeijer},
  \bibinfo{person}{Martino Luca}, \bibinfo{person}{Peter~W. O'Hearn},
  \bibinfo{person}{Irene Papakonstantinou}, \bibinfo{person}{Jim Purbrick},
  {and} \bibinfo{person}{Dulma Rodriguez}.} \bibinfo{year}{2015}\natexlab{}.
\newblock \showarticletitle{Moving Fast with Software Verification}. In
  \bibinfo{booktitle}{{\em NASA International Symposium on Formal Methods
  (NFM)}}. \bibinfo{pages}{3--11}.
\newblock


\bibitem[\protect\citeauthoryear{Chin, David, Nguyen, and Qin}{Chin
  et~al\mbox{.}}{2012}]%
        {ChinDNQ12}
\bibfield{author}{\bibinfo{person}{Wei-Ngan Chin}, \bibinfo{person}{Cristina
  David}, \bibinfo{person}{Huu~Hai Nguyen}, {and} \bibinfo{person}{Shengchao
  Qin}.} \bibinfo{year}{2012}\natexlab{}.
\newblock \showarticletitle{Automated verification of shape, size and bag
  properties via user-defined predicates in {Separation Logic}}.
\newblock \bibinfo{journal}{{\em Science of Computer Programming (SCP)\/}}
  \bibinfo{volume}{77}, \bibinfo{number}{9} (\bibinfo{year}{2012}),
  \bibinfo{pages}{1006--1036}.
\newblock


\bibitem[\protect\citeauthoryear{Chu, Jaffar, and Trinh}{Chu
  et~al\mbox{.}}{2015}]%
        {ChuJT15}
\bibfield{author}{\bibinfo{person}{Duc-Hiep Chu}, \bibinfo{person}{Joxan
  Jaffar}, {and} \bibinfo{person}{Minh-Thai Trinh}.}
  \bibinfo{year}{2015}\natexlab{}.
\newblock \showarticletitle{{Automatic induction proofs of data-structures in
  imperative programs}}. In \bibinfo{booktitle}{{\em Conference on Programming
  Language Design and Implementation (PLDI)}}. \bibinfo{pages}{457--466}.
\newblock


\bibitem[\protect\citeauthoryear{Col{\'{o}}n, Sankaranarayanan, and
  Sipma}{Col{\'{o}}n et~al\mbox{.}}{2003}]%
        {ColonSS03}
\bibfield{author}{\bibinfo{person}{Michael Col{\'{o}}n},
  \bibinfo{person}{Sriram Sankaranarayanan}, {and} \bibinfo{person}{Henny
  Sipma}.} \bibinfo{year}{2003}\natexlab{}.
\newblock \showarticletitle{Linear Invariant Generation Using Non-linear
  Constraint Solving}. In \bibinfo{booktitle}{{\em International Conference on
  Computer Aided Verification (CAV)}}. \bibinfo{pages}{420--432}.
\newblock


\bibitem[\protect\citeauthoryear{Cook, Haase, Ouaknine, Parkinson, and
  Worrell}{Cook et~al\mbox{.}}{2011}]%
        {CookHOPW11}
\bibfield{author}{\bibinfo{person}{Byron Cook}, \bibinfo{person}{Christoph
  Haase}, \bibinfo{person}{Jo{\"{e}}l Ouaknine}, \bibinfo{person}{Matthew~J.
  Parkinson}, {and} \bibinfo{person}{James Worrell}.}
  \bibinfo{year}{2011}\natexlab{}.
\newblock \showarticletitle{{Tractable Reasoning in a Fragment of Separation
  Logic}}. In \bibinfo{booktitle}{{\em International Conference on Concurrency
  Theory (CONCUR)}}. \bibinfo{pages}{235--249}.
\newblock


\bibitem[\protect\citeauthoryear{Distefano and Parkinson}{Distefano and
  Parkinson}{2008}]%
        {DistefanoP08}
\bibfield{author}{\bibinfo{person}{Dino Distefano} {and}
  \bibinfo{person}{Matthew~J. Parkinson}.} \bibinfo{year}{2008}\natexlab{}.
\newblock \showarticletitle{jStar: towards practical verification for java}.
  \bibinfo{pages}{213--226}.
\newblock


\bibitem[\protect\citeauthoryear{Enea, Leng{\'{a}}l, Sighireanu, and
  Vojnar}{Enea et~al\mbox{.}}{2014}]%
        {EneaLSV14}
\bibfield{author}{\bibinfo{person}{Constantin Enea}, \bibinfo{person}{Ondrej
  Leng{\'{a}}l}, \bibinfo{person}{Mihaela Sighireanu}, {and}
  \bibinfo{person}{Tom{\'{a}}s Vojnar}.} \bibinfo{year}{2014}\natexlab{}.
\newblock \showarticletitle{{Compositional Entailment Checking for a Fragment
  of Separation Logic}}. In \bibinfo{booktitle}{{\em Asian Symposium on
  Programming Languages and Systems (APLAS)}}. \bibinfo{pages}{314--333}.
\newblock


\bibitem[\protect\citeauthoryear{Enea, Sighireanu, and Wu}{Enea
  et~al\mbox{.}}{2015}]%
        {EneaSW15}
\bibfield{author}{\bibinfo{person}{Constantin Enea}, \bibinfo{person}{Mihaela
  Sighireanu}, {and} \bibinfo{person}{Zhilin Wu}.}
  \bibinfo{year}{2015}\natexlab{}.
\newblock \showarticletitle{On Automated Lemma Generation for Separation Logic
  with Inductive Definitions}. In \bibinfo{booktitle}{{\em International
  Symposium on Automated Technology for Verification and Analysis (ATVA)}}.
  \bibinfo{pages}{80--96}.
\newblock


\bibitem[\protect\citeauthoryear{Iosif, Rogalewicz, and Sim{\'{a}}cek}{Iosif
  et~al\mbox{.}}{2013}]%
        {IosifRS13}
\bibfield{author}{\bibinfo{person}{Radu Iosif}, \bibinfo{person}{Adam
  Rogalewicz}, {and} \bibinfo{person}{Jiri Sim{\'{a}}cek}.}
  \bibinfo{year}{2013}\natexlab{}.
\newblock \showarticletitle{{The Tree Width of Separation Logic with Recursive
  Definitions}}. In \bibinfo{booktitle}{{\em International Conference on
  Automated Deduction (CADE)}}. \bibinfo{pages}{21--38}.
\newblock


\bibitem[\protect\citeauthoryear{Iosif, Rogalewicz, and Vojnar}{Iosif
  et~al\mbox{.}}{2014}]%
        {IosifRV14}
\bibfield{author}{\bibinfo{person}{Radu Iosif}, \bibinfo{person}{Adam
  Rogalewicz}, {and} \bibinfo{person}{Tom{\'{a}}s Vojnar}.}
  \bibinfo{year}{2014}\natexlab{}.
\newblock \showarticletitle{{Deciding Entailments in Inductive Separation Logic
  with Tree Automata}}. In \bibinfo{booktitle}{{\em International Symposium on
  Automated Technology for Verification and Analysis (ATVA)}}.
  \bibinfo{pages}{201--218}.
\newblock


\bibitem[\protect\citeauthoryear{Le, Sun, and Chin}{Le et~al\mbox{.}}{2016}]%
        {LeSC16}
\bibfield{author}{\bibinfo{person}{Quang~Loc Le}, \bibinfo{person}{Jun Sun},
  {and} \bibinfo{person}{Wei{-}Ngan Chin}.} \bibinfo{year}{2016}\natexlab{}.
\newblock \showarticletitle{Satisfiability Modulo Heap-Based Programs}. In
  \bibinfo{booktitle}{{\em International Conference on Computer Aided
  Verification (CAV)}}. \bibinfo{pages}{382--404}.
\newblock


\bibitem[\protect\citeauthoryear{Moura and Bj{\o}rner}{Moura and
  Bj{\o}rner}{2008}]%
        {MouraB08}
\bibfield{author}{\bibinfo{person}{Leonardo Mendon{\c{c}}a~De Moura} {and}
  \bibinfo{person}{Nikolaj Bj{\o}rner}.} \bibinfo{year}{2008}\natexlab{}.
\newblock \showarticletitle{{{Z3:} An Efficient {SMT} Solver}}. In
  \bibinfo{booktitle}{{\em International Conference on Tools and Algorithms for
  Construction and Analysis of Systems (TACAS)}}. \bibinfo{pages}{337--340}.
\newblock


\bibitem[\protect\citeauthoryear{Nguyen and Chin}{Nguyen and Chin}{2008}]%
        {NguyenC08}
\bibfield{author}{\bibinfo{person}{Huu~Hai Nguyen} {and}
  \bibinfo{person}{Wei-Ngan Chin}.} \bibinfo{year}{2008}\natexlab{}.
\newblock \showarticletitle{{Enhancing Program Verification with Lemmas}}. In
  \bibinfo{booktitle}{{\em International Conference on Computer Aided
  Verification (CAV)}}. \bibinfo{pages}{355--369}.
\newblock


\bibitem[\protect\citeauthoryear{O'Hearn, Reynolds, and Yang}{O'Hearn
  et~al\mbox{.}}{2001}]%
        {OHearnRY01}
\bibfield{author}{\bibinfo{person}{Peter~W. O'Hearn}, \bibinfo{person}{John~C.
  Reynolds}, {and} \bibinfo{person}{Hongseok Yang}.}
  \bibinfo{year}{2001}\natexlab{}.
\newblock \showarticletitle{{Local Reasoning about Programs that Alter Data
  Structures}}. In \bibinfo{booktitle}{{\em International Conference on
  Computer Science Logic (CSL)}}. \bibinfo{pages}{1--19}.
\newblock


\bibitem[\protect\citeauthoryear{P{\'{e}}rez and Rybalchenko}{P{\'{e}}rez and
  Rybalchenko}{2011}]%
        {PerezR11}
\bibfield{author}{\bibinfo{person}{Juan Antonio~Navarro P{\'{e}}rez} {and}
  \bibinfo{person}{Andrey Rybalchenko}.} \bibinfo{year}{2011}\natexlab{}.
\newblock \showarticletitle{{Separation Logic + Superposition Calculus = Heap
  Theorem Prover}}. In \bibinfo{booktitle}{{\em Conference on Programming
  Language Design and Implementation (PLDI)}}. \bibinfo{pages}{556--566}.
\newblock


\bibitem[\protect\citeauthoryear{P{\'{e}}rez and Rybalchenko}{P{\'{e}}rez and
  Rybalchenko}{2013}]%
        {PerezR13}
\bibfield{author}{\bibinfo{person}{Juan Antonio~Navarro P{\'{e}}rez} {and}
  \bibinfo{person}{Andrey Rybalchenko}.} \bibinfo{year}{2013}\natexlab{}.
\newblock \showarticletitle{{Separation Logic Modulo Theories}}. In
  \bibinfo{booktitle}{{\em Asian Symposium on Programming Languages and Systems
  (APLAS)}}. \bibinfo{pages}{90--106}.
\newblock


\bibitem[\protect\citeauthoryear{Piskac, Wies, and Zufferey}{Piskac
  et~al\mbox{.}}{2013}]%
        {PiskacWZ13}
\bibfield{author}{\bibinfo{person}{Ruzica Piskac}, \bibinfo{person}{Thomas
  Wies}, {and} \bibinfo{person}{Damien Zufferey}.}
  \bibinfo{year}{2013}\natexlab{}.
\newblock \showarticletitle{{Automating Separation Logic Using {SMT}}}. In
  \bibinfo{booktitle}{{\em International Conference on Computer Aided
  Verification (CAV)}}. \bibinfo{pages}{773--789}.
\newblock


\bibitem[\protect\citeauthoryear{Piskac, Wies, and Zufferey}{Piskac
  et~al\mbox{.}}{2014}]%
        {PiskacWZ14}
\bibfield{author}{\bibinfo{person}{Ruzica Piskac}, \bibinfo{person}{Thomas
  Wies}, {and} \bibinfo{person}{Damien Zufferey}.}
  \bibinfo{year}{2014}\natexlab{}.
\newblock \showarticletitle{{Automating Separation Logic with Trees and Data}}.
  In \bibinfo{booktitle}{{\em International Conference on Computer Aided
  Verification (CAV)}}. \bibinfo{pages}{711--728}.
\newblock


\bibitem[\protect\citeauthoryear{Pugh}{Pugh}{1991}]%
        {Pugh91}
\bibfield{author}{\bibinfo{person}{William Pugh}.}
  \bibinfo{year}{1991}\natexlab{}.
\newblock \showarticletitle{{The Omega Test: a fast and practical integer
  programming algorithm for dependence analysis}}. In \bibinfo{booktitle}{{\em
  Proceedings Supercomputing '91, Albuquerque, NM, USA, November 18-22, 1991}}.
  \bibinfo{pages}{4--13}.
\newblock


\bibitem[\protect\citeauthoryear{Qiu, Garg, Stefanescu, and Madhusudan}{Qiu
  et~al\mbox{.}}{2013}]%
        {QiuGSM13}
\bibfield{author}{\bibinfo{person}{Xiaokang Qiu}, \bibinfo{person}{Pranav
  Garg}, \bibinfo{person}{Andrei Stefanescu}, {and}
  \bibinfo{person}{Parthasarathy Madhusudan}.} \bibinfo{year}{2013}\natexlab{}.
\newblock \showarticletitle{{Natural proofs for structure, data, and
  separation}}. In \bibinfo{booktitle}{{\em Conference on Programming Language
  Design and Implementation (PLDI)}}. \bibinfo{pages}{231--242}.
\newblock


\bibitem[\protect\citeauthoryear{Reynolds}{Reynolds}{2002}]%
        {Reynolds02}
\bibfield{author}{\bibinfo{person}{John~C. Reynolds}.}
  \bibinfo{year}{2002}\natexlab{}.
\newblock \showarticletitle{{Separation Logic: {A} Logic for Shared Mutable
  Data Structures}}. In \bibinfo{booktitle}{{\em Symposium on Logic in Computer
  Science (LICS)}}. \bibinfo{pages}{55--74}.
\newblock


\bibitem[\protect\citeauthoryear{Reynolds}{Reynolds}{2008}]%
        {Reynolds08}
\bibfield{author}{\bibinfo{person}{John~C. Reynolds}.}
  \bibinfo{year}{2008}\natexlab{}.
\newblock \showarticletitle{{An Introduction to Separation Logic}}.
\newblock \bibinfo{publisher}{{Lecture Notes for the PhD Fall School on Logics
  and Semantics of State, Copenhagen 2008. Retrieved on 2017, March 16th}}.
\newblock
\showURL{%
\url{http://www.cs.cmu.edu/~jcr/copenhagen08.pdf}}


\bibitem[\protect\citeauthoryear{Schrijver}{Schrijver}{1986}]%
        {Schrijver1986}
\bibfield{author}{\bibinfo{person}{Alexander Schrijver}.}
  \bibinfo{year}{1986}\natexlab{}.
\newblock \bibinfo{booktitle}{{\em Theory of Linear and Integer Programming}}.
\newblock \bibinfo{publisher}{John Wiley \& Sons, Inc.}, \bibinfo{address}{New
  York, NY, USA}.
\newblock
\showISBNx{0-471-90854-1}


\bibitem[\protect\citeauthoryear{Sighireanu and Cok}{Sighireanu and
  Cok}{2016}]%
        {SighireanuC14}
\bibfield{author}{\bibinfo{person}{Mihaela Sighireanu} {and}
  \bibinfo{person}{David~R. Cok}.} \bibinfo{year}{2016}\natexlab{}.
\newblock \showarticletitle{{Report on SL-COMP 2014}}.
\newblock \bibinfo{journal}{{\em Journal on Satisfiability, Boolean Modeling
  and Computation\/}}  \bibinfo{volume}{9} (\bibinfo{year}{2016}),
  \bibinfo{pages}{173--186}.
\newblock


\bibitem[\protect\citeauthoryear{Ta, Le, Khoo, and Chin}{Ta
  et~al\mbox{.}}{2016}]%
        {TaLKC16}
\bibfield{author}{\bibinfo{person}{Quang{-}Trung Ta},
  \bibinfo{person}{Ton~Chanh Le}, \bibinfo{person}{Siau{-}Cheng Khoo}, {and}
  \bibinfo{person}{Wei{-}Ngan Chin}.} \bibinfo{year}{2016}\natexlab{}.
\newblock \showarticletitle{Automated Mutual Explicit Induction Proof in
  Separation Logic}. In \bibinfo{booktitle}{{\em International Symposium on
  Formal Methods (FM)}}. \bibinfo{pages}{659--676}.
\newblock


\bibitem[\protect\citeauthoryear{Whitehead and Russell}{Whitehead and
  Russell}{1912}]%
        {WhiteheadR12}
\bibfield{author}{\bibinfo{person}{Alfred~North Whitehead} {and}
  \bibinfo{person}{Bertrand Russell}.} \bibinfo{year}{1912}\natexlab{}.
\newblock \bibinfo{booktitle}{{\em Principia Mathematica}}.
\newblock \bibinfo{publisher}{University Press}.
\newblock


\end{thebibliography}

\iftechreport

\appendix

\clearpage
\section{Soundness proof}
\label{append:SoundnessProof}

\subsection{Soundness of inference rules }
\label{sec:SoundnessLogicalRules}

\setcounter{myequationNo}{0}

We prove soundness of inference rules in Section \ref{sec:ProofSystem} by
showing that if entailments in their premises are valid, and their side
conditions, if any, are satisfied, then goal entailments in their
conclusions are also valid.

%% ---------------
\textbf{1. Axiom rules $\rulePureEntail, \ruleFalseLeftOne,\ruleFalseLeftTwo$}

\begin{center}
  \small
  \begin{tabular}{ccc}
    \minipagePureEntail &
    \minipageFalseLeftOne &
    \minipageFalseLeftTwo
  \end{tabular}
\end{center}

\begin{itemize}[label=--, leftmargin=1.2em]

\item When the rule $\rulePureEntail$ is applied, SMT solvers or theorem
  provers, such as Z3 \cite{MouraB08} or Omega Calculator \cite{Pugh91}, can
  be invoked to check the side condition: $\Pi_1 \rightarrow \Pi_2$. If this
  side condition holds, then clearly the entailment $\Pi_1 \entails \Pi_2$ is
  valid.

\item For the two rules $\ruleFalseLeftOne,\ruleFalseLeftTwo$, it is easy
  to verify their goal entailments' antecedents are unsatisfiable since
  they either contain an overlapped singleton heap
  ($\formst{\varSort_1}{u}{\vec{v}} * \formst{\varSort_2}{u}{\vec{t}}$ in
  the rule $\ruleFalseLeftOne$), or a contradiction ($\Pi_1 \rightarrow
  \false$ in the rule $\ruleFalseLeftOne$). Therefore, these entailments are
  evidently valid.

\end{itemize}

\textbf{2. Rule $\ruleEqualLeft, \ruleExistsLeft, \ruleExistsRight$}

\begin{center}
  \small
  \begin{tabular}{ccc}
    \minipageEqualLeft &
    \minipageExistsLeft &
    \minipageExistsRight
  \end{tabular}
\end{center}

\begin{itemize}[label=--, leftmargin=1.2em]

\item For the rule $\ruleEqualLeft$, consider an arbitrary model $s,h$ of
  the goal entailment's antecedent. Then $s,h \satisfies F_1 \wedge u{=}v$.
  It follows that $\evalVar{u}{s} = \evalVar{v}{s}$, therefore $s,h
  \satisfies F_1[u/v]$. Since the entailment $F_1[u/v] \entails F_2 [u/v]$ in
  the rule's premise is valid, it implies $s,h \satisfies F_2[u/v]$. This
  means $s,h \satisfies F_2$, due to $\evalVar{u}{s} = \evalVar{v}{s}$.
  Therefore, $F_1 \wedge u{=}v \entails F_2$ is valid.

\item To prove correctness of the rule $\ruleExistsLeft$, we consider an
  arbitrary model $s,h$ such that $s,h \satisfies \exists x. F_1$. By
  semantics of the $\exists$ quantification, there is an integer value $v \in
  \sInt$ such that $s', h \satisfies F_1$, with $s' = \modelExtOne{s}{x}{v}$.
  Then $s'', h \satisfies F_1[u/x]$ with $s''$ is extended from $s'$ such
  that $s''(u) = s'(x)$. Since $s' = \modelExtOne{s}{x}{v}$, it follows that
  $s'' = \modelExtOne{s}{u}{v}$. On the other hand, given that entailment in
  the rule's premise is valid, then $s'',h \satisfies F_2$. It implies that
  $\modelExtOne{s}{u}{v} \satisfies F_2$. In addition $u \not\in
  \freevars{F_2}$. Therefore $s,h \satisfies F_2$. Since $s,h$ is chosen
  arbitrarily, it follows that the rule $\ruleExistsLeft$ is correct.

\item Correctness of $\ruleExistsRight$ is straight forward. Suppose that
  $s,h$ is an arbitrary model such that $s,h \satisfies F_1$. Since
  entailment in the rule's premise is valid, then $s,h \satisfies F_2[e/x]$.
  It follows that $s,h$ also satisfies $\exists x. F_2$, by simply choosing
  value $v$ of $x$ such that $v = \evalForm{e}{s}$. Since $s,h$ is chosen
  arbitrarily, it follows that the entailment in the rule's conclusion is
  valid. Therefore $\ruleExistsRight$ is valid.

\end{itemize}

\textbf{3. Rule $\ruleEmpLeft$ and $\ruleEmpRight$}

\begin{center}
  \small
  \begin{tabular}{cc}
    \minipageEmpLeft & \hspace*{8em} \minipageEmpRight
  \end{tabular}
\end{center}

\begin{itemize}[label=--, leftmargin=1.2em]

\item It is evident that two formulas $F_1 * \emp$ and $F_1$ in the rule
  $\ruleEmpLeft$ are semantically equivalent. In addition, $F_2 * \emp$ and
  $F_2$ in the rule $\ruleEmpRight$ are also semantically equivalent. It
  follows that both the two rules $\ruleEmpLeft$ and $\ruleEmpRight$ are
  correct.

\end{itemize}

\newpage
\textbf{4. Rule $\ruleCaseAnalysis$}

\begin{center}
  \small
  \begin{tabular}{c}
    \minipageCaseAnalysis
  \end{tabular}
\end{center}

\begin{itemize}[label=--, leftmargin=1.2em]

\item The rule's premises provide that both the two entailments $F_1 \wedge
  \Pi \entails F_2$ and $F_1 \wedge \neg\Pi \entails F_2$ are valid. For an
  arbitrary model $s,h$ such that $s,h \satisfies F_1$, we consider two
  cases as follows.

  \begin{itemize}[leftmargin=1.5em]

  \item[(1)] $s,h \satisfies F_1 \wedge \Pi$. Since $F_1 \wedge \Pi \entails
    F_2$ is valid, it follows that $s,h \satisfies F_2$.

  \item[(2)] $s,h \not\satisfies F_1 \wedge \Pi$. Since $s,h \satisfies
    F_1$, it follows that $s,h \not\satisfies \Pi$. Consequently, $s,h
    \satisfies \neg\Pi$, by the semantics of separation logic formulas. We
    have shown that $s,h \satisfies F_1$ and $s,h \satisfies \neg\Pi$, it
    implies that $s,h \satisfies F_1 \wedge \neg\Pi$. Since $F_1 \wedge
    \neg\Pi \entails F_2$ is valid, it follows that $s,h \satisfies F_2$.
  \end{itemize}

  In both the above cases, we show that for an arbitrary model $s,h$, if
  $s,h \satisfies F_1$, then $s,h \satisfies F_2$. Therefore, the
  entailment $F_1 \entails F_2$ in the conclusion of the rule
  $\ruleCaseAnalysis$ is valid.
\end{itemize}

\textbf{5. Rule $\ruleStarData$ and $\ruleStarPred$}

\begin{center}
  \small
  \begin{tabular}{cc}
    \minipageStarData &\hspace*{2em} \minipageStarPred
  \end{tabular}
\end{center}

\begin{itemize}[label=--, leftmargin=1.2em]

\item To prove the rule $\ruleStarData$, we consider an arbitrary model
  $s,h$ such that $s,h \satisfies F_1 * \formst{\varSort}{u}{\vec{v}}$.
  Then, there exists $h_1 \disjoins h_2$ such that $h = h_1 \hunions h_2$,
  and $s,h_1 \satisfies F_1$, and $s,h_2 \satisfies
  \formst{\varSort}{u}{\vec{v}}$. On one hand, entailment in the rule's
  premise is valid, it follows that $s,h_1 \satisfies \exists \vec{x}. (F_2
  \wedge u{=}t \wedge \vec{v}{=}\vec{w})$. By semantics of $\exists$
  quantification, $s',h_1 \satisfies F_2 \wedge u{=}t \wedge
  \vec{v}{=}\vec{w}$, with $s'$ is a model extended from $s$ with integer
  values of $\vec{x}$. On the other hand, the rule's side condition gives
  $u \not\in \vec{x}$, and $\vec{v} \disjoins \vec{x}$, and $s'$ is extend
  from $s$ with values of $\vec{x}$, and $s,h_2 \satisfies
  \formst{\varSort}{u}{\vec{v}}$, it follows that $s',h_2 \satisfies
  \formst{\varSort}{u}{\vec{v}}$. Combining these two hands, with the fact
  that $h_1 \disjoins h_2$ and $h = h_1 \hunions h_2$, the following holds:
  $s',h \satisfies F_2 * \formst{\varSort}{u}{\vec{v}} \wedge u{=}t \wedge
  \vec{v}{=}\vec{w}$. By semantics of equality $(=)$, the following also
  holds: $s',h \satisfies F_2 * \formst{\varSort}{t}{\vec{w}} \wedge u{=}t
  \wedge \vec{v}{=}\vec{w}$. By weakening this formula via dropping the
  condition $u{=}t \wedge \vec{v}{=}\vec{w}$, it is evident that $s',h
  \satisfies F_2 * \formst{\varSort}{t}{\vec{w}}$. Since $s'$ is extended
  from $s$ with values of $\vec{x}$, it is evident that $s,h \satisfies
  \exists \vec{x}. (F_2 * \formst{\varSort}{t}{\vec{w}})$. Recall that
  $s,h$ is chosen arbitrarily, this implies that the rule $\ruleStarData$
  is sound.

\item Soundness of $\ruleStarPred$ can be proved in the similar way with
  the rule $\ruleStarData$ above.

\end{itemize}

\textbf{6. Rule $\rulePredRight$}

\begin{center}
  \small
  \begin{tabular}{c}
    \minipagePredIntroRight
  \end{tabular}
\end{center}

\begin{itemize}[label=--, leftmargin=1.2em]

\item Consider an arbitrary model $s,h$ such that $s,h \satisfies F_1$. Since
  entailment in the rule's premise is valid, it follows that $s,h \satisfies
  \exists \vec{x}. (F_2 * \formp{\formF^{\P}_i}{\vec{u}})$. In addition, the
  rule's side condition that $\formp{\formF^{\P}_i}{\vec{u}}$ is one of the
  definition cases of $\formp{\P}{\vec{u}}$ clearly implies that $s,h
  \satisfies \exists \vec{x}. (F_2 * \formp{\P}{\vec{u}})$. Since $s,h$ is
  chosen arbitrarily, it follows that entailment in the rule's conclusion is
  valid.

\end{itemize}

\newpage
\textbf{7. Rule $\ruleInduction$}

\begin{center}
  \small
  \begin{tabular}{c}
    \minipageInduction
  \end{tabular}
\end{center}

\begin{itemize}[label=--, leftmargin=1.2em]

\item We show that if all of the entailments $F_1 *
  \formp{\formF^{\P}_1}{\vec{u}} \entails F_2$, \ldots, $F_1 *
  \formp{\formF^{\P}_m}{\vec{u}} \entails F_2$ in the rule premise are
  valid, then so is the entailment $F_1 * \formp{\P}{\vec{u}} \entails
  F_2$ in the conclusion.

  Indeed, consider an arbitrary model $s,h$ such that $s,h \satisfies F_1 *
  \formp{\P}{\vec{u}}$. The side condition of the rule gives that
  $\formp{\P}{\vec{u}} \triangleq \formp{\formF^{\P}_1}{\vec{u}} \vee \ldots
  \vee \formp{\formF^{\P}_m}{\vec{u}}$, i.e.,
  $\formp{\formF^{\P}_1}{\vec{u}}$, \ldots, $\formp{\formF^{\P}_m}{\vec{u}}$
  are all definition cases of $\formp{\P}{\vec{u}}$. Since $s,h \satisfies
  F_1 * \formp{\P}{\vec{u}}$, it follows that $s,h \satisfies F_1 *
  \formp{\formF^{\P}_i}{\vec{u}}$, for all $i\,{=\,}1 \ldots m$. On the other
  hand, $F_1 * \formp{\formF^{\P}_1}{\vec{u}}$, \ldots, $F_1 *
  \formp{\formF^{\P}_m}{\vec{u}}$ are antecedents of all entailments in this
  rule's premises, and these entailments have the same consequent $F_2$.
  Therefore, $s,h \satisfies F_2$. Since $s,h$ is chosen arbitrarily, it
  follows that entailment in the rule's conclusion is valid. This confirms
  soundness of the rule $\ruleInduction$.

\end{itemize}

\textbf{8. Rule $\ruleHypo$:}

\begin{center}
  \small
  \begin{tabular}{c}
    \minipageHypothesis
  \end{tabular}
\end{center}

\begin{itemize}[label=--, leftmargin=1.2em]

\item The rule's side conditions $\Sigma_1 \,{*}\, \formp{\P}{\vec{u}} \synequiv
  \Sigma_3\theta \,{*}\, \formp{\P'}{\vec{v}}\theta \,{*}\, \Sigma$ and $
  \Pi_1 {\rightarrow} \Pi_3\theta$ imply that the entailment $\Sigma_1
  \,{*}\, \formp{\P}{\vec{u}} \wedge \Pi_1 \entails \Sigma_3\theta \,{*}\,
  \formp{\P'}{\vec{v}}\theta \,{*}\, \Sigma \wedge \Pi_3\theta \wedge \Pi_1$
  is valid.

  By applying Theorem \ref{thm:SubstitutionLawEntailments}, the induction
  hypothesis $\Sigma_3 * \formr{\P'}{\vec{v}} \wedge \Pi_3 \entails F_4$
  implies that $\Sigma_3\theta * \formr{\P'}{\vec{v}}\theta \wedge
  \Pi_3\theta \entails F_4\theta$ is valid. It follows that the following
  entailment is also valid: $\Sigma_3\theta * \formr{\P'}{\vec{v}}\theta *
  \Sigma \wedge \Pi_3\theta \wedge \Pi_1 \entails F_4\theta * \Sigma \wedge
  \Pi_1$.

  We have shown that the two entailments $\Sigma_1 \,{*}\,
  \formp{\P}{\vec{u}} \wedge \Pi_1 \entails \Sigma_3\theta \,{*}\,
  \formp{\P'}{\vec{v}}\theta \,{*}\, \Sigma \wedge \Pi_3\theta \wedge \Pi_1$
  and $\Sigma_3\theta * \formr{\P'}{\vec{v}}\theta * \Sigma \wedge
  \Pi_3\theta \wedge \Pi_1 \entails F_4\theta * \Sigma \wedge \Pi_1$ are
  valid. In addition, the rule's premise gives that $F_4\theta \,{*}\, \Sigma
  \,{\wedge}\, \Pi_1 \entails F_2$ is valid. It follows that the entailment
  $\Sigma_1 * \formp{\P}{\vec{u}} \,{\wedge}\, \Pi_1 \,{\entails}\, F_2$ in
  the rule's conclusion is valid as well. Therefore, the rule $\ruleHypo$ is
  correct.
\end{itemize}

\textbf{9. Rule $\ruleLemmaLeft, \ruleLemmaRight$:}

\begin{center}
  \small
  \begin{tabular}{l}
    \hspace{-3.5em}\minipageLemmaLeft\\[1.8em]
    \hspace{-3.5em}\minipageLemmaRight
  \end{tabular}
\end{center}

\begin{itemize}[label=--, leftmargin=1.2em]

\item Soundness of the lemma application rule $\ruleLemmaLeft$ can be proved
  similar to $\ruleHypo$. Specifically, this rule's side conditions $\Sigma_1
  \synequiv \Sigma_3\theta * \Sigma$ and $\Pi_1 {\pureImply} \Pi_3\theta$
  imply that the entailment $\Sigma_1 \wedge \Pi_1 \entails \Sigma_3\theta *
  \Sigma \wedge \Pi_3\theta \wedge \Pi_1$ is valid. On the other hand, by
  applying Theorem \ref{thm:SubstitutionLawEntailments}, the lemma $\Sigma_3
  \wedge \Pi_3 \entails F_4$ implies that the entailment $\Sigma_3\theta
  \wedge \Pi_3\theta \entails F_4\theta$ is valid. It follows that the
  following entailment is also valid: $\Sigma_3\theta * \Sigma \wedge
  \Pi_3\theta \wedge \Pi_1 \entails F_4\theta * \Sigma \wedge \Pi_1$. We have
  shown that both the two entailments $\Sigma_1 \wedge \Pi_1 \entails
  \Sigma_3\theta * \Sigma \wedge \Pi_3\theta \wedge \Pi_1$ and
  $\Sigma_3\theta * \Sigma \wedge \Pi_3\theta \wedge \Pi_1 \entails F_4\theta
  * \Sigma \wedge \Pi_1$ are valid. In addition, this rule's premise provides
  that the entailment $F_4\theta * \Sigma \wedge \Pi_1 \entails F_2$ is also
  valid. Consequently, the goal entailment $\Sigma_1 \wedge \Pi_1 \entails
  F_2$ is valid as well. Therefore, soundness of the rule $\ruleLemmaLeft$ is
  claimed.

\item For the rule $\ruleLemmaRight$, the lemma $F_3 \entails \exists
  \vec{w}.(\Sigma_4 \wedge \Pi_4)$ implies that $F_3 \entails \exists
  \vec{w}.\Sigma_4$ is valid, therefore $F_3\theta \entails \exists
  \vec{w}\theta.\Sigma_4\theta$ is also valid by Theorem
  \ref{thm:SubstitutionLawEntailments}. It follows that $F_3\theta * \Sigma
  \wedge \Pi_2 \entails \exists \vec{w}\theta.(\Sigma_4\theta * \Sigma
  \wedge \Pi_2)$ is also valid. In addition, this rule's side condition
  provides that $\vec{w}\theta \subseteq \vec{x}$, therefore $F_3\theta *
  \Sigma \wedge \Pi_2 \entails \exists \vec{x}.(\Sigma_4\theta * \Sigma
  \wedge \Pi_2)$ is valid. Furthermore, the rule's premise provides that
  the entailment $F_1 \entails \exists \vec{x}.(F_3\theta * \Sigma \wedge
  \Pi_2)$ is valid. Hence, by Theorem \ref{thm:EntailmentTransitivity}, the
  entailment $F_1 \entails \exists \vec{x}.(\Sigma_4\theta * \Sigma \wedge
  \Pi_2)$ is valid. On the other hand, the rule's side condition
  $\Sigma_4\theta * \Sigma \synequiv \Sigma_2$ implies that the entailment
  $\Sigma_4\theta * \Sigma \wedge \Pi_2 \entails \Sigma_2 \wedge \Pi_2$ is
  valid. We have shown that both the two entailments $F_1 \entails \exists
  \vec{x}.(\Sigma_4\theta * \Sigma \wedge \Pi_2)$ and $\Sigma_4\theta *
  \Sigma \wedge \Pi_2 \entails \Sigma_2 \wedge \Pi_2$
  are valid, it follows by Theorem \ref{thm:EntailmentTransitivity} again
  that the goal entailment $F_1 \entails \exists \vec{x}.(\Sigma_2 \wedge
  \Pi_2)$ is also valid. Correctness of the rule $\ruleLemmaRight$,
  therefore, is claimed.
\end{itemize}

\subsection{Soundness of the synthesis rules}
\label{sec:SoundnessSynthesisRules}

We now present soundness proof the synthesis rules, presented in Figure
\ref{fig:SynthesisRules}, by showing that if all {\em assumptions} and
entailments in their premises are valid, and their side conditions, if any,
are satisfied, then goal entailments in their conclusions are also valid.

\textbf{1. Rule $\ruleSynPiOne$ and $\ruleSynPiTwo$}

\begin{center}
  \small
  \begin{tabular}{cc}
    \minipagePiOne & \hspace*{9em} \minipagePiTwo
  \end{tabular}
\end{center}

\begin{itemize}[label=--, leftmargin=1.2em]

\item Consider an arbitrary model $s,h$ such that $s,h \satisfies \Pi_1
  \wedge \formr{\rUnk}{\vec{x}}$. Since the unknown assumption $\Pi_1
  \wedge \formr{\rUnk}{\vec{x}} \rightarrow \Pi_2$ is assumed valid, then
  it follows from the semantics of $\rightarrow$ that $s,h \satisfies
  \Pi_2$. Since $s,h$ is chosen arbitrarily, it follows that the goal
  entailment $\Pi_1 \wedge \formr{\rUnk}{\vec{x}} \entails \Pi_2$ is also
  valid.

\item The proof of the rule $\ruleSynPiTwo$ is similar to the proof of
  $\ruleSynPiOne$.

\end{itemize}

\textbf{2. Rule $\ruleSynSigmaOne$ and $\ruleSynSigmaTwo$}

\begin{center}
  \small
  \begin{tabular}{cc}
    \minipageSigmaOne & \hspace*{3em} \minipageSigmaTwo
  \end{tabular}
\end{center}

\begin{itemize}[label=--, leftmargin=1.2em]

\item If the unknown assumption $\Pi_1 \wedge \formr{\rUnk} {\vec{x}}
  \rightarrow \false$ in the premise of the rule $\ruleSynSigmaOne$ is
  valid, then it is evident that $\Pi_1 \wedge \formr{\rUnk} {\vec{x}}
  \equiv \false$, since only $\false$ can prove $\false$. It follows that
  $\Sigma_1 \wedge \Pi_1 \wedge \formr{\rUnk} {\vec{x}} \equiv \false$.
  Consequently, the goal entailment $\Sigma_1 \wedge \Pi_1 \wedge
  \formr{\rUnk}{\vec{x}} \entails \Pi_2$ is valid, since its antecedent is
  a contradiction.

\item Soundness proof the rule $\ruleSynSigmaTwo$ is similar where the
  assumption $\Pi_1 \wedge \formr{\rUnk} {\vec{x}} \rightarrow \false$ in
  the rule's premise also implies that the goal entailment's antecedent is
  a contradiction ($\Pi_1 \wedge \formr{\rUnk}{\vec{x}} \equiv \false$)

\end{itemize}

\textbf{3. Rule $\ruleSynHypo$}

\begin{center}
  \small
  \begin{tabular}{l}
    \minipageSynHypo
  \end{tabular}
\end{center}

\begin{itemize}[label=--, leftmargin=1.2em]

\item Since the rule $\ruleSynHypo$ applies an induction hypothesis, which
  contains an unknown relation, to prove another unknown entailment, the
  soundness of this rule can be proved in a similar manner to the normal
  induction hypothesis application rule $\ruleHypo$. Note that in the proof
  of $\ruleSynHypo$, although the side condition $\Pi_1 \,{\wedge}\,
  \formr{\rUnk}{\vec{x}} \rightarrow \Pi_3\theta \,{\wedge}\,
  \formr{\rUnk}{\vec{y}}\theta$ contains unknown relations
  $\formr{\rUnk}{\vec{x}}$ and $\formr{\rUnk}{\vec{y}}\theta$, validity of
  this side condition is clearly implied by the assumption $\Assumpts
  {\triangleq} \{\Pi_1 \,{\wedge}\, \formr{\rUnk} {\vec{x}} \pureImply
  \Pi_3\theta \,{\wedge}\, \formr{\rUnk} {\vec{y}}\theta\}$ in the rule's
  premises.

\end{itemize}

\subsection{Soundness proofs of Propositions
  \ref{thm:SoundnessProofSystemUnknown},
  \ref{thm:SoundnessProofSystemClassic},
  \ref{thm:SoundnessProofSystemLemma}}

\TheoremSoundnessProofSystemUnknown*

\TheoremSoundnessProofSystemClassic*

\TheoremSoundnessProofSystemLemma*

\textsc{Proof of Propositions \ref{thm:SoundnessProofSystemUnknown},
    \ref{thm:SoundnessProofSystemClassic},
    \ref{thm:SoundnessProofSystemLemma}.}
  We have argued in Sections \ref{sec:SoundnessLogicalRules} and
  \ref{sec:SoundnessSynthesisRules} that all of our inference rules and
  synthesis rules are sound. In addition, our proof system, \emph{without
    the lemma synthesis component}, is clearly an instance of Noetherian
  induction where the sub-structural relation is proved to be well-founded
  (Theorem \ref{thm:WellFoundedness}) and is utilized by the induction
  hypothesis application rule $\ruleHypo$. Therefore, soundness of the
  proof system, when the lemma synthesis is disabled ($\Prove$ is invoked
  with $\ProofMode = \NoSynLemma$), is guaranteed by the Noetherian
  induction principle and the soundness of all inference and synthesis
  rules. Now, we consider each proposition as follows.

  \begin{enumerate}[label=--, leftmargin=1em]
  \item In Proposition \ref{thm:SoundnessProofSystemUnknown}, the proof
    system is invoked in the lemma-synthesis-disabled mode ($\NoSynLemma$).
    Based on our argument above about the proof system's soundness in this
    mode, it is clear that if $\Prove$ returns $\ValidWitness{\_}$ and
    generates a set of assumptions $\Assumpts$ when proving the unknown
    entailment $E$, then $E$ is semantically valid given that all unknown
    assumptions in $\Assumpts$ are valid.

  \item In Proposition \ref{thm:SoundnessProofSystemClassic}, the proof
    system is also invoked in the lemma-synthesis-disabled mode
    ($\NoSynLemma$). Since the goal entailment does not contain any unknown
    relation, there is no unknown assumption introduced while proving $E$.
    Hence, if $\Prove$ returns $\ValidWitness{\_}$, then $E$ is
    semantically valid.

  \item In Proposition \ref{thm:SoundnessProofSystemLemma}, the proof system
    can synthesize a set of new lemma $\LemmasSyn$ when proving $E$. Hence,
    we rely on the soundness of the lemma synthesis framework (Section
    \ref{sec:LemmaSynthesis}) to argue about the soundness of the proof
    system. Details about the lemma synthesis's soundness are presented
    in Theorem \ref{thm:SoundnessLemmaSynthesis}. Now, suppose that the lemma
    synthesis is already proved sound, then all lemmas synthesized in
    $\LemmasSyn$ are semantically valid. Following that, we can separate the
    proof derivation of $E$ into multiple phases, in which the auxiliary
    phases are used to synthesize all lemmas in $\LemmasSyn$, and the main
    phase is to prove $E$, using both the input lemmas in $\Lemmas$ or the
    new synthesized lemmas in $\LemmasSyn$. Certainly, this main phase
    reflects the structural induction aspect of our proof system, without the
    lemma synthesis component. In addition, all lemmas given in $\Lemmas$ and
    $\LemmasSyn$ are valid. Therefore, if $\Prove$ returns
    $\ValidWitness{\_}$ when proving $E$, then $E$ is semantically valid.
    \qed
  \end{enumerate}

%%% Local Variables:
%%% mode: latex
%%% TeX-master: "main"
%%% End:

\clearpage
\section{Formal verification of the proof system}
\label{append:HoareProof}

\begin{figure}[H]
\begin{center}
  \small
  \begin{tabular}{c}
    \minipageGeneralRule
  \end{tabular}
\end{center}
\vspace{-0.3em}
\caption{General form of inference or synthesis rules}
\label{fig:RuleGeneralForm}
\end{figure}

In Figure \ref{fig:RuleGeneralForm}, we present a general form of an
inference rule or a synthesis rule $\vRule$ where $\Hypos \sepAnte\, \Lemmas
\sepAnte\, F_{1} \entails F_{2}$ is the rule's conclusion; $(\Hypos_1
\sepAnte\, \Lemmas_1 \sepAnte\, F_{11} \entails F_{21})\,,\, \ldots,\,
(\Hypos_n \sepAnte\, \Lemmas_n \sepAnte\, F_{1n} \entails F_{2n})$ and
$\Assumpts$ are premises corresponding to the derived sub-goal entailments
and the unknown assumption set; and $\dagger$ is the rule's side condition.
We define the retrieval functions \assumptions{\vRule}, \premises{\vRule},
and \conclusion{\vRule} to extract these components of the rule $\vRule$,
as follows:

\begin{adjustwidth}{2em}{}
  \begin{tabular}{lll}
    \assumptions{\vRule}
    & $\triangleq\hspace{0.5em}$
    & $\{\Pi_1 \pureImply \Pi_2\}$\\[0.3em]

    \premises{\vRule}
    & $\triangleq$
    & $\{\Hypos_i, \Lemmas_i, F_{1i} {\entails} F_{2i}\}_{i{=}1}^n$\\[0.3em]

    \conclusion{\vRule}
    & $\triangleq$
    & $(\Hypos, \Lemmas, F_{1} {\entails} F_{2})$
  \end{tabular}
\end{adjustwidth}

Note that the first two functions $\assumptions{\vRule}$ and $\premises{\vRule}$
may return an empty set. Moreover, with respect to the set $\sRule$ of all
inference and synthesis rules given in Figures \ref{fig:LogicalRules},
\ref{fig:InductionRules}, \ref{fig:LemmaRules}, and \ref{fig:SynthesisRules},
we always have $\Lemmas = \Lemmas_i$.

Due to the soundness of these rules, we have:

\begin{center}
$\validlem{\Assumpts} \wedge
\bigwedge_{i{=}1}^n
  \validindt{\Hypos_i, \Lemmas_i, F_{1i} {\entails} F_{2i}}
\rightarrow
\validindt{\Hypos, \Lemmas, F_1 {\entails} F_2}$.
\end{center}

From this, we define the predicate \sound{\vRule} to denote the soundness of
a rule $\vRule \in \sRule$ as follows:

\begin{center}
$
\sound{\vRule} \triangleq
\validlem{\assumptions{\vRule}} \wedge \validindt{\premises{\vRule}}
\rightarrow \validindt{\conclusion{\vRule}}$.
\end{center}

The soundness of each inference or synthesis rule $\vRule$,
which is proved in Appendix \ref{sec:SoundnessLogicalRules}
or \ref{sec:SoundnessSynthesisRules}, is always
assured at the beginning of the formal verification for each procedure
in our framework. Therefore, this assurance clearly connects the soundness
of our procedures to the soundness of these rules. Such guarantee can be
considered as an {\em invariant} of the set of all rules $\sRule$ exploited
in these procedures. We denote the invariant as
$ \sound{\sRule} \triangleq \forall \vRule \in \sRule.\, \sound{\vRule}$,
which specifies that every rule $\vRule \in \sRule$ is sound.

Below are the specifications of two retrieval procedures $\Assumptions(\vRule)$
and $\Premises(\vRule)$ to get the set of assumptions and the set of premises
from $\vRule$.

\vspace{0.5em}

\begin{figure}[H]
  \begin{small}
    \begin{adjustwidth}{-0.5em}{}
      \begin{tabular}{l}
        {\bf Procedure} \Call{\Assumptions}{\vRule}
        // Retrieve assumptions on unknown relations in $\vRule$\\
        \Summary{Requires:} $\true$\\
        \Summary{Ensures:}
        $res \,{=}\, \assumptions{\vRule}$\\%[0.5em]
      \end{tabular}
    \end{adjustwidth}
  \end{small}
  \label{fig:AssumptionsSpec}
\end{figure}

\vspace{0.5em}

\begin{figure}[H]
  \begin{small}
    \begin{adjustwidth}{-0.5em}{}
      \begin{tabular}{l}
        {\bf Procedure} \Call{\Premises}{\vRule}
        // Retrieve premises of $\vRule$\\
        \Summary{Requires:} $\true$\\
        \Summary{Ensures:}
        $res \,{=}\, \premises{\vRule}$\\%[0.5em]
      \end{tabular}
    \end{adjustwidth}
  \end{small}
  \label{fig:PremisesSpec}
\end{figure}

In addition, we also give the specifications of two auxiliary procedures
$\IsAxiomRule(\vRule)$ and $\IsSynthesisRule(\vRule)$ which
returns $\true$ if the rule $\vRule$ is an axiom rule
%($\vRuleInst \in \setenum{
%  \rulePureEntail, \ruleFalseLeftOne, \ruleFalseLeftTwo,
%  \ruleSynPiOne, \ruleSynPiTwo, \ruleSynSigmaOne,
%  \ruleSynSigmaTwo}$)
or a synthesis rule.
%($\vRuleInst \in \setenum{
%  \ruleSynPiOne, \ruleSynPiTwo,
%  \ruleSynSigmaOne, \ruleSynSigmaTwo, \ruleSynHypo}$).

\vspace{0.5em}
\begin{figure}[H]
\begin{small}
    \begin{adjustwidth}{-0.5em}{}
      \begin{tabular}{l}
        {\bf Procedure} \Call{\IsAxiomRule}{\vRule}
        // Check if $\vRule$ is an axiom rule\\
        \Summary{Requires:} $\vRule \in \setenum{
          \rulePureEntail, \ruleFalseLeftOne, \ruleFalseLeftTwo,
          \ruleSynPiOne, \ruleSynPiTwo, \ruleSynSigmaOne, \ruleSynSigmaTwo}$\\
        \Summary{Ensures:}
        $(res=\true) \wedge (\premises{\vRule} = \setempty)$\\[0.3em]

        \Summary{Requires:} $\vRule \not\in \setenum{
          \rulePureEntail, \ruleFalseLeftOne, \ruleFalseLeftTwo,
          \ruleSynPiOne, \ruleSynPiTwo, \ruleSynSigmaOne,
          \ruleSynSigmaTwo}$\\
        \Summary{Ensures:} $res = \false$\\%[0.5em]
      \end{tabular}
    \end{adjustwidth}
  \end{small}
  \label{fig:IsAxiomRuleSpec}
\end{figure}

The above specifications of $\IsAxiomRule(\vRule)$ specify two possible
cases of the input $\vRule$, respectively when $\vRule$ is an axiom
rule ($\vRule \in \setenum{\rulePureEntail, \ruleFalseLeftOne,
\ruleFalseLeftTwo, \ruleSynPiOne, \ruleSynPiTwo, \ruleSynSigmaOne,
\ruleSynSigmaTwo}$), or otherwise. In the former case, we can ensure that
there are no premises in $\vRule$ (i.e., $\premises{\vRule} = \setempty$).

\vspace{0.5em}
\begin{figure}[H]
  \begin{small}
    \begin{adjustwidth}{-0.5em}{}
      \begin{tabular}{l}
        {\bf Procedure} \Call{\IsSynthesisRule}{\vRule}
        // Check if $\vRule$ is a synthesis rule\\
        \Summary{Requires:} $\vRule \in \setenum{
          \ruleSynPiOne, \ruleSynPiTwo,
          \ruleSynSigmaOne, \ruleSynSigmaTwo, \ruleSynHypo}$\\
        \Summary{Ensures:} $res = \true$\\[0.3em]
        \Summary{Requires:} $\vRule \not\in \setenum{
          \ruleSynPiOne, \ruleSynPiTwo,
          \ruleSynSigmaOne, \ruleSynSigmaTwo, \ruleSynHypo}$\\
        \Summary{Ensures:}
        $(res = \false) \wedge (\assumptions{\vRule} = \setempty)$\\%[0.5em]
      \end{tabular}
    \end{adjustwidth}
  \end{small}
  \label{fig:IsSynthesisRuleSpec}
\end{figure}

The specifications of the procedure $\IsSynthesisRule$ denote two cases when
$\vRule$ is a synthesis rule or not. In the latter case, it is obvious that
the assumption set of $\vRule$ is empty
(i.e., $\assumptions{\vRule} = \setempty$).

In the following subsections, we illustrate the formal verification of the
main proof search procedure $\Prove$ in Figure \ref{fig:ProofSearch} on its
three specifications, depicted by pairs of $\Summary{Requires/Ensures}$.
We perform a Hoare-style forward verification, in which $\longrightarrow$
is a weakening on program states.

The formal verification of other lemma synthesis procedures in Section
\ref{sec:LemmaSynthesis}, i.e., $\SynthesizeLemma$, $\RefineAnte$,
$\RefineConseq$, $\Preprocess$, and $\FineTuneConseq$ in Figures
\ref{fig:SynthesizeLemmas}, \ref{fig:FineTuneAnte}, \ref{fig:RefineConseq},
\ref{fig:PreprocessTemplate}, and \ref{fig:FineTuneConseq} are
straightforward.

\subsection{Verification of \emph{the first specification} of
  the proof search procedure in Figure \ref{fig:ProofSearch}}

\vspace{0.5em}
\begin{figure}[H]
  \algrenewcommand\algorithmicindent{1em}
  \begin{algorithm}[H]
    \small
    \caption*{\small{\bf Procedure} \Call{\Prove}{
        \Hypos, \Lemmas, F_1 \entails F_2, \ProofMode}}
    \begin{flushleft}
      \Summary{Requires:}
      $(\ProofMode \,{=}\, \NoSynLemma)
      \wedge \hasUnk{F_1 {\entails} F_2}
      \wedge \validlem{\Lemmas}$\\
      \Summary{Ensures:} $(\res \,{=}\, (\Unknown, \setempty, \setempty))
      ~\vee~
      \exists\,\xi,\Assumpts.\,
      (\res \,{=}\, (\ValidWitness{\xi}, \setempty, \Assumpts)
      \wedge (\validlem{\Assumpts} \rightarrow
      \validindt{\Hypos, \Lemmas, F_1 {\entails} F_2}))$\\
      \Summary{Invariant:} $\sound{\sRule}$\\
    \end{flushleft}
    \begin{algorithmic}[1]
      \def\indent{\hspace{1.2em}}
      \CommentLnColor{blue}{$\{\,
        (\ProofMode \,{=}\, \NoSynLemma)
          \wedge \hasUnk{F_1 {\entails} F_2}
          \wedge \validlem{\Lemmas}
          \wedge \sound{\sRule} \,\}$}
      \State \Assign{\sRuleSelected}{\setenum{
          \Call{\Unify}{\vRule, (\Hypos, \Lemmas, F_1 \entails F_2)}
          ~|~ \vRule \in \sRule}}

      \CommentLnColor{blue}{$\{\,
        (\ProofMode \,{=}\, \NoSynLemma)
          \wedge \hasUnk{F_1 {\entails} F_2}
          \wedge \validlem{\Lemmas}
          \wedge \sound{\sRuleSelected} \,\}$}
      \Comment{$\sound{\sRule} \wedge \sRuleSelected \subseteq \sRule
              \rightarrow \sound{\sRuleSelected}$}

      \State \Assign{\LemmasSyn}{\setempty}
      \CommentLnColor{blue}{$\{\,
        (\ProofMode \,{=}\, \NoSynLemma)
          \wedge \hasUnk{F_1 {\entails} F_2}
          \wedge \validlem{\Lemmas}
          \wedge \LemmasSyn = \setempty
          \wedge \sound{\sRuleSelected} \,\}$}

      \If{\ProofMode = \SynLemma \Keyword{and}
          \Call{\NeedLemmas}{F_1 \entails F_2, \sRuleSelected}}

        \State \Assign{\LemmasSyn}{
          \Call{\SynthesizeLemma}{\Lemmas,\,F_1 \entails F_2}}
        \Comment{unreachable code}

        \State \Assign{\sRuleSelected}{
          \sRuleSelected \cup \setenum{\Call{\Unify}{
              \vRule, (\Hypos, \LemmasSyn, F_1 \entails F_2)}
            ~|~ \vRule \in \{\ruleLemmaLeft, \ruleLemmaRight\}}}

      \EndIf

      \CommentLnColor{blue}{$\{\,
        (\ProofMode \,{=}\, \NoSynLemma)
          \wedge \hasUnk{F_1 {\entails} F_2}
          \wedge \validlem{\Lemmas}
          \wedge \LemmasSyn = \setempty
          \wedge \sound{\sRuleSelected} \,\}$}

      \For{\Keyword{each} \vRuleInst \Keyword{in} \sRuleSelected}
        \Comment{$\sound{\sRuleSelected} \wedge \vRuleInst \in \sRuleSelected
                        \rightarrow \sound{\vRuleInst}$}
        \CommentLnIndentColor{blue}{$\{\,
          (\ProofMode \,{=}\, \NoSynLemma)
            \wedge \hasUnk{F_1 {\entails} F_2}
            \wedge \validlem{\Lemmas}
            \wedge \LemmasSyn = \setempty
            \wedge \sound{\vRuleInst} \,\}$}

        \State \Assign{\Assumpts}{\setempty}

        \If{\Call{IsSynthesisRule}{\vRuleInst}}
          \Assign{\Assumpts}{\Call{\Assumptions}{\vRuleInst}}
        \EndIf

        \CommentLnColor{blue}{$\{\,
          (\ProofMode \,{=}\, \NoSynLemma)
            \wedge \hasUnk{F_1 {\entails} F_2}
            \wedge \validlem{\Lemmas}
            \wedge \LemmasSyn = \setempty
            \wedge \sound{\vRuleInst}
            \wedge \Assumpts = \assumptions{\vRuleInst} \,\}$}

        \If{\Call{IsAxiomRule}{\vRuleInst}} %

        \CommentLnIndentColor{blue}{$\{\,
            \ldots
            \wedge \LemmasSyn = \setempty
            \wedge \sound{\vRuleInst}
            \wedge \assumptions{\vRuleInst} = \Assumpts
            \wedge \premises{\vRuleInst} = \setempty \,\}$}
        \CommentLnIndentColor{blue}{$\longrightarrow \{\,
          \LemmasSyn = \setempty
           \wedge (\validlem{\Assumpts} \rightarrow
             \validindt{\Hypos, \Lemmas, F_1 {\entails} F_2}) \,\}$}
        \Comment{$\validindt{\setempty} = \true$}

          \State \Assign{\xi}{\Call{\CreateWitnessProofTree}{
              F_1 \entails F_2, \vRuleInst, \setempty}}
          \State \ReturnTuple{\ValidWitness{\xi}, \LemmasSyn, \Assumpts}

          \CommentLnColor{blue}{$\{\,
              \LemmasSyn = \setempty
              \wedge (\validlem{\Assumpts} \rightarrow
                \validindt{\Hypos, \Lemmas, F_1 {\entails} F_2})
              \wedge \res \,{=}\, (\ValidWitness{\xi}, \LemmasSyn, \Assumpts)
              \,\}$}

          %% \CommentLnColor{blue}{$\longrightarrow$}

          \CommentLnColor{blue}{$\longrightarrow \{\,
            \exists\,\xi,\Assumpts.\,
                (\res \,{=}\, (\ValidWitness{\xi}, \setempty, \Assumpts)
                \wedge (\validlem{\Assumpts} \rightarrow
                \validindt{\Hypos, \Lemmas, F_1 {\entails} F_2})) \,\}$}
        \EndIf{}

        \State \Assign{
          (\Hypos_i, \Lemmas_i, F_{1i} \entails F_{2i})_{i\,=\,1 \ldots n}}{
          \Call{\Premises}{\vRuleInst}}

        \CommentLnColor{blue}{$\{\,
          (\ProofMode \,{=}\, \NoSynLemma)
            \wedge \hasUnk{F_1 {\entails} F_2}
            \wedge \validlem{\Lemmas}
            \wedge \LemmasSyn = \setempty
            \wedge \sound{\vRuleInst}$}
        \CommentLnColor{blue}{$\qquad
          \wedge\, \assumptions{\vRuleInst} = \Assumpts
          \wedge\, \premises{\vRuleInst} =
            \{\Hypos_i, \Lemmas_i, F_{1i} \entails F_{2i}\}_{i{=}1}^n
          \wedge \bigwedge_{i{=}1}^n (\Lemmas_i = \Lemmas) \,\}$}

        \State \Assign{(\vR_i,
          \Lemmas^i_{\mathrm{syn}},\Assumpts_i)_{i\,=\,1 \ldots n}}{
          \Call{\Prove}{
            \Hypos_i,\, \Lemmas \cup \LemmasSyn \cup \Lemmas_i,\,F_{1i} \entails
            F_{2i}, \ProofMode}_{i\,=\,1 \ldots n}}

        \CommentLnColor{blue}{$\{\,
          \ldots
          \wedge \LemmasSyn = \setempty
          \wedge (
            \validlem{\Assumpts} \wedge
            \bigwedge_{i{=}1}^n \validindt{\Hypos_i, \Lemmas_i, F_{1i} \entails F_{2i}}
            \rightarrow \validindt{\Hypos, \Lemmas, F_1 {\entails} F_2})
          \,\wedge $}
        \CommentLnColor{blue}{$\qquad
          \bigwedge_{i{=}1}^n (
            \Lemmas^i_{\mathrm{syn}} = \setempty
            \wedge ((\vR_i \,{=}\, \Unknown \wedge \Assumpts_i = \setempty)$}

        \CommentLnColor{blue}{$\qquad \qquad
            \vee\, (\hasUnk{F_{1i} {\entails} F_{2i}}
              \wedge (\exists \xi_i.\, (\vR_i \,{=}\, \ValidWitness{\xi_i}
                        \wedge (\validlem{\Assumpts_i} \rightarrow
                                 \validindt{\Hypos_i, \Lemmas_i, F_{1i}
                                   {\entails} F_{2i}}))))$}
        \CommentLnColor{blue}{$\qquad \qquad
            \vee\, (\neg\hasUnk{F_{1i} {\entails} F_{2i}}
              \wedge (\exists \xi_i.\, (\vR_i \,{=}\, \ValidWitness{\xi_i}
                        \wedge \Assumpts_i = \setempty
                        \wedge \validindt{\Hypos_i, \Lemmas_i, F_{1i} {\entails} F_{2i}})))) \,\}$}

        \CommentLnColor{blue}{$\longrightarrow \{\,
          \LemmasSyn = \setempty
          \wedge (
            \validlem{\Assumpts} \wedge
            \bigwedge_{i{=}1}^n \validindt{\Hypos_i, \Lemmas_i, F_{1i} \entails F_{2i}}
            \rightarrow \validindt{\Hypos, \Lemmas, F_1 {\entails} F_2})
          \,\wedge$}
        \CommentLnColor{blue}{$\qquad
          \bigwedge_{i{=}1}^n (
            \Lemmas^i_{\mathrm{syn}} = \setempty
            \wedge ((\vR_i \,{=}\, \Unknown \wedge \Assumpts_i = \setempty)$}
        \CommentLnColor{blue}{$\qquad \qquad
            \vee\, (\exists \xi_i.\, (\vR_i \,{=}\, \ValidWitness{\xi_i}
                      \wedge (\validlem{\Assumpts_i} \rightarrow
                        \validindt{\Hypos_i, \Lemmas_i, F_{1i} {\entails}
                        F_{2i}})))) \,\}$}
        \Comment{$\validlem{\setempty} = \true$}
        \algstore{myalg}
    \end{algorithmic}
  \end{algorithm}
  \vspace{-0.5em}
  \caption{Verification of the proof search procedure $\Prove$
    on the first specification (Part 1)}
  \label{fig:ProofSearchHoare1a}
\end{figure}

\vspace{0.5em}
\begin{figure}[H]
  \algrenewcommand\algorithmicindent{1em}
  \begin{verify}
    \small
    \begin{algorithmic}[1]
        \algrestore{myalg}

    		\CommentLnColor{blue}{$\{\,
          \LemmasSyn = \setempty
          \wedge (
            \validlem{\Assumpts} \wedge
            \bigwedge_{i{=}1}^n \validindt{\Hypos_i, \Lemmas_i, F_{1i} \entails
            F_{2i}}
            \rightarrow \validindt{\Hypos, \Lemmas, F_1 {\entails} F_2})
          \,\wedge$}
        \CommentLnColor{blue}{$\qquad
          \bigwedge_{i{=}1}^n (
            \Lemmas^i_{\mathrm{syn}} = \setempty
            \wedge ((\vR_i \,{=}\, \Unknown \wedge \Assumpts_i = \setempty)$}
        \CommentLnColor{blue}{$\qquad \qquad
            \vee\, (\exists \xi_i.\, (\vR_i \,{=}\, \ValidWitness{\xi_i}
                      \wedge (\validlem{\Assumpts_i} \rightarrow
                        \validindt{\Hypos_i, \Lemmas_i, F_{1i} {\entails}
                        F_{2i}})))) \,\}$}

        \If{\vR_i = \ValidWitness{\xi_i}\Keyword{for~all~}i = 1 \ldots n}
        \CommentLnIndentColor{blue}{$\{\,
          \LemmasSyn = \setempty
          \wedge (
            \validlem{\Assumpts} \wedge
            \bigwedge_{i{=}1}^n \validindt{\Hypos_i, \Lemmas_i, F_{1i} \entails
            F_{2i}}
            \rightarrow \validindt{\Hypos, \Lemmas, F_1 {\entails} F_2})
          \,\wedge $}
        \CommentLnIndentColor{blue}{$\qquad
          \bigwedge_{i{=}1}^n
            (\vR_i \,{=}\, \ValidWitness{\xi_i}
              \wedge \Lemmas^i_{\mathrm{syn}} = \setempty
              \wedge (\validlem{\Assumpts_i} \rightarrow
              \validindt{\Hypos_i, \Lemmas_i, F_{1i} {\entails} F_{2i}})) \,\}$}

          \State \Assign{\xi}{\Call{\CreateWitnessProofTree}{
              F_1 \entails F_2, \vRuleInst, \setenum{\xi_1, \ldots, \xi_n}}}

          \State \ReturnTuple{\ValidWitness{\xi},~
            \LemmasSyn \cup \Lemmas^1_{\mathrm{syn}} \cup \ldots
            \cup \Lemmas^n_{\mathrm{syn}},~
            \Assumpts \cup \Assumpts_1 \cup \ldots \cup \Assumpts_n}
          \label{line:Spec1PostState}

          \CommentLnColor{blue}{$\{\,
            \LemmasSyn = \setempty
            \wedge (
              \validlem{\Assumpts} \wedge
              \bigwedge_{i{=}1}^n \validindt{\Hypos_i, \Lemmas_i, F_{1i}
              \entails F_{2i}}
              \rightarrow \validindt{\Hypos, \Lemmas, F_1 {\entails} F_2})
            \,\wedge $}
          \CommentLnColor{blue}{$\qquad
            \bigwedge_{i{=}1}^n
              (\Lemmas^i_{\mathrm{syn}} = \setempty
                \wedge (\validlem{\Assumpts_i} \rightarrow
                \validindt{\Hypos_i, \Lemmas_i, F_{1i} {\entails} F_{2i}}))
            $}
          \CommentLnColor{blue}{$\qquad
            \wedge\, \res \,{=}\, (\ValidWitness{\xi}, \setempty,
                \Assumpts \cup \Assumpts_1 \cup \ldots \cup \Assumpts_n)
            \,\}$}

        \EndIf

      \EndFor{}

      \State{\ReturnTuple{\Unknown, \setempty, \setempty}}

      \CommentLnColor{blue}{$\{\,
        \res \,{=}\, (\Unknown, \setempty, \setempty) \,\}$}

    \end{algorithmic}
  \end{verify}
  \vspace{-0.5em}
  \caption{Verification of the proof search procedure $\Prove$
    on the first specification (Part 2)}
  \label{fig:ProofSearchHoare1}
\end{figure}

Given the post-state after line \ref{line:Spec1PostState} in Figure
\ref{fig:ProofSearchHoare1}, we prove the postcondition of $\Prove$ as
follows, where $\longrightarrow$ is a weakening on program states.
%\vspace{-0.5em}
\[\arraycolsep=1.4pt
  \begin{array}{ll}
    & \emph{// the program state at line 16}\\
    & \LemmasSyn = \setempty
    \wedge (
    \validlem{\Assumpts} \wedge
    \bigwedge_{i{=}1}^n \validindt{\Hypos_i, \Lemmas_i, F_{1i}
      \entails F_{2i}}
    \rightarrow \validindt{\Hypos, \Lemmas, F_1 {\entails} F_2})
    \\
    & \wedge~ \bigwedge_{i{=}1}^n
    (\Lemmas^i_{\mathrm{syn}} = \setempty
    \wedge (\validlem{\Assumpts_i} \rightarrow
    \validindt{\Hypos_i, \Lemmas_i, F_{1i} {\entails} F_{2i}}))
    \\
    & \wedge~ \res \,{=}\, (\ValidWitness{\xi}, \setempty,
    \Assumpts \cup \Assumpts_1 \cup \ldots \cup \Assumpts_n)
    \,\}\\

    \longrightarrow & (
    \validlem{\Assumpts} \wedge
    \bigwedge_{i{=}1}^n \validindt{\Hypos_i, \Lemmas_i, F_{1i} \entails F_{2i}}
    \rightarrow \validindt{\Hypos, \Lemmas, F_1 {\entails} F_2})
    \\
    & \wedge~ \bigwedge_{i{=}1}^n (\validlem{\Assumpts_i} \rightarrow
    \validindt{\Hypos_i, \Lemmas_i, F_{1i} {\entails} F_{2i}})
    \\
    & \wedge~ \res \,{=}\, (\ValidWitness{\xi}, \setempty,
    \Assumpts \cup \Assumpts_1 \cup \ldots \cup \Assumpts_n)
    \\
    \longrightarrow
    & (\validlem{\Assumpts} {\wedge}
    \!\bigwedge_{i{=}1}^n \validlem{\Assumpts_i}
    \rightarrow \validindt{\Hypos, \Lemmas, F_1 {\entails} F_2})
    \wedge \res {=} (\ValidWitness{\xi}, \setempty,
    \Assumpts {\cup} \Assumpts_1 {\cup} \ldots {\cup} \Assumpts_n)
    \\
    \longrightarrow
    & (\validlem{\Assumpts {\cup} \Assumpts_1 {\cup} \ldots {\cup} \Assumpts_n}
    \rightarrow \validindt{\Hypos, \Lemmas, F_1 {\entails} F_2})
    \wedge \res {=} (\ValidWitness{\xi}, \setempty,
    \Assumpts {\cup} \Assumpts_1 {\cup} \ldots {\cup} \Assumpts_n)
    \\
    \longrightarrow
    & \exists\,\xi,\Assumpts.\,
    (\res \,{=}\, (\ValidWitness{\xi}, \setempty, \Assumpts)
    \wedge (\validlem{\Assumpts} \rightarrow
    \validindt{\Hypos, \Lemmas, F_1 {\entails} F_2}) \\
    \longrightarrow
    & (\res \,{=}\, (\Unknown, \setempty, \setempty))
    ~\vee~
    \exists\,\xi,\Assumpts.\,
    (\res \,{=}\, (\ValidWitness{\xi}, \setempty, \Assumpts)
    \wedge (\validlem{\Assumpts} \rightarrow
    \validindt{\Hypos, \Lemmas, F_1 {\entails} F_2}))
    \\
    & \emph{// the post condition in the first specification of $\Prove$}
  \end{array}
\]

\newpage

\subsection{Verification of \emph{the second specification} of
  the proof search procedure in Figure \ref{fig:ProofSearch}}

\vspace{-0.5em}
\begin{figure}[H]
  \algrenewcommand\algorithmicindent{1em}
  \begin{algorithm}[H]
    \small
    \caption*{\small{\bf Procedure} \Call{\Prove}{
        \Hypos, \Lemmas, F_1 \entails F_2, \ProofMode}}
    \begin{flushleft}
      \Summary{Requires:}
        $(\ProofMode \,{=}\, \NoSynLemma)
        \wedge \neg\hasUnk{F_1 {\entails} F_2}
        \wedge \validlem{\Lemmas}$\\
      \Summary{Ensures:} $(\res \,{=}\, (\Unknown, \setempty, \setempty))
        ~\vee~
        \exists\,\xi.\,
        (\res \,{=}\, (\ValidWitness{\xi}, \setempty, \setempty)
        \wedge \validindt{\Hypos, \Lemmas, F_1 {\entails} F_2})$\\
      \Summary{Invariant:} $\sound{\sRule}$\\[0.5em]
    \end{flushleft}
    \begin{algorithmic}[1]
      \def\indent{\hspace{1.2em}}

      \CommentLnColor{blue}{$\{\,
        (\ProofMode \,{=}\, \NoSynLemma)
          \wedge \neg\hasUnk{F_1 {\entails} F_2}
          \wedge \validlem{\Lemmas}
          \wedge \sound{\sRule} \,\}$}

      \State \Assign{\sRuleSelected}{\setenum{
          \Call{\Unify}{\vRule, (\Hypos, \Lemmas, F_1 \entails F_2)}
          ~|~ \vRule \in \sRule}}

      \CommentLnColor{blue}{$\{\,
        (\ProofMode \,{=}\, \NoSynLemma)
          \wedge \neg\hasUnk{F_1 {\entails} F_2}
          \wedge \validlem{\Lemmas}
          \wedge \sound{\sRuleSelected}$}
      \Comment{$\sound{\sRule} \wedge \sRuleSelected \subseteq \sRule
              \rightarrow \sound{\sRuleSelected}$}
      \CommentLnColor{blue}{$\qquad
      	\wedge\, \sRuleSelected \cap
        	\setenum{\ruleSynPiOne, \ruleSynPiTwo,
          	\ruleSynSigmaOne, \ruleSynSigmaTwo, \ruleSynHypo}
          = \setempty \,\}$}
      \Comment{since $\neg\hasUnk{F_1 {\entails} F_2}$}

      \State \Assign{\LemmasSyn}{\setempty}
      \CommentLnColor{blue}{$\{\,
        (\ProofMode \,{=}\, \NoSynLemma)
          \wedge \neg\hasUnk{F_1 {\entails} F_2}
          \wedge \validlem{\Lemmas}
          \wedge \LemmasSyn = \setempty$}
      \CommentLnColor{blue}{$\qquad
      		\wedge\, \sound{\sRuleSelected}
          \wedge \sRuleSelected \cap
          	\setenum{\ruleSynPiOne, \ruleSynPiTwo,
          	\ruleSynSigmaOne, \ruleSynSigmaTwo, \ruleSynHypo}
          	= \setempty \,\}$}

      \If{\ProofMode = \SynLemma \Keyword{and}
          \Call{\NeedLemmas}{F_1 \entails F_2, \sRuleSelected}}

        \State \Assign{\LemmasSyn}{
          \Call{\SynthesizeLemma}{\Lemmas,\,F_1 \entails F_2}}
        \Comment{unreachable code}

        \State \Assign{\sRuleSelected}{
          \sRuleSelected \cup \setenum{\Call{\Unify}{
              \vRule, (\Hypos, \LemmasSyn, F_1 \entails F_2)}
            ~|~ \vRule \in \{\ruleLemmaLeft, \ruleLemmaRight\}}}

      \EndIf

      \CommentLnColor{blue}{$\{\,
        (\ProofMode \,{=}\, \NoSynLemma)
          \wedge \neg\hasUnk{F_1 {\entails} F_2}
          \wedge \validlem{\Lemmas}
          \wedge \LemmasSyn = \setempty$}
      \CommentLnColor{blue}{$\qquad
      		\wedge\, \sound{\sRuleSelected}
          \wedge \sRuleSelected \cap
          	\setenum{\ruleSynPiOne, \ruleSynPiTwo,
          	\ruleSynSigmaOne, \ruleSynSigmaTwo, \ruleSynHypo}
          	= \setempty \,\}$}

      \For{\Keyword{each} \vRuleInst \Keyword{in} \sRuleSelected}
      \CommentLnIndentColor{blue}{$\{\,
        (\ProofMode \,{=}\, \NoSynLemma)
          \wedge \neg\hasUnk{F_1 {\entails} F_2}
          \wedge \validlem{\Lemmas}
          \wedge \LemmasSyn = \setempty$}
      \CommentLnIndentColor{blue}{$\qquad
      		\wedge\, \sound{\vRuleInst}
          \wedge \vRuleInst \not\in \setenum{
            \ruleSynPiOne, \ruleSynPiTwo,
            \ruleSynSigmaOne, \ruleSynSigmaTwo, \ruleSynHypo} \,\}$}
      \Comment{since $\vRuleInst \in \sRuleSelected$}

        \State \Assign{\Assumpts}{\setempty}
        \CommentLnColor{blue}{$\{\,
          (\ProofMode \,{=}\, \NoSynLemma)
            \wedge \neg\hasUnk{F_1 {\entails} F_2}
            \wedge \validlem{\Lemmas}
            \wedge \LemmasSyn = \setempty
            \wedge \Assumpts = \setempty$}
        \CommentLnColor{blue}{$\qquad
        		\wedge\, \sound{\vRuleInst}
            \wedge \vRuleInst \not\in \setenum{
              \ruleSynPiOne, \ruleSynPiTwo,
              \ruleSynSigmaOne, \ruleSynSigmaTwo, \ruleSynHypo} \,\}$}
        \If{\Call{IsSynthesisRule}{\vRuleInst}}
          \Assign{\Assumpts}{\Call{\Assumptions}{\vRuleInst}}
          \Comment{unreachable since $\vRuleInst$ is not a synthesis rule}
        \EndIf

        \CommentLnColor{blue}{$\{\,
          \ldots
            \wedge (\ProofMode \,{=}\, \NoSynLemma)
            \wedge \neg\hasUnk{F_1 {\entails} F_2}
            \wedge \validlem{\Lemmas}
            \wedge \LemmasSyn = \setempty
            \wedge \Assumpts = \setempty
            \wedge \assumptions{\vRuleInst} = \setempty
            \wedge \sound{\vRuleInst} \,\}$}

        \If{\Call{IsAxiomRule}{\vRuleInst}} %
        \Comment{$\validlem{\setempty} {=} \true$ and
          $\validindt{\setempty} {=} \true$}
        \CommentLnIndentColor{blue}{$\{\,
          \ldots
            \wedge \LemmasSyn \,{=}\, \setempty
            \wedge \Assumpts \,{=}\, \setempty
            \wedge \assumptions{\vRuleInst} \,{=}\, \setempty
            \wedge \premises{\vRuleInst} \,{=}\, \setempty
            \wedge \validindt{\Hypos, \Lemmas, F_1 {\entails} F_2} \,\}$}

          \State \Assign{\xi}{\Call{\CreateWitnessProofTree}{
              F_1 \entails F_2, \vRuleInst, \setempty}}
          \State \ReturnTuple{\ValidWitness{\xi}, \LemmasSyn, \Assumpts}
          \CommentLnColor{blue}{$\{\,
            \ldots
              \wedge \LemmasSyn \,{=}\, \setempty
              \wedge \Assumpts \,{=}\, \setempty
              \wedge \validindt{\Hypos, \Lemmas, F_1 {\entails} F_2})
              \wedge \res \,{=}\, (\ValidWitness{\xi}, \LemmasSyn, \Assumpts)
              \,\}$}
          \CommentLnColor{blue}{$\longrightarrow \{\,\exists\,\xi.\,
                (\res \,{=}\, (\ValidWitness{\xi}, \setempty, \setempty)
                \wedge \validindt{\Hypos, \Lemmas, F_1 {\entails} F_2}) \,\}$}
        \EndIf{}

        \State \Assign{
          (\Hypos_i, \Lemmas_i, F_{1i} \entails F_{2i})_{i\,=\,1 \ldots n}}{
          \Call{\Premises}{\vRuleInst}}
        \CommentLnColor{blue}{$\{\,
            \ldots
            \wedge (\ProofMode \,{=}\, \NoSynLemma)
            \wedge \validlem{\Lemmas}
            \wedge \LemmasSyn = \setempty
            \wedge \Assumpts = \setempty
            \wedge \neg\hasUnk{F_{1} {\entails} F_{2}}
            \wedge \bigwedge_{i{=}1}^n (\Lemmas_i = \Lemmas)$}
        \CommentLnColor{blue}{$\qquad
          \wedge\, \assumptions{\vRuleInst} = \setempty
          \wedge \premises{\vRuleInst} =
            \{\Hypos_i, \Lemmas_i, F_{1i} \entails F_{2i}\}_{i{=}1}^n
          \wedge \sound{\vRuleInst} \,\}$}
        \CommentLnColor{blue}{$\longrightarrow$}
        \Comment{since
            $\neg\hasUnk{F_{1} {\entails} F_{2}}
            \rightarrow
            \bigwedge_{i{=}1}^n (\neg\hasUnk{F_{1i} {\entails} F_{2i}}) $}
        \CommentLnColor{blue}{$\{\,
          (\ProofMode \,{=}\, \NoSynLemma)
            \wedge \validlem{\Lemmas}
            \wedge \LemmasSyn = \setempty
            \wedge \Assumpts = \setempty
            \wedge \bigwedge_{i{=}1}^n
              (\neg\hasUnk{F_{1i} {\entails} F_{2i}}
              \wedge \Lemmas_i = \Lemmas)$}
        \CommentLnColor{blue}{$\qquad
          \wedge\, \assumptions{\vRuleInst} = \setempty
          \wedge \premises{\vRuleInst} =
            \{\Hypos_i, \Lemmas_i, F_{1i} \entails F_{2i}\}_{i{=}1}^n
          \wedge \sound{\vRuleInst} \,\}$}

        \State \Assign{(\vR_i,
          \Lemmas^i_{\mathrm{syn}},\Assumpts_i)_{i\,=\,1 \ldots n}}{
          \Call{\Prove}{
            \Hypos_i,\, \Lemmas \cup \LemmasSyn \cup \Lemmas_i,\,F_{1i} \entails
            F_{2i}, \ProofMode}_{i\,=\,1 \ldots n}}
        \CommentLnColor{blue}{$\{\,
          \ldots
          \wedge \LemmasSyn = \setempty
          \wedge \Assumpts = \setempty
          \wedge (\bigwedge_{i{=}1}^n
            \validindt{\Hypos_i, \Lemmas, F_{1i} \entails F_{2i}}
            \rightarrow \validindt{\Hypos, \Lemmas, F_1 {\entails} F_2})
          \,\wedge$}
        \CommentLnColor{blue}{$\quad
          \bigwedge_{i{=}1}^n
            (((\vR_i \,{=}\, \Unknown) ~\vee~
            (\exists \xi_i.\,
              (\vR_i \,{=}\, \ValidWitness{\xi_i}
              \wedge \validindt{\Hypos_i, \Lemmas, F_{1i} {\entails} F_{2i}})))
              \wedge \Lemmas^i_{\mathrm{syn}} {=} \setempty
              \wedge \Assumpts_i {=} \setempty) \,\}$}
        \CommentLnColor{blue}{$\longrightarrow$}
        \CommentLnColor{blue}{$\{\,
          \LemmasSyn = \setempty
          \wedge \Assumpts = \setempty
          \wedge (\bigwedge_{i{=}1}^n
            \validindt{\Hypos_i, \Lemmas, F_{1i} \entails F_{2i}}
            \rightarrow \validindt{\Hypos, \Lemmas, F_1 {\entails} F_2})
          \,\wedge$}
        \CommentLnColor{blue}{$\quad
          \bigwedge_{i{=}1}^n
            (((\vR_i \,{=}\, \Unknown) ~\vee~
            (\exists \xi_i.\,
              (\vR_i \,{=}\, \ValidWitness{\xi_i}
              \wedge \validindt{\Hypos_i, \Lemmas, F_{1i} {\entails} F_{2i}})))
              \wedge \Lemmas^i_{\mathrm{syn}} {=} \setempty
              \wedge \Assumpts_i {=} \setempty) \,\}$}
				\algstore{myalg}
  	\end{algorithmic}
  \end{algorithm}
  \vspace{-0.5em}
  \caption{Verification of the proof search procedure $\Prove$
    on the second specification (Part 1)}
  \label{fig:ProofSearchHoare2a}
\end{figure}

\vspace{-0.5em}
\begin{figure}[H]
  \algrenewcommand\algorithmicindent{1em}
  \begin{verify}
    \small
    \begin{algorithmic}[1]
			\algrestore{myalg}
        \CommentLnColor{blue}{$\{\,
          \LemmasSyn = \setempty
          \wedge \Assumpts = \setempty
          \wedge (\bigwedge_{i{=}1}^n
            \validindt{\Hypos_i, \Lemmas, F_{1i} \entails F_{2i}}
            \rightarrow \validindt{\Hypos, \Lemmas, F_1 {\entails} F_2})
          \,\wedge$}
        \CommentLnColor{blue}{$\quad
          \bigwedge_{i{=}1}^n
            (((\vR_i \,{=}\, \Unknown) ~\vee~
            (\exists \xi_i.\,
              (\vR_i \,{=}\, \ValidWitness{\xi_i}
              \wedge \validindt{\Hypos_i, \Lemmas, F_{1i} {\entails} F_{2i}})))
              \wedge \Lemmas^i_{\mathrm{syn}} {=} \setempty
              \wedge \Assumpts_i {=} \setempty) \,\}$}

        \If{\vR_i = \ValidWitness{\xi_i}\Keyword{for~all~}i = 1 \ldots n}
        \CommentLnIndentColor{blue}{$\{\,
          \LemmasSyn = \setempty
          \wedge \Assumpts = \setempty
          \wedge (\bigwedge_{i{=}1}^n
            \validindt{\Hypos_i, \Lemmas, F_{1i} \entails F_{2i}}
            \rightarrow \validindt{\Hypos, \Lemmas, F_1 {\entails} F_2})
          \,\wedge$}
        \CommentLnIndentColor{blue}{$\quad
          \bigwedge_{i{=}1}^n (
              \vR_i \,{=}\, \ValidWitness{\xi_i}
              \wedge \validindt{\Hypos_i, \Lemmas, F_{1i} {\entails} F_{2i}}
              \wedge \Lemmas^i_{\mathrm{syn}} {=} \setempty
              \wedge \Assumpts_i {=} \setempty) \,\}$}

          \State \Assign{\xi}{\Call{\CreateWitnessProofTree}{
              F_1 \entails F_2, \vRuleInst, \setenum{\xi_1, \ldots,
                \xi_n}}}

          \State \ReturnTuple{\ValidWitness{\xi},~
            \LemmasSyn \cup \Lemmas^1_{\mathrm{syn}} \cup \ldots
            \cup \Lemmas^n_{\mathrm{syn}},~
            \Assumpts \cup \Assumpts_1 \cup \ldots \cup \Assumpts_n}

          \CommentLnColor{blue}{$\{\,
            \ldots
            \wedge \res \,{=}\, (\ValidWitness{\xi}, \setempty, \setempty)
            \wedge \bigwedge_{i{=}1}^n
              \validindt{\Hypos_i, \Lemmas, F_{1i} {\entails} F_{2i}}
            \,\wedge$}
          \CommentLnColor{blue}{$\quad
             (\bigwedge_{i{=}1}^n
              \validindt{\Hypos_i, \Lemmas, F_{1i} \entails F_{2i}}
              \rightarrow \validindt{\Hypos, \Lemmas, F_1 {\entails} F_2})
            \,\}$}
          \CommentLnColor{blue}{$\longrightarrow \{\,
            (\res \,{=}\, (\ValidWitness{\xi}, \setempty, \setempty)
            \wedge \validindt{\Hypos, \Lemmas, F_{1} {\entails} F_{2}})) \,\}$}

        \EndIf

      \EndFor{}

      \State{\ReturnTuple{\Unknown, \setempty, \setempty}
      \CommentLnColor{blue}{$
            \{\, \res \,{=}\, (\Unknown, \setempty, \setempty)
            \,\}$}}
  	\end{algorithmic}
  \end{verify}
  \vspace{-0.5em}
  \caption{Verification of the proof search procedure $\Prove$
    on the second specification (Part 2)}
  \label{fig:ProofSearchHoare2}
\end{figure}

\newpage
\subsection{Verification of \emph{the third specification} of
  the proof search procedure in Figure \ref{fig:ProofSearch}}

\begin{figure}[H]
  \algrenewcommand\algorithmicindent{1em}

  \begin{algorithm}[H]
    \small
    \caption*{\small{\bf Procedure} \Call{\Prove}{
        \Hypos, \Lemmas, F_1 \entails F_2, \ProofMode}}
    \begin{flushleft}
      \Summary{Requires:}
        $(\ProofMode \,{=}\, \SynLemma)
        \wedge \neg\hasUnk{F_1 {\entails} F_2}
        \wedge \validlem{\Lemmas}$\\
      \Summary{Ensures:} $(\res \,{=}\, (\Unknown, \setempty, \setempty))
        \vee
        \exists\,\xi,\LemmasSyn.
        (\res \,{=}\, (\ValidWitness{\xi}, \LemmasSyn, \setempty)
        \wedge \validlem{\LemmasSyn}
        \wedge \validindt{\Hypos, \Lemmas, F_1 {\entails} F_2})$\\
      \Summary{Invariant:} $\sound{\sRule}$\\[0.2em]
    \end{flushleft}
    \begin{algorithmic}[1]
      \def\indent{\hspace{1.2em}}

      \CommentLnColor{blue}{$\{\,
        (\ProofMode \,{=}\, \SynLemma)
          \wedge \neg\hasUnk{F_1 {\entails} F_2}
          \wedge \validlem{\Lemmas}
          \wedge \sound{\sRule} \,\}$}
      \State \Assign{\sRuleSelected}{\setenum{
          \Call{\Unify}{\vRule, (\Hypos, \Lemmas, F_1 \entails F_2)}
          ~|~ \vRule \in \sRule}}

      \CommentLnColor{blue}{$\{\,
        (\ProofMode \,{=}\, \SynLemma)
          \wedge \neg\hasUnk{F_1 {\entails} F_2}
          \wedge \validlem{\Lemmas}
          \wedge \sound{\sRuleSelected}$}
      \Comment{$\sound{\sRule} \wedge \sRuleSelected \subseteq \sRule
              \rightarrow \sound{\sRuleSelected}$}
      \CommentLnColor{blue}{$\qquad
      	\wedge\, \sRuleSelected \cap
        	\setenum{\ruleSynPiOne, \ruleSynPiTwo,
          	\ruleSynSigmaOne, \ruleSynSigmaTwo, \ruleSynHypo}
          = \setempty \,\}$}
      \Comment{since $\neg\hasUnk{F_1 {\entails} F_2}$}

      \State \Assign{\LemmasSyn}{\setempty}
      \CommentLnColor{blue}{$\{\,
        (\ProofMode \,{=}\, \SynLemma)
          \wedge \neg\hasUnk{F_1 {\entails} F_2}
          \wedge \validlem{\Lemmas}
          \wedge \LemmasSyn = \setempty$}
      \CommentLnColor{blue}{$\qquad
      		\wedge\, \sound{\sRuleSelected}
          \wedge \sRuleSelected \cap
          	\setenum{\ruleSynPiOne, \ruleSynPiTwo,
          	\ruleSynSigmaOne, \ruleSynSigmaTwo, \ruleSynHypo}
          	= \setempty \,\}$}

      \If{\ProofMode = \SynLemma \Keyword{and}
          \Call{\NeedLemmas}{F_1 \entails F_2, \sRuleSelected}}

        \State \Assign{\LemmasSyn}{
          \Call{\SynthesizeLemma}{\Lemmas,\,F_1 \entails F_2}}

        \State \Assign{\sRuleSelected}{
          \sRuleSelected \cup \setenum{\Call{\Unify}{
              \vRule, (\Hypos, \LemmasSyn, F_1 \entails F_2)}
            ~|~ \vRule \in \{\ruleLemmaLeft, \ruleLemmaRight\}}}

      \EndIf

      \CommentLnColor{blue}{$\{\,
        (\ProofMode \,{=}\, \SynLemma)
          \wedge \neg\hasUnk{F_1 {\entails} F_2}
          \wedge \validlem{\Lemmas}
          \wedge \validlem{\LemmasSyn}$}
      \Comment{w.r.t. the spec of $\SynthesizeLemma$}
      \CommentLnColor{blue}{$\qquad
      		\wedge\, \sound{\sRuleSelected}
          \wedge \sRuleSelected \cap
          	\setenum{\ruleSynPiOne, \ruleSynPiTwo,
          	\ruleSynSigmaOne, \ruleSynSigmaTwo, \ruleSynHypo}
          	= \setempty \,\}$}

      \For{\Keyword{each} \vRuleInst \Keyword{in} \sRuleSelected}
      \CommentLnIndentColor{blue}{$\{\,
        (\ProofMode \,{=}\, \SynLemma)
          \wedge \neg\hasUnk{F_1 {\entails} F_2}
          \wedge \validlem{\Lemmas}
          \wedge \validlem{\LemmasSyn}$}
       \CommentLnIndentColor{blue}{$\qquad
          \wedge\, \sound{\vRuleInst}
          \wedge \vRuleInst \not\in \setenum{
          	\ruleSynPiOne, \ruleSynPiTwo,
            \ruleSynSigmaOne, \ruleSynSigmaTwo, \ruleSynHypo}
     			\,\}$}
     	 \Comment{since $\vRuleInst \in \sRuleSelected$}

        \State \Assign{\Assumpts}{\setempty}

        \CommentLnColor{blue}{$\{\,
          (\ProofMode \,{=}\, \SynLemma)
            \wedge \neg\hasUnk{F_1 {\entails} F_2}
            \wedge \validlem{\Lemmas}
            \wedge \validlem{\LemmasSyn}
            \wedge \Assumpts {=} \setempty$}
       \CommentLnColor{blue}{$\qquad
       		\wedge\, \sound{\vRuleInst}
       		\wedge \vRuleInst \not\in \setenum{
       			\ruleSynPiOne, \ruleSynPiTwo,
       		  \ruleSynSigmaOne, \ruleSynSigmaTwo, \ruleSynHypo}
          \,\}$}
        \If{\Call{IsSynthesisRule}{\vRuleInst}}
          \Assign{\Assumpts}{\Call{\Assumptions}{\vRuleInst}}
          \Comment{unreachable since $\vRuleInst$ is not a synthesis rule}
        \EndIf
        \CommentLnColor{blue}{$\{\,
          \ldots
            \wedge (\ProofMode \,{=}\, \SynLemma)
            \wedge \neg\hasUnk{F_1 {\entails} F_2}
            \wedge \validlem{\Lemmas}
            \wedge \validlem{\LemmasSyn}
            \wedge \Assumpts {=} \setempty
            \wedge \assumptions{\vRuleInst} {=} \setempty
            \wedge \sound{\vRuleInst} \,\}$}

        \If{\Call{IsAxiomRule}{\vRuleInst}} %
        \Comment{$\validlem{\setempty} {=} \true$ and
          $\validindt{\setempty} {=} \true$}
          \CommentLnColor{blue}{$\{\,
            \ldots
            \wedge \validlem{\LemmasSyn}
            \wedge \Assumpts = \setempty
            \wedge \assumptions{\vRuleInst} = \setempty
            \wedge \premises{\vRuleInst} = \setempty
            \wedge \validindt{\Hypos, \Lemmas, F_1 {\entails} F_2} \,\}$}

          \State \Assign{\xi}{\Call{\CreateWitnessProofTree}{
              F_1 \entails F_2, \vRuleInst, \setempty}}
          \State \ReturnTuple{\ValidWitness{\xi}, \LemmasSyn, \Assumpts}

          \CommentLnColor{blue}{$\{\,
            \ldots
            \wedge \validlem{\LemmasSyn}
            \wedge \Assumpts = \setempty
            \wedge \validindt{\Hypos, \Lemmas, F_1 {\entails} F_2}
            \wedge \res \,{=}\, (\ValidWitness{\xi}, \LemmasSyn, \Assumpts)
            \,\}$}

          \CommentLnColor{blue}{$\longrightarrow \{\, \exists\,\xi.\,
                (\res \,{=}\, (\ValidWitness{\xi}, \LemmasSyn, \setempty)
                \wedge \validlem{\LemmasSyn}
                \wedge \validindt{\Hypos, \Lemmas, F_1 {\entails} F_2}) \,\}$}
        \EndIf{}

        \State \Assign{
          (\Hypos_i, \Lemmas_i, F_{1i} \entails F_{2i})_{i\,=\,1 \ldots n}}{
          \Call{\Premises}{\vRuleInst}}
        \CommentLnColor{blue}{$\{\,
          \ldots
            \wedge (\ProofMode \,{=}\, \SynLemma)
            \wedge \validlem{\Lemmas}
            \wedge \validlem{\LemmasSyn}
            \wedge \Assumpts = \setempty
            \wedge \neg\hasUnk{F_{1} {\entails} F_{2}}
            \wedge \bigwedge_{i{=}1}^n (\Lemmas_i = \Lemmas)$}
        \CommentLnColor{blue}{$\qquad
          \wedge\, \assumptions{\vRuleInst} = \setempty
          \wedge \premises{\vRuleInst} =
            \{\Hypos_i, \Lemmas_i, F_{1i} \entails F_{2i}\}_{i{=}1}^n
          \wedge \sound{\vRuleInst} \,\}$}
        \CommentLnColor{blue}{$\longrightarrow$}
        \Comment{since
            $\neg\hasUnk{F_{1} {\entails} F_{2}}
            \rightarrow
            \bigwedge_{i{=}1}^n (\neg\hasUnk{F_{1i} {\entails} F_{2i}}) $}
        \CommentLnColor{blue}{$\{\,
            (\ProofMode \,{=}\, \SynLemma)
            \wedge \validlem{\Lemmas}
            \wedge \validlem{\LemmasSyn}
            \wedge \Assumpts = \setempty
            \wedge \bigwedge_{i{=}1}^n
              (\neg\hasUnk{F_{1i} {\entails} F_{2i}}
              \wedge \Lemmas_i = \Lemmas)$}
        \CommentLnColor{blue}{$\qquad
          \wedge\, \assumptions{\vRuleInst} = \setempty
          \wedge \premises{\vRuleInst} =
            \{\Hypos_i, \Lemmas_i, F_{1i} \entails F_{2i}\}_{i{=}1}^n
          \wedge \sound{\vRuleInst}\,\}$}
        \algstore{myalg}
    \end{algorithmic}
  \end{algorithm}
  \vspace{-1em}
  \caption{Verification of the proof search procedure $\Prove$
    on the third specification (Part 1)}
  \label{fig:ProofSearchHoare3a}
\end{figure}

\begin{figure}[H]
  \algrenewcommand\algorithmicindent{1em}
  \begin{verify}
    \small
    \begin{algorithmic}[1]
        \algrestore{myalg}
        \CommentLnColor{blue}{$\{\,
            (\ProofMode \,{=}\, \SynLemma)
            \wedge \validlem{\Lemmas}
            \wedge \validlem{\LemmasSyn}
            \wedge \Assumpts = \setempty
            \wedge \bigwedge_{i{=}1}^n
              (\neg\hasUnk{F_{1i} {\entails} F_{2i}}
              \wedge \Lemmas_i = \Lemmas)$}
        \CommentLnColor{blue}{$\qquad
          \wedge\, \assumptions{\vRuleInst} = \setempty
          \wedge \premises{\vRuleInst} =
            \{\Hypos_i, \Lemmas_i, F_{1i} \entails F_{2i}\}_{i{=}1}^n
          \wedge \sound{\vRuleInst}\,\}$}
        \State \Assign{(\vR_i,
          \Lemmas^i_{\mathrm{syn}},\Assumpts_i)_{i\,=\,1 \ldots n}}{
          \Call{\Prove}{
            \Hypos_i,\, \Lemmas \cup \LemmasSyn \cup \Lemmas_i,\,F_{1i} \entails
            F_{2i}, \ProofMode}_{i\,=\,1 \ldots n}}
        \CommentLnColor{blue}{$\{\,
          \ldots
          \wedge \validlem{\LemmasSyn}
          \wedge \Assumpts = \setempty
          \wedge (\bigwedge_{i{=}1}^n
            \validindt{\Hypos_i, \Lemmas, F_{1i} \entails F_{2i}}
            \rightarrow \validindt{\Hypos, \Lemmas, F_1 {\entails} F_2})
          \,\wedge$}
        \CommentLnColor{blue}{$\qquad
          \bigwedge_{i{=}1}^n
            ((\vR_i \,{=}\, \Unknown
                \wedge \Lemmas^i_{\mathrm{syn}} = \setempty
                \wedge \Assumpts_i = \setempty)$}
        \CommentLnColor{blue}{$\qquad \qquad
          \vee~ (\exists \xi_i.\,
              (\vR_i \,{=}\, \ValidWitness{\xi_i}
              \wedge \validindt{\Hypos_i, \Lemmas, F_{1i} {\entails} F_{2i}}
              \wedge \validlem{\Lemmas^i_{\mathrm{syn}}}
              \wedge \Assumpts_i = \setempty))) \,\}$}
				\CommentLnColor{blue}{$\longrightarrow$}
        \CommentLnColor{blue}{$\{\,
          \validlem{\LemmasSyn}
          \wedge \Assumpts = \setempty
          \wedge (\bigwedge_{i{=}1}^n
            \validindt{\Hypos_i, \Lemmas, F_{1i} \entails F_{2i}}
            \rightarrow \validindt{\Hypos, \Lemmas, F_1 {\entails} F_2})
          \,\wedge$}
        \CommentLnColor{blue}{$\qquad
          \bigwedge_{i{=}1}^n
            ((\vR_i \,{=}\, \Unknown
                \wedge \Lemmas^i_{\mathrm{syn}} = \setempty
                \wedge \Assumpts_i = \setempty)$}
        \CommentLnColor{blue}{$\qquad \qquad
          \vee~ (\exists \xi_i.\,
              (\vR_i \,{=}\, \ValidWitness{\xi_i}
              \wedge \validindt{\Hypos_i, \Lemmas, F_{1i} {\entails} F_{2i}}
              \wedge \validlem{\Lemmas^i_{\mathrm{syn}}}
              \wedge \Assumpts_i = \setempty))) \,\}$}
        \CommentLnColor{blue}{$\{\,
          \validlem{\LemmasSyn}
          \wedge \Assumpts = \setempty
          \wedge (\bigwedge_{i{=}1}^n
            \validindt{\Hypos_i, \Lemmas, F_{1i} \entails F_{2i}}
            \rightarrow \validindt{\Hypos, \Lemmas, F_1 {\entails} F_2})
          \,\wedge$}
        \CommentLnColor{blue}{$\qquad
          \bigwedge_{i{=}1}^n
            ((\vR_i \,{=}\, \Unknown
                \wedge \Lemmas^i_{\mathrm{syn}} = \setempty
                \wedge \Assumpts_i = \setempty)$}
        \If{\vR_i = \ValidWitness{\xi_i}\Keyword{for~all~}i = 1 \ldots n}
        \CommentLnColor{blue}{$\{\,
          \validlem{\LemmasSyn}
          \wedge \Assumpts = \setempty
          \wedge (\bigwedge_{i{=}1}^n
            \validindt{\Hypos_i, \Lemmas, F_{1i} \entails F_{2i}}
            \rightarrow \validindt{\Hypos, \Lemmas, F_1 {\entails} F_2})
          \,\wedge$}

        \CommentLnColor{blue}{$\qquad
          \bigwedge_{i{=}1}^n
            (\vR_i \,{=}\, \ValidWitness{\xi_i}
              \wedge \validlem{\Lemmas^i_{\mathrm{syn}}}
              \wedge \Assumpts_i = \setempty
              \wedge \validindt{\Hypos_i, \Lemmas, F_{1i} {\entails} F_{2i}})
              \,\}$}

          \State \Assign{\xi}{\Call{\CreateWitnessProofTree}{
              F_1 \entails F_2, \vRuleInst, \setenum{\xi_1, \ldots,
                \xi_n}}}

          \State \ReturnTuple{\ValidWitness{\xi},~
            \LemmasSyn \cup \Lemmas^1_{\mathrm{syn}} \cup \ldots
            \cup \Lemmas^n_{\mathrm{syn}},~
            \Assumpts \cup \Assumpts_1 \cup \ldots \cup \Assumpts_n}

          \CommentLnColor{blue}{$\{\,
            \ldots
            \wedge \res \,{=}\, (\ValidWitness{\xi}, \LemmasSyn \cup
                \Lemmas^1_{\mathrm{syn}} \cup \ldots \cup
                \Lemmas^n_{\mathrm{syn}}, \setempty)
             \wedge \validlem{\LemmasSyn}
             \wedge \bigwedge_{i{=}1}^n \validlem{\Lemmas^i_{\mathrm{syn}}}
             \,\wedge$}
          \CommentLnColor{blue}{$\qquad
            \bigwedge_{i{=}1}^n
              \validindt{\Hypos_i, \Lemmas, F_{1i} {\entails} F_{2i}} \,\}
          \wedge (\bigwedge_{i{=}1}^n
            \validindt{\Hypos_i, \Lemmas, F_{1i} {\entails} F_{2i}}
            {\rightarrow} \validindt{\Hypos, \Lemmas, F_1 {\entails} F_2}) \,$}
          \CommentLnColor{blue}{$\longrightarrow \{\,
            \res \,{=}\, (\ValidWitness{\xi}, \LemmasSyn {\cup}
                \Lemmas^1_{\mathrm{syn}} \cup \ldots \cup
                \Lemmas^n_{\mathrm{syn}}, \setempty)$}
          \CommentLnColor{blue}{$\qquad
            \wedge\, \validlem{\LemmasSyn \cup
                \Lemmas^1_{\mathrm{syn}} \cup \ldots \cup
                \Lemmas^n_{\mathrm{syn}}}
            \wedge \validindt{\Hypos{,} \Lemmas{,} F_{1} {\entails} F_{2}}
            \,\}$}

        \EndIf

      \EndFor{}

      \State{\ReturnTuple{\Unknown, \setempty, \setempty}}
      \CommentLnColor{blue}{$
            \{\, \res \,{=}\, (\Unknown, \setempty, \setempty)
            \,\}$}

    \end{algorithmic}
  \end{verify}
  \vspace{-0.5em}
  \caption{Verification of the proof search procedure $\Prove$
    on the third specification (Part 2)}
  \label{fig:ProofSearchHoare3}
\end{figure}

\fi

\end{document}